\documentclass[12pt]{article}%
\usepackage{cite}
\usepackage{graphicx}
\usepackage{multicol}
\usepackage{amsfonts}
\usepackage{amssymb}
\usepackage{amsmath}
\usepackage{heck}
\usepackage{setspace}
\usepackage{verbatim}
\usepackage{color}
\usepackage{epsfig}
\usepackage[margin=1in]{geometry}%
\providecommand{\U}[1]{\protect\rule{.1in}{.1in}}
\numberwithin{equation}{section}

\newcommand{\ba}{\begin{eqnarray}}
\newcommand{\ea}{\end{eqnarray}}

\newcommand{\cF}{\mathcal{F}}

\newcommand{\cH}{\mathcal{H}}

\newcommand{\cM}{\mathcal M}

\newcommand{\cO}{\mathcal{O}}

\newcommand{\cT}{\mathcal{T}}
\newcommand{\cV}{\mathcal{V}}

\newcommand\ZZ{\mathbf{Z}}

\newcommand\Gr{\mathrm{Gr}}
\newcommand\im{\mathrm{Im}}

\begin{document}

\date{October 2013}

\title{T-Branes and Geometry}

\institution{VTECH}{\centerline{${}^{1}$Department of Physics, Virginia Tech, Blacksburg, VA 24061, USA}}

\institution{HARVARD}{\centerline{${}^{2}$Jefferson Physical Laboratory, Harvard University, Cambridge, MA 02138, USA}}

\institution{UIUC}{\centerline{${}^{3}$Department of Mathematics, University of Illinois, Urbana, IL 61801, USA}}

\authors{Lara B. Anderson\worksat{\VTECH, \HARVARD}\footnote{e-mail: {\tt lara.anderson@vt.edu}},
Jonathan J. Heckman\worksat{\HARVARD}\footnote{e-mail: {\tt jheckman@physics.harvard.edu}}, and
Sheldon Katz \worksat{\UIUC}\footnote{e-mail: {\tt katz@math.uiuc.edu}}}

\abstract{
T-branes are a non-abelian generalization
of intersecting branes in which the matrix of normal deformations is nilpotent
along some subspace. In this paper we study the geometric remnant of this open string data
for six-dimensional F-theory vacua. We show that in the dual M-theory / IIA
compactification on a \textit{smooth} Calabi-Yau threefold $X_{\text{smth}}$, the geometric
remnant of T-brane data translates to periods of the three-form potential valued in the
intermediate Jacobian of $X_{\text{smth}}$. Starting from a smoothing of a singular Calabi-Yau,
we show how to track this data in singular limits using the theory of limiting mixed Hodge structures,
which in turn directly points to an emergent Hitchin-like system coupled to defects. We argue that the
physical data of an F-theory compactification on a singular threefold involves
specifying both a geometry as well as the remnant of three-form potential moduli and
flux which is localized on the discriminant. We give examples of T-branes
in compact F-theory models with heterotic duals, and comment on the extension of our results
to four-dimensional vacua.}

\maketitle

\tableofcontents

\enlargethispage{\baselineskip}

\setcounter{tocdepth}{2}

\newpage

\section{Introduction}

Recently there has been renewed interest in F-theory, both as a starting point for
model building efforts, as well as for addressing more conceptual questions connected with
non-perturbative phenomena in string compactifications. In F-theory,
the IIB\ axio-dilaton is interpreted as a modulus of a
twelve-dimensional geometry, thus providing a
geometric characterization of vacua with order one string coupling as well as
non-perturbative bound states of seven-branes \cite{VafaFTHEORY}.

From a model building standpoint, an attractive feature of
F-theory is that it can simultaneously accommodate the
flexibility of intersecting brane constructions
with the promising GUT\ phenomenology of exceptional gauge symmetries.
In a suitable decoupling limit, the task of constructing realistic string
vacua in F-theory reduces to the specification of intersection patterns for
seven-branes. In this way, GUT\ model building \cite{BHVI, BHVII, DWI,
DWII}, flavor physics \cite{HVCKM, Font:2008id, BHSV, Randall:2009dw, HVCP,
Dudas:2009hu, FGUTSNC, Marchesano:2009rz, Font:2012wq, Pawelczyk:2013tza, Font:2013ida} and
various string-motivated scenarios for physics beyond the Standard Model can
all be accommodated in a single local formulation. For recent reviews on
F-theory model building, see e.g. \cite{HVLHC, Heckman:2010bq, Weigand:2010wm,
Maharana:2012tu, Wijnholt:2012fx}.

This allows a division of labor in building up phenomenologically viable models.
The basic outline in this programme (c.f. \cite{Berenstein:2001nk,
Antoniadis:2002qm, UrangaBottomUp, VerlindeWijnholtBottomUp, BHVI, DWI, FUZZ})
is to first identify the local aspects of an intersecting seven-brane gauge theory necessary to realize
the gauge theoretic data of a field theory, i.e. an \textquotedblleft open
string sector\textquotedblright. Second, there is the recoupling to gravity,
i.e. the \textquotedblleft closed string sector\textquotedblright\ of the
model. For recent efforts in constructing global F-theory compactifications, see e.g.
\cite{DWII, DWIII, Collinucci:2008zs, Collinucci:2009uh, Marsano:2009gv, Blumenhagen:2009up, Blumenhagen:2009yv,
Grimm:2009yu, Marsano:2009wr, Marsano:2010ix, Marsano:2011hv, Esole:2011sm,
Esole:2011cn, Marsano:2012yc, Cvetic:2012ts, Marsano:2012bf, Donagi:2012ts, Braun:2013cb,
Braun:2013yti}.\footnote{As a brief aside
on the philosophy of the \textquotedblleft local to global\textquotedblright\ programme
of F-theory model building, we note that there could in principle be several global
completions of a local model, and these global completions may not even be
geometric. Rather, the global model serves more as a proof of concept, i.e. UV consistency. From this perspective,
the relevant question is how well one might expect to distinguish these different
choices of UV completion. For recent perspectives on this issue, see for example
\cite{Heckman:2013kza, Hebecker:2013eba}.}

A powerful tool in understanding the open string sector is the effective field
theory of a seven-brane coupled to defects \cite{BershadskyFOURD, BHVI, DWI}.
In this field theory, an adjoint-valued complex field $\Phi$ controls the
position of the stack of seven-branes in the local geometry normal to the brane. Of
particular significance for flavor physics models are configurations where $\Phi$ has position dependent
eigenvalues with branch cuts \cite{Hayashi:2009ge, Hayashi:2010zp, BHSV, EPOINT, TBRANES, glueI, glueII}. In
a holomorphic presentation of $\Phi$ i.e. without branch cuts, this means $[\Phi, \Phi^{\dag}] \neq 0$ (see \cite{TBRANES}).
The non-zero commutator means the seven-brane has puffed up to a dielectric nine-brane, and the eigenvalues
of $\Phi$ do not fully characterize the configuration. A T-brane is
any such configuration where $\Phi$ is nilpotent (i.e. upper or lower triangular as a matrix)
along some subspace of the worldvolume of the brane. Such nilpotent
Higgs fields in bound states of branes were first introduced in \cite{Donagi:2003hh}, and in the
context of F-theory in \cite{TBRANES} (see also \cite{Chiou:2011js, glueI, glueII, Font:2013ida}).

To move forward with the second stage of F-theory model building where gravity
is recoupled, it is necessary to match the data of the seven-brane
gauge theory back to the geometry of an F-theory compactification. However, since
T-brane data is not visible in holomorphic Casimir
invariants, it has remained an open and surprisingly basic question as to
how to identify its geometric remnant in global models.\footnote{See \cite{Marsano:2012bf,Braun:2013cb} for some recent work.}

In this paper we show how to identify the geometric remnants of T-branes. We
view the open string sector and closed string sectors as defining overlapping
patches for the full moduli space of an F-theory compactification. Our aim
will be to determine the \textquotedblleft transition
functions\textquotedblright\ which interpolate between these two coordinate
systems.\footnote{We view this as developing the dictionary entries
in a gauge / gravity correspondence involving gauge theory on stack(s) of seven-brane(s), and
the non-singular locus of an F-theory base manifold. This is in line with the interpretation
of local F-theory model building given in \cite{FUZZ}.}

Our focus in this paper will be on T-branes in six-dimensional F-theory vacua with
eight real supercharges.\footnote{In supersymmetric compactifications to eight dimensions,
T-branes do not exist because the seven-brane equation of motion reduces to $[\Phi,\Phi^{\dag}]=0$.}
In this case, the internal dynamics of the seven-brane gauge
theory is governed by a Hitchin-like system coupled to point-like defects.
These defects are often associated with localized matter fields, but can also reflect couplings to a
theory of tensionless strings.

To find the geometric remnants of T-brane data, we first show how to identify the data of the Hitchin
system with defects in geometric terms. Locally, we model this by a
curve of ADE\ singularities with possible higher order singularities
at some marked points. We view this local threefold $X$ as the limit of a family of smoothings
$X_{t}\rightarrow X$ so that as $t\rightarrow0$ we recover the singular space
$X$. These smoothings physically correspond to moving the stack of
seven-branes around to more general positions. We find that in the smoothing,
the remnant of T-brane data is captured by the intermediate Jacobian of
$X_{t}$:%
\begin{equation}
J\left(  X_{t}\right)  =H^{3}(X_{t},\mathbb{R})/H^{3}(X_{t}%
,\mathbb{Z})
\end{equation}
which fibers over the complex structure moduli $M_{\mathrm{cplx}}$. Indeed, after compactifying on a further
$T^2$, there is a dual description in terms of IIA supergravity. There, the intermediate Jacobian
and complex structure moduli (along with the universal axion and dilaton) combine
to form the hypermultiplet moduli space. The holomorphic Casimir
invariants of the Higgs field give coordinates on the complex structure moduli, while
the remnants of T-brane data corresponds in the dual IIA / M-theory\ picture to integrating
the three-form potential over three-cycles.

However, this description is inadequate in the singular limit $t\rightarrow0$, since it is not
even possible to speak of the classical intermediate Jacobian. It is nevertheless possible
to study the limiting behavior of these constructs using the theory of limiting mixed Hodge structures (LMHS),
which directly points to the Hitchin system coupled to defects. In other words, we are going to use the
theory of limiting mixed Hodge structures to construct the transition functions
between the ``open string patch'' and ``closed string patch''.

This begs the question: What is the defining data of a 6D F-theory model at large volume? The standard
procedure in much of the literature is to start with a singular Calabi-Yau
$X$, and to view it as the limit of either a blowdown $\widetilde
{X}\rightarrow X$, or as the limit of a smoothing $X_{t}\rightarrow X$. In
fact, our analysis shows that just specifying $X$ is ambiguous from a physics
standpoint: The T-brane data must also be included. On the smoothing side,
this is reflected in the three-form potential moduli valued in the intermediate Jacobian.

Turning to the characterization as a blowdown, we find that T-branes
\textit{obstruct} some K\"{a}hler resolutions, simply because the theory is
still at a non-trivial point of the Higgs branch. In the low energy effective theory, this is the statement that
the vevs of the three-form moduli continue to give a mass to states in the theory.

\begin{figure}
[t!]
\begin{center}
\includegraphics[
height=1.7521in,
width=3.7152in
]%
{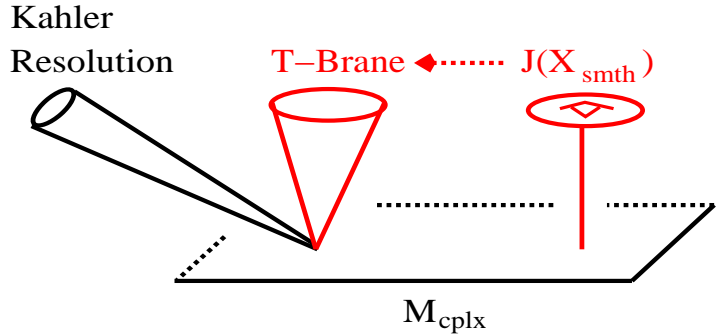}%
\caption{Depiction of the moduli space of F-theory in six dimensions. The intermediate Jacobian
$J(X_{\mathrm{smth}})$ fibers over the complex structure moduli $M_{\mathrm{cplx}}$. At singular points of the complex structure,
the classical intermediate Jacobian description breaks down and is replaced by an emergent Hitchin system which
captures the T-brane data. At the singular point in moduli space where both the complex structure and T-brane
data are switched off, one can instead perform a K\"{a}hler resolution of the geometry,
moving onto the Coulomb branch of the low energy theory. The two branches only meet at
singular loci in the moduli space.}%
\label{fig:tbrane}%
\end{center}
\end{figure}

To repair this ambiguity, we propose to supplement the definition of
compactification of F-theory on a singular threefold $X$ by appropriate
\textquotedblleft T-brane data\textquotedblright. This data consists of the
singular limit of an intermediate Jacobian, as well as abelian flux data of the
Hitchin system. In configurations which admit a K\"{a}hler resolution, this would translate to
a four-form flux in the dual M-theory description. Now, in compactifications to six dimensions on \textit{smooth}
Calabi-Yau threefolds, such fluxes are inconsistent with the supergravity equations of
motion. However, in singular limits, such fluxes \textit{can} be activated. In fact, it is known that for a
smooth Calabi-Yau, the three-form moduli, and flux data are naturally packaged in terms of a single object, the
Deligne cohomology $H_{\mathcal{D}}^{2}(X_{\text{smth}},\mathbb{Z}(2))$ \cite{Donagi:1998vw}. From this perspective, the emergent
Hitchin-like system gives a definition of this object in certain singular limits. See figure \ref{fig:tbrane} for a
depiction of the moduli space.

As a check of our proposal, we present examples of compact F-theory models which contain
T-branes. We focus on the specific case of F-theory compactified on a Hirzebruch base
$\mathbb{F}_{n}$ for $n \in \mathbb{Z}$ and $-12 \leq n \leq 12$.  These F-theory vacua have a dual
description in terms of the $E_{8}\times E_{8}$ heterotic string compactified
on a K3 surface with $(12+n,12-n)$ instantons in each vector bundle factor (see e.g. \cite{MorrisonVafaI,
MorrisonVafaII, BershadskyPLUS}).\footnote{When
we write $\mathbb{F}_n$ with $n<0$,  we mean $\mathbb{F}_{|n|}$ with the
sections of self-intersection $\pm n$ switched.}

Quite remarkably, some of the simplest heterotic string compactifications
are dual to T-branes in F-theory! For example, the singular local geometry:
\begin{equation}
y^{2}=x^{3}+g_{12 + n}(z^{\prime})z^{5}%
\end{equation}
is ambiguous, and can actually refer either to a theory with small instantons, or to a \textit{smooth}
vector bundle which happens to have a \textit{singular} spectral cover. An important
example of this type is the standard embedding of the spin connection in one of the $E_8$ factors. This is
a perfectly smooth vector bundle which has a singular spectral cover. This ambiguity
was noted in \cite{Aspinwall:1998he}, and was recently revisited in \cite{glueII}. More generally,
heterotic theory abounds with examples of smooth vector bundles with singular spectral covers \cite{Bershadsky:1997zv}
(see also \cite{Anderson:2007nc, Anderson:2008uw, Anderson:2009mh}). Our
plan will be to show how to identify the geometric remnants of T-brane data
in these and related situations.

As a final remark, we note that the main aim of our work is to carefully track the
behavior of the relevant geometric structures in singular limits. Indeed, though some aspects of our discussion, especially in
relation to spectral covers, are well-known for \textit{smooth} spectral covers
and their associated threefold duals in F-theory, comparatively far less is
known in singular limits.

The rest of this paper is organized as follows. First, in section
\ref{sec:REVIEW}, we review two complementary perspectives of F-theory vacua
based on a global \textquotedblleft closed string\textquotedblright%
\ description and the local \textquotedblleft open string\textquotedblright%
\ description. Next, in section \ref{sec:EFT}, we study the effective field
theory of a seven-brane using the associated Hitchin system with defects. In
section \ref{sec:REMNANT} we show how the limiting behavior of a local curve of singularities
directly points to an emergent Hitchin system, and in section \ref{sec:PRESCRIPTION}
we give a revised prescription for how to analyze the effective field
theory associated with F-theory on a singular threefold $X$. We provide
realizations of T-branes in compact F-theory models in
section \ref{sec:COMPACT}. We conclude in section \ref{sec:CONC}. Some
additional review and technical details are deferred to the Appendices.

\section{Global and Local Models in 6D\ F-theory \label{sec:REVIEW}}

In preparation for our later discussions, in this section we review two
complementary approaches to 6D\ F-theory compactification based on the local
geometry of a stack of seven-branes, and the closed string description of such vacua.

F-theory \cite{VafaFTHEORY} provides a non-perturbative generalization of type
IIB\ superstring theory in which the axio-dilaton:%
\begin{equation}
\tau=C_{0}+\frac{i}{g_{s}}%
\end{equation}
is now interpreted as the modular parameter of an auxiliary elliptic curve:%
\begin{equation}
y^{2}=x^{3}+ f x + g  \text{,}%
\end{equation}
where $f$ and $g$ can have non-trivial dependence on the positions of the ten
spacetime dimensions. In physical terms, the position dependence in the
axio-dilaton is induced by the presence of non-perturbative bound states of seven-branes.

Supersymmetric compactifications of F-theory to flat space can be obtained as
follows. We split up the ten dimensional spacetime of string theory as the
product $M_{10-2m}\times B$, where $M_{10-2m}$ designates the $10-2m$
uncompactified directions and $B$ is a compact manifold of complex
dimension $m$. This can be arranged by assuming that the elliptic fibration
only depends on $B$ so that the Weierstrass model defines an elliptic
fibration with section $\mathbb{E}\rightarrow X\rightarrow B$, where $X$ is a
Calabi-Yau $(m+1)$-fold. In this geometric description,
seven-branes are associated with the locations where the elliptic fibration
degenerates. This occurs when some of the roots of the cubic in $x$ collide,
i.e. at the vanishing of the discriminant:%
\begin{equation}
\Delta=4f^{3}+27g^{2}\text{.}%
\end{equation}
The components of the discriminant locus $(\Delta=0)\subset B$ define
complex codimension one subspaces inside $B$, which should be thought of
as submanifolds which are wrapped by seven-branes.  This defines the location of the ``open string sector''. The
``closed string string sector'' is associated with the complement, i.e. the non-singular parts of the elliptic fibration.
One of the remarkable features of F-theory is the close interplay between the open and closed string sectors.
Our primary interest in this paper will be the special case of compactifications to six dimensions, i.e.
$m=2$, with $X$ a Calabi-Yau threefold.

In the remainder of this section we describe two complementary descriptions of
such six-dimensional vacua. First, we consider the closed string description, focussing on the
M-theory / IIA dual compactifications on $X$ a smooth Calabi-Yau threefold. In the limit of
moduli where $X$ develops singularities, the supergravity description breaks
down and must be supplemented by additional light degrees of freedom. In local
patches of $B$ it is then fruitful to switch to a second description based
on the gauge theory of a seven-brane. It is in this second patch that we will
see the appearance of T-branes. Part of our aim in this work will be to
provide a patchwork of different coordinate systems for the moduli space, and
to then describe how these different patches fit together in overlapping
regions of validity.

\subsection{Closed String Description}

One way to characterize some compactifications of F-theory is in terms of a
dual description in M-theory. The lift of T-duality between IIA\ and
IIB\ vacua then becomes a statement about duality between M-theory on a smooth
Calabi-Yau threefold $X_{\text{smth}}$ and F-theory compactified on
$S^{1}\times X$. \ A further circle compactification then relates IIA string
theory compactified on $X_{\text{smth}}$ to F-theory compactified on
$T^{2}\times X$. In both the IIA\ and M-theory descriptions, we reach the
F-theory limit by shrinking the volume of the elliptic fiber to zero size.

The vacua of IIA compactified on a Calabi-Yau threefold has been extensively studied, so we shall
be brief. For additional review, see for example the review \cite{Alexandrov:2013yva}
and references therein. The effective theory in four dimensions has eight real supercharges,
i.e. $\mathcal{N}=2$ supersymmetry. In the IIA supergravity, we get gauge
fields from integrating the three-form potential over the $h^{1,1}%
(X_{\text{smth}})$ independent two-cycles. The vector multiplet moduli are
controlled by the special geometry of the complexified K\"{a}hler moduli.
There are $(1/2)h^{3}(X)$ hypermultiplet moduli, i.e. $2h^{3}(X)$ real degrees of
freedom. Working about a fixed choice of holomorphic three-form $\Omega$,
there are $h^{2,1}(X_{\text{smth}})$ complex structure moduli, and two
additional real moduli from the dilaton and the universal axion (from
dualization of $B_{\mu\nu}$ in the $\mathbb{R}^{3,1}$ directions). The other
$h^{3}(X)$ real degrees of freedom come from periods of the
Ramond-Ramond three-form potential over the A- and B-cycles of $X_{\text{smth}%
}$. Since the dilaton sits in a hypermultiplet, the geometry of the
hypermultiplet moduli space will generically receive various perturbative and
non-perturbative corrections in string theory.

For this paper, our main focus will be on the geometric moduli space specified by the
intermediate Jacobian $J(X_{\mathrm{smth}})$, and its fibration
over $M_{\mathrm{cplx}}$, the complex structure moduli. The three-form potential moduli take values in the
fiber $J(X_{\mathrm{smth}})$, i.e. the intermediate Jacobian:
\begin{equation}
J(X_{\mathrm{smth}})=H^{3}(X_{\text{smth}},\mathbb{R})/H^{3}(X_{\text{smth}}%
,\mathbb{Z}),
\end{equation}
that is, we view $H^{3}(X_{\text{smth}},\mathbb{R})$ as a vector space,
and then mod out by the integral lattice $H^{3}%
(X_{\text{smth}},\mathbb{Z})$.\footnote{Here we are ignoring possible
torsional contributions to $H^3(X,\mathbb{Z})$.} In the IIA supergravity, the
hypermultiplet moduli space is given by a $\mathbb{C}^{\ast}$ bundle
which is fibered over the intermediate Jacobian introduced by Griffiths
$\mathcal{J}_{G}(X_{\mathrm{smth}})$, i.e. the total space of the
fibration $J(X_{\mathrm{smth}}) \rightarrow
\mathcal{J}_{G}(X_{\mathrm{smth}}) \
\rightarrow M_{\mathrm{cplx}}$, with $\mathcal{J}_{G}(X_{\mathrm{smth}})=
(H^{3,0}(X)\oplus H^{2,1}(X))^*/H_3(X,\mathbb{Z})$.  This intermediate Jacobian
specifies the correct complex structure on the real torus
(2.4).\footnote{Note that $H^{3,0}(X)$ is the tangent space to the
moduli of the dilaton and $H^{2,1}(X)$ is the tangent space to the moduli
of complex structures, so the tangent space to the intermediate Jacobian
naturally pairs with the tangent space to the complex structure plus
dilaton, as must be the case for the complex halves of hypermultiplet
moduli.  This is not the case for the Weil intermediate Jacobian, which
is sometimes incorrectly used in the physics literature as a description
of the RR moduli.}
In the low energy effective field theory, we can expand around some point of the quaternionic Kahler moduli space,
with fluctuations taking values in the tangent space. For example, the infinitesimal variations of complex structure moduli live in
$H^{2,1}(X) \equiv Def(X)$, and the $SU(2)$ R-symmetry acts on the tangent space, rotating the different
components of the hypermultiplet fluctuations into one another.

The supergravity description requires all volumes on $X$ to be large compared with the Planck scale.
Away from this limit additional light degrees enter the effective theory.
Starting from a smooth $X_{\text{smth}}$, we can reach a singular geometry
either by tuning the K\"ahler moduli or by tuning complex structure moduli of
$X_{\text{smth}}$. By abuse of notation, we
shall refer to this singular Calabi-Yau as $X$. The main condition to have an
F-theory compactification is that $X$ admits an elliptic fibration with
section. From this perspective, there is more than one possible
$X_{\text{smth}}$ which could degenerate to $X$. We can approach $X$ either
via a blowdown $\widetilde{X}\rightarrow X$, or by switching off a family of
smoothings $X_{t}\rightarrow X$.

\subsection{Open String Description}

In local regions of the twofold base $B$, there is a complementary description of an
F-theory compactification in terms of the gauge theory of a seven-brane. In the local
description, a seven-brane wraps a genus $g$ curve $C$. The curve comes with
some number of marked points $p_{1},...,p_{k}$ corresponding to locations
where other components of the discriminant intersect $C$.

In M-theory terms, some of the field content of the effective field theory on
the seven-brane descends from variations in the complex structure
$\delta\Omega_{(2,1)}$, and variations in the three-form potential $\delta
C_{3}$. Assuming we have a curve of ADE singularities, we can, at a heuristic
level, consider a basis of $(1,1)$ forms $\omega^{\alpha}$ for the fiber, and
then locally decompose these modes as:%
\begin{equation}
\delta\Omega_{(2,1)}\sim\underset{\alpha}{\sum}\Phi_{(1,0)}^{\alpha}%
\wedge\omega^{\alpha}\text{ \ \ and \ \ }\delta C_{3}\sim\underset{\alpha
}{\sum}A^{\alpha}\wedge\omega^{\alpha},
\end{equation}
where $\Phi_{(1,0)}^{\alpha}$ is a collection of $(1,0)$ forms on $C$,
and $A^{\alpha}$ is a connection on $C$. These fields should be viewed as
taking values in the abelian subalgebra of some non-abelian gauge algebra $\frak{g}$.
The rest of the states in the adjoint representation come from M2-branes
wrapped over collapsing cycles.

Taking these additional degrees of freedom into account, the effective theory
for this system is controlled by a Hitchin-like system coupled to defects
\cite{BershadskyFOURD, BHVI, DWI}. For a seven-brane with gauge group $G$,
this is given by a principal $G$ bundle $E$, and an adjoint-valued
$(1,0)$-form Higgs field $\Phi$, which satisfy the equations of motion:
\begin{equation}
F + \left[  \Phi,\Phi^{\dag}\right]  =\underset{p}{\sum}\delta
_{(p)}\mu_{\mathbb{R}}^{(p)}\text{ \ \ and \ \ }\overline{\partial}_{A}%
\Phi=\underset{p}{\sum}\delta_{(p)}\mu_{\mathbb{C}}^{(p)},
\end{equation}
where here, $F=dA+A\wedge A$ is the curvature of $E$, $\delta_{p}$ is a delta
function (i.e. $(1,1)$ current) localized at the point $p\in C$, and
$\mu_{\mathbb{R}}^{(p)}$ and $\mu_{\mathbb{C}}^{(p)}$ form a triplet of
adjoint valued moment maps. In most cases, these moment maps can be associated
with bilinears in the vevs of ``bifundamentals''.

Some of the moduli of the Hitchin system have a clear geometric interpretation
in the local geometry of the F-theory compactification. It is helpful to
illustrate this unfolding by way of example. To this end, consider a curve of singularities:
\begin{equation}
y^{2}=x^{3}+z^{5}\label{e8curve}%
\end{equation}
so that $x = y = z = 0$ denotes the location of the curve $C$. We view this as a
seven-brane with gauge group $E_{8}$ wrapping the curve $C$. A Higgs field
such as:%
\begin{equation}
\Phi=\left[
\begin{array}
[c]{cc}%
\phi & \\
& -\phi
\end{array}
\right]
\end{equation}
valued in the $\mathfrak{su}(2)$ factor of $\mathfrak{e}_{7}\times
\mathfrak{su}(2)\subset\mathfrak{e}_{8}$ corresponds to the unfolding of
$E_{8}$ to $E_{7}$:%
\begin{equation}
y^{2}=x^{3}+z^{5} + \phi^{2}xz^{3},
\end{equation}
where $\phi^{2}$ is the quadratic Casimir of the $\mathfrak{su}(2)$ valued Higgs
field. More generally, the holomorphic Casimir invariants $\mathcal{C}%
_{(i)}\left(  \Phi\right)  $ of $\Phi$ give gauge invariant
coordinates on the moduli space of the Hitchin system via the map:%
\begin{equation}
\Phi\rightarrow\mathcal{C}_{(i)}\left(  \Phi\right)  .\label{HitchinMap}%
\end{equation}
In the curve of singularities, such deformations show up as unfolding parameters.

A natural way to characterize the local neighborhood of the seven-brane is in terms of the total space
$\mathcal{O}(K_{C}) \rightarrow C$. Then, after activating a vev for the Higgs field, the location of the
seven-brane is captured by the spectral equation:
\begin{equation}
\det(s - \Phi) = 0
\end{equation}
where $s$ is a section of $\mathcal{O}(K_{C}) \otimes \mathbf{Id}$. The key point is that in the spectral equation, the
holomorphic Casimir invariants of line (\ref{HitchinMap}) directly show up.

But the holomorphic geometry fails to capture all of the moduli of the
Hitchin system. We define \textquotedblleft T-brane data\textquotedblright%
\ \cite{TBRANES} as any gauge invariant data of the Hitchin system which is
absent from the map (\ref{HitchinMap}). Roughly speaking, such data is
associated with a nilpotent Higgs field and cannot show up as a complex
structure deformation.

To give an example, consider again the Hitchin system defined by the local
model of equation (\ref{e8curve}). We assume the Higgs field again takes
values in the $\mathfrak{su}(2)$ factor of $\mathfrak{e}_{7}\times
\mathfrak{su}(2)\subset\mathfrak{e}_{8}$. Then, a nilpotent Higgs field such
as:%
\begin{equation}
\Phi=\left[
\begin{array}
[c]{cc}%
0 & 1\\
0 & 0
\end{array}
\right]  ,
\end{equation}
will satisfy the equations of motion on a genus $g>1$ curve \cite{HitchinSelf}.
To solve the Hitchin system of equations, we need
a non-trivial flux, corresponding to the rank two
bundle $E = K_{C}^{1/2}\oplus K_{C}^{-1/2}$. In this case, all of the $\frak{su}(2)$
gauge symmetry is broken. Observe, however, that all holomorphic Casimir
invariants vanish. So, the complex geometry remains as:%
\begin{equation}
y^{2}=x^{3}+z^{5}\text{,}%
\end{equation}
which would suggest a singular geometry with $\frak{e}_{8}$ gauge symmetry,
even though the $\frak{su}(2)$ factor in $\frak{e}_7 \times \frak{su}(2) \subset \frak{e}_8$
has already been broken.  Let us note that instead of taking
$g > 1$ we could instead add multiple punctures to
a genus zero curve. We will encounter the latter situation repeatedly in
the context of F-theory models.

In the related context of four-dimensional compactifications of F-theory, such
T-brane data can also be detected through D3-brane probes \cite{Funparticles,
FCFT, D3gen, HVW, DSSM, HoloHiggs, Heckman:2012jm}. In that context, the resulting superconformal
fixed points are determined by the Jordan block structure of the Higgs field.

\section{Seven-Branes Wrapping Riemann Surfaces\label{sec:EFT}}

In preparation for our later geometric analysis, we now discuss in more detail the
low energy effective field theory associated with a seven-brane wrapping a
Riemann surface which is coupled to defect modes. To set the stage, here we
study the low energy effective field theory defined
by a seven-brane with gauge group $G$ wrapping a genus $g$ curve which is
coupled to matter localized at points of the curve. For each point $p\in C$,
let $R_{(p)}$ denote the representation of the matter fields. This system has been studied
from various perspectives, including \cite{Katz:1996ht, BershadskyPLUS, BHVI}.

Let us now turn to the six-dimensional effective theory defined by this
system. For ease of exposition, we assume that no background fields have been
switched on. For details on less trivial backgrounds, see \cite{BHVI, TBRANES}.
The massless field content is given by all zero mode
fluctuations around this background. First of all, we have a single 6D vector
multiplet with gauge group $G$. Additionally, we have a number of
hypermultiplet moduli. The zero modes of the Higgs field satisfy the equation
of motion $\overline{\partial}\Phi=0$, and are associated with global sections
of the canonical bundle, which has complex dimension $h^{0}(C,K_{C})=g$.
Additionally, the holonomies of the gauge field around the $2g$ one-cycles
of $C$ produce another $2g$ real degrees of freedom. We therefore conclude
that the zero mode content of the system produces $g$ adjoint-valued
hypermultiplets. Additionally, we have the localized hypermultiplets
at points of $C$.

It is also instructive to compactify on a $T^{2}$ to reach a four-dimensional
system. This yields a theory with $\mathcal{N}=2$ supersymmetry, with the same
number of hypermultiplets. The only difference in the mode content is that now
the vector multiplet also contains a complex adjoint-valued scalar $\varphi$.
In $\mathcal{N} = 1$ language, the associated superpotential for this system is:%
\begin{equation}
W=\underset{i=1}{\overset{g}{\sum}}\text{Tr}_{\frak{g}}\left(  \varphi \cdot \left[
X_{i},\widetilde{X}_{i}\right]  \right)  + \underset{p}{\sum}\widetilde
{q}_{(p)}\cdot\varphi\cdot q_{(p)},\label{superpot}%
\end{equation}
and the D-term constraint is:
\begin{equation}
\underset{i=1}{\overset{g}{\sum}}\text{Tr}_{\frak{g}} \left(T\cdot\left[  X_{i},X_{i}^{\dag
}\right]  \right) + \underset{p}{\sum}q_{(p)}^{\dag}\cdot T\cdot q_{(p)}-\underset
{p}{\sum}\widetilde{q}_{(p)}\cdot T\cdot\widetilde{q}_{(p)}^{\dag}=0
\end{equation}
modulo gauge transformations. Here, $\text{Tr}_{\frak{g}}$ is shorthand for
contraction using the Killing form of the Lie algebra $\frak{g}$,
$T$ is a generator of the Lie algebra $\frak{g}$,
$X_{i}\oplus\widetilde{X}_{i}$ denotes the $g$ adjoint valued
hypermultiplets, and $q_{(p)}\oplus\widetilde{q}_{(p)}$ denotes a
hypermultiplet transforming in a representation $R_{(p)}\oplus R_{(p)}^{\ast}$,
which descends from a localized point $p\in C$.

The Higgs branch corresponds to activating vevs for the hypermultiplets, and the Coulomb branch corresponds to
giving vevs to the complex scalars in the vector multiplet. There can also be mixed Coulomb / Higgs branches. In
an $\mathcal{N} = 2$ theory, these ``mixed'' pieces still factorize (see e.g. \cite{Argyres:1996eh}). Jumping ahead to a geometric
characterization, the Coulomb branch corresponds to activating a K\"ahler resolution, while the Higgs branch corresponds
to activating a smoothing and non-trivial periods for the three-form moduli.

The total dimension of the hypermultiplet moduli space is given by adding up
the dimensions and representations of all points, and modding out by the
hyperkahler quotient defined by $G$. The complex dimension
of the hypermultiplet moduli space is:%
\begin{equation}
\dim_{\mathbb{C}}M_{\text{hyper}}=2(g-1)\cdot\dim G+\underset{p}{\sum}2\dim
R_{(p)}.
\end{equation}

\subsection{$SU(N)$ Example \label{ssec:SUNexamp}}

Let us illustrate some additional aspects of the above remarks in the case
$G=SU(N)$. For simplicity, we assume there are $k$ localized hypermultiplets
in the fundamental representation. Then, the dimension of the hypermultiplet
moduli space is:%
\begin{equation}
\dim_{\mathbb{C}}M_{\text{hyper}}=2(g-1)\cdot(N^{2}-1)+2kN.
\end{equation}
We can geometrically engineer an example of this type with a curve of
$A_{N-1}$\ singularities:%
\begin{equation}
y^{2}=x^{2}+z^{N+1}+\alpha_{k}(z^{\prime})z^{N},
\end{equation}
where $z^{\prime}$ denotes a local coordinate on $C$, and $\alpha_{k}(z')$ has $k$ zeroes,
indicating the points of localized matter.

Once we switch on background values for the defects, we induce delta function
supported curvature in the gauge field flux, and poles in the Higgs field. A
meromorphic Higgs field of the form:
\begin{equation}
\Phi=\text{diag}(\phi_{1},...,\phi_{k},\underset{N-k}{\underbrace{-\frac
{1}{N-k}\Sigma\phi_{i},...,-\frac{1}{N-k}\Sigma\phi_{i}}}),
\end{equation}
where:%
\begin{equation}
\phi_{i}=\frac{\mu_{i}dz^{\prime}}{z^{\prime}- p_{i}}%
\end{equation}
corresponds to activating a bifundamental vev for the localized matter fields
at the roots $p_{i}$ of $\alpha_{k}(z^{\prime})$. As explained in \cite{BHVI},
the $A_{N-1}$ singularity then unfolds to:
\begin{equation}
y^{2} = x^{2}+z^{N+1}+z^{N-k}\left(  \alpha_{k}(z^{\prime})z^{k}+\alpha
_{k-1}(z^{\prime})z^{k-1}+...+\alpha_{0}(z^{\prime})z^{0}\right)  .
\end{equation}
The $\alpha_{i}$ for $i=0,...,k-1$ correspond to holomorphic Casimir
invariants of $SU(k)$ built from the vevs of the defect modes. So in other words, we can capture
all the effects of unfolding by a suitable $SU(k)$ Higgs field with poles along the
localized matter of a parent $SU(N)$ gauge theory.

\section{Geometric Remnants of T-Branes \label{sec:REMNANT}}

In this section we argue that T-brane data should be viewed in the dual M-theory and type IIA\ descriptions
as the remnants of three-form potential moduli. This is in
accord with $\mathcal{N}=(1,0)$ supersymmetry in six dimensions, where the
three-form potential moduli fiber over the complex structure moduli. By abuse
of notation, we shall sometimes refer to the three-form potential moduli simply as ``RR moduli''.

Since we are now asking how to go from the closed string moduli space back to the open string moduli of
a T-brane configuration, we will assume that we have started with a compact
Calabi-Yau threefold $X_{\text{cpct}}$, and that we have taken a local limit
for this system which consists of a curve of ADE\ singularities possibly coupled to defects.
In this section, we shall therefore confine our discussion to a family of non-compact
smooth Calabi-Yau threefolds $X_{t}$ such that the $t\rightarrow0$ limit
defines a curve of ADE\ singularities. In fact, since these local models can
be embedded in F-theory, it will be enough for us to consider the related
compactification of IIA string theory on the background $X_{t}$.

The plan should now be clear: We shall study the degeneration of the
hypermultiplet moduli space, and in particular the intermediate Jacobian
$J(X_{t})$ as $t\rightarrow0$. There are at least two difficulties
with this approach. First, there is no well-defined notion of a limit of a
family of complex tori, and second, naive definitions of this limit do not
match the expected properties of the RR moduli space, and in particular the
hypermultiplet moduli space of the low energy effective theory.

Both issues are symptoms of the fact that near the singular points of the moduli space,
some of the three-cycles are about to disappear from the classical geometry. This is
also reflected in monodromies in the basis of three-cycles around the singular points in moduli space.
To study this behavior, we apply the machinery of limiting mixed Hodge structures. Using this, we
can introduce a trajectory dependent notion of the intermediate Jacobian as it approaches a singular point in moduli space.
This will allow us to establish that as $t\rightarrow0$, the geometric characterization becomes
singular, but that the geometry points to the appearance of a Hitchin system. We
stress that in this discussion we do not put the Hitchin system in
\textquotedblleft by hand\textquotedblright\ ahead of time. Rather, we will
see this structure \textit{emerge}!

Our plan in this section will be to describe in more precise terms this
limiting operation. The main mathematical tool we use is known as the theory
of \textquotedblleft limiting mixed Hodge structures\textquotedblright\ (LMHS).
Since this may be unfamiliar to some readers, we illustrate some elementary
aspects of this theory in the case of the conifold.
We then apply this machinery to the cases of interest
for geometric engineering in string theory. We illustrate how to
recognize the appearance of a Hitchin system for an isolated curve of ADE
singularities, building on the work of \cite{Diaconescu:2005jw,
Diaconescu:2006ry}. Then, we illustrate how to extend this to geometries which
include collisions with other singularities. In physical terms, this
corresponds to a situation where the seven-brane is coupled to charged defect
modes. When these modes get a vev, we induce poles in the gauge field and
Higgs field of the Hitchin system. We show that our analysis of limiting mixed
Hodge structures extends to this case as well, and that the remnants of
the RR\ moduli are accounted for by a parabolic Hitchin system.

\subsection{Limiting Mixed Hodge Structures}

Before launching into our analysis, we would first like to give some general background on why the theory of
limiting mixed Hodge structures is the relevant machinery for our analysis.
For additional details, we refer the interested reader to Appendix \ref{app:LIMITS}.

Our basic claim is that T-brane data comes about from the limiting behavior of the intermediate Jacobian. The idea here is
to consider not a single Calabi-Yau threefold, but rather, a whole family of them $X_t$, with $t$ a smoothing parameter. The
singular limit corresponds to taking $t \rightarrow 0$. The straightforward approach would be to simply compute $J(X_t)$ for each non-zero
$t$ and to then extrapolate to the singular limit.

Now, for \textit{generic} points of complex structure moduli, there is indeed a natural
notion of parallel transport for the complex structure given by the Gauss-Manin connection. We can extend this
to the intermediate Jacobian, so we can also naturally consider how this space responds to variations in the complex structure.

Subtleties materialize in singular limits. Indeed, at a singular point in complex structure moduli,
some elements in the basis of three-cycles will start to disappear. One way to see this is by considering monodromies
in the parameter $t$ around the singular point $t = 0$. Such monodromies can reshuffle the original basis of three-cycles,
and can also mix the spaces $H^{p,q}(X)$ of the K\"ahler manifold.

To isolate the effects of monodromy we need a refined notion of how to define a family of intermediate Jacobians. The basic idea
is to have a notion of complex structure moduli and intermediate Jacobian in which we can ``untwist'' the effects of monodromy, i.e.
by going to a suitable cover. This is where the
theory of mixed Hodge structure, and in particular limiting mixed Hodge structure comes to the fore. To have a natural notion of a
holomorphic family of Hodge structures, we can take $H^{p,q}(X)$, and add to it all other spaces which can
mix with it under parallel transport with respect to the Gauss-Manin connection. In the case of
a K\"ahler manifold, a well-known example of this type is the Hodge filtration:
\begin{equation}
F^pH^k(X,\mathbb{C})=\bigoplus_{p'\ge p}H^{p',k-p'}(X).
\end{equation}
This has the property that $F^{p} \subset F^{p'}$ for $p > p'$, i.e. it defines a decreasing filtration. It is common to refer to the
entire filtration as $F^{\bullet}$. By a similar token, we can also introduce an increasing filtration $W_{\bullet}$, known as the
``weight filtration''.

Now, the utility of introducing this additional formalism is that as we move around in the parameter $t$, this filtration structure
remains intact, so we have a well-defined notion of parallel transport on the vector spaces. We can then define various limits of this Hodge structure
by a choice of trajectory $t \rightarrow 0$. To this degeneration is associated
a limiting mixed Hodge structure which is denoted by $H^{3}_{\mathrm{lim}}$.
Details are reviewed in Appendix~\ref{limmhs}.

Since the intermediate Jacobian fibers over the complex structure moduli, we can
extend these considerations fiberwise. Along these lines,
we can introduce a generalized notion of the intermediate Jacobian for a
cohomology theory $H$ by quotienting $H_{\mathbb{C}}$, the
complexification of the integral cohomology:
\begin{equation}
J^p H = H_\mathbb{C} /(F^{p}H_{\mathbb{C}} + H_\mathbb{Z} ).
\end{equation}
where $H_\mathbb{Z}$ refers to the cohomology theory with integer coefficients.
When $p = 2$ and $X$ is smooth, we get back our previous notion of the intermediate Jacobian $J(X_t)$, though
now, we can track the behavior of this space as we move towards a singular limit. In what follows, we shall refer to this
limiting behavior as $J^{2}(H^{3}_{\mathrm{lim}})$, in the obvious notation.

By now, it should hopefully be clear that the theory of limiting mixed Hodge structure
is the appropriate tool for tracking the behavior of the intermediate Jacobian as we move towards
singular limits. We now illustrate how this machinery works in progressively more involved examples, and
show that it matches to the expected behavior of T-branes.

\subsection{Warmup: The Conifold}

By way of example, in this subsection we show how to calculate the limiting
behavior of the intermediate Jacobian in the case of conifold singularities.
See Appendix \ref{app:LIMITS} for additional details.

Before we describe the deformation theory of conifold singularities, we
digress by reviewing the deformation theory of a general affine hypersurface
singularity. Given any affine hypersurface $X$ defined by $f(x_{1}%
,\ldots,x_{r})=0$ in $\mathbb{C}^{r}$, the space of first order deformations
of $X$ is identified with the vector space underlying the Jacobian ring
\begin{equation}
\mathcal{J}_{f}=\mathbb{C}\Big[x_{1},\ldots,x_{r}\Big]\Big\slash
\left(  \frac{\partial
f}{\partial x_{i}}\right)  . \label{jacobianring}%
\end{equation}
Now, the first order deformations of an arbitrary variety $X$ are given by the
vector space $\mathrm{Ext}^{1}(\Omega_{X}^{1},\mathcal{O}_{X})$. The local to
global spectral sequence leads to%

\begin{equation}
0\rightarrow H^{1}(X,\underline{\mathrm{Ext}}^{0}(\Omega_{X}^{1}%
,\mathcal{O}_{X}))\rightarrow\mathrm{Ext}^{1}(\Omega_{X}^{1},\mathcal{O}%
_{X})\rightarrow H^{0}(X,\underline{\mathrm{Ext}}^{1}(\Omega_{X}%
^{1},\mathcal{O}_{X})). \label{localtoglobal}%
\end{equation}
Here, $\underline{\mathrm{Ext}}^i$ denotes ``local Ext'' and is a sheaf on $X$.
In particular given sheaves $F$ and $G$ on $X$, the sheaf
$\underline{\mathrm{Ext}}^0(F,G)$ is the sheaf whose sections on sufficiently small
open sets are just the local homomorphisms from $F$ to $G$.

Since $\Omega_{X}^{1}$ is locally free on the smooth locus of $X$, it follows
that the sheaf $\underline{\mathrm{Ext}}^{1}(\Omega_{X}^{1},\mathcal{O}_{X})$ vanishes
there, hence is supported on the singular locus of $X$. We put
\begin{equation}
\mathcal{T}^{1}=\underline{\mathrm{Ext}}^{1}(\Omega_{X}^{1},\mathcal{O}_{X})
\label{T1def}%
\end{equation}
and the last map in (\ref{localtoglobal}) describes the restriction of a
first order deformation of $X$ to a first order deformation of the singular
locus of $X$. Note that
this map need not be surjective. In more detail, it is natural to study the
limiting mixed Hodge structure $H_{\mathrm{lim}}^{3}$ of a generic
degeneration, to
which a Jacobian is associated:
\begin{equation}
J^{2}(H_{\mathrm{lim}}^{3})=H_{\mathrm{lim}}^{3}/\left(  F^{2}H_{\mathrm{lim}%
}^{3}+H_{\mathrm{lim},\mathbb{Z}}^{3}\right)  ,
\end{equation}
where $F^{2}H_{\mathrm{lim}}^{3}$ denotes a piece of the Hodge filtration and
$H_{\mathrm{lim},\mathbb{Z}}^{3}$ denotes the integer lattice of the mixed
Hodge structure $H_{\mathrm{lim}}^{3}$.

Let us now illustrate these aspects for the case of a conifold singularity.
From (\ref{jacobianring}), a single conifold singularity
\begin{equation}
w^{2}+x^{2}+y^{2}+z^{2}=0,
\end{equation}
has a ring of first order deformations:%
\begin{equation}
\mathbb{C}[w,x,y,z]/\left(  w,x,y,z\right)  \simeq\mathbb{C}, \label{jacrcon}%
\end{equation}
which tells us that the first order deformations of the conifold are all given
by
\begin{equation}
w^{2}+x^{2}+y^{2}+z^{2}=t,
\end{equation}
as is well-known.

More generally, let us now consider the case of $X$ a Calabi-Yau threefold
with only conifold singularities. The calculation (\ref{jacrcon}) shows that
$\mathcal{T}^{1}$ is a skyscraper sheaf supported at the conifolds whose
stalks are all isomorphic to $\mathbb{C}$. Returning to the exact sequence of
(\ref{localtoglobal}):%
\begin{equation}
0\rightarrow H^{1}(X,\underline{\mathrm{Ext}}^{0}(\Omega_{X}^{1}%
,\mathcal{O}_{X}))\rightarrow\mathrm{Ext}^{1}(\Omega_{X}^{1},\mathcal{O}%
_{X})\rightarrow H^{0}(X,\underline{\mathrm{Ext}}^{1}(\Omega_{X}%
^{1},\mathcal{O}_{X})),
\end{equation}
\begin{sloppypar}
\noindent
the term $H^{1}(X,\underline{\mathrm{Ext}}^{0}(\Omega_{X}^{1},\mathcal{O}%
_{X}))$ is the space of first-order deformations which preserve the conifold
singularities, $\mathrm{Ext}^{1}(\Omega_{X}^{1},\mathcal{O}_{X})$ is the space
of all first order deformations, and $H^{0}(X,\underline{\mathrm{Ext}}%
^{1}(\Omega_{X}^{1},\mathcal{O}_{X}))$ is the space of local deformations of
the conifolds. By abuse of notation, we shall refer to a smoothing of $X$ by
$X_{t}$, so that $t\rightarrow0$ denotes the singular case.
\end{sloppypar}

For expository purposes, we assume that $X$ has a small K\"{a}hler resolution
$\widetilde{X}$, in which case a smoothing $X_{t}$ of $X$ completes a conifold transition. We
now show in this case that the intermediate Jacobian $J(X_{t})$ has a canonical
limiting behavior as $t\rightarrow0$.

Basically, we need to find a basis of three-cycles to calculate the periods of
the RR\ moduli, and then track the behavior of these periods as $t\rightarrow
0$. Associated with each conifold point is a vanishing cycle in $H^{3}%
(X_{t},\mathbb{Z})$, Poincar\'{e} dual to the class in $H_{3}(X,\mathbb{Z})$
of the familiar vanishing $S^{3}$ near the conifold point. In general, if
there are $m$ such conifold points they will typically only span a lattice
$\mathcal{W}$ of rank $m-r$ for some $0\leq r<m$. This number $r$ is nothing
other than the jump in the Hodge numbers in passing from the deformed side of
the conifold to the resolved side \cite{Clemens, Friedman, Tian}:
\begin{equation}
h^{1,1}(\widetilde{X})=h^{1,1}(X_{t})+r,\qquad h^{2,1}(\widetilde{X})=h^{2,1}%
(X_{t})+r-m.
\end{equation}
We also have in this case that $\dim
H^{0}(X,\underline{\mathrm{Ext}}^{1}(\Omega_{X}%
^{1},\mathcal{O}_{X}))=m-r$.

For example, in the conifold transition of the quintic acquiring 16 conifolds,
we have $h^{1,1}(X_t) = 1,h^{2,1}(X_t) = 101,m=16,r=1$. So the Hodge numbers of $\widetilde{X}$
in that case are $h^{1,1}(\widetilde{X}) = 2$ and $h^{2,1}(\widetilde{X}) = 86$.

Next, let $H_{\mathrm{lim}}^{3}$ denote the limiting mixed Hodge
structure of the smoothing of a conifold (which is a mixed Hodge structure on
$H^{3}(X_{t})$). For now, all we need to know is that a mixed Hodge structure
$H$ has an increasing filtration, the weight filtration, which we denote by
$W_{i}H_{\mathbb{Q}}$ since it is a vector space defined over the rationals.
Letting $Gr_{k}H_{\mathbb{Q}}=W_{k}H_{\mathbb{Q}}/W_{k-1}H_{\mathbb{Q}}$ be
the associated graded vector space, we compute in
Appendix \ref{app:LIMITS} that
\begin{equation}%
\begin{array}
[c]{ccl}%
Gr_{4}H_{\mathrm{lim}}^{3} & \simeq & \mathcal{W}_{\mathbb{Q}}\\
Gr_{3}H_{\mathrm{lim}}^{3} & \simeq & H^{3}(\widetilde{X},\mathbb{Q})\\
Gr_{2}H_{\mathrm{lim}}^{3} & \simeq & \mathcal{W}_{\mathbb{Q}},%
\end{array}
\end{equation}
where $\mathcal{W}_{\mathbb{Q}}=\mathcal{W}\otimes{\mathbb{Q}}$.

Now, recall that to any mixed Hodge structure $H$, we can associate for each $p$ a
generalized Jacobian
\begin{equation}
J^{p}H \equiv H_{\mathbb{C}}/\left(  F^{p}H_{\mathbb{C}}+H_\mathbb{Z}\right)  .
\end{equation}
So, returning to the smoothing $X_{t}$ from the conifold transition, instead of
examining the limiting mixed Hodge structure, which is a limit of third
cohomologies, for the purpose of computing RR moduli we study the limit of the
intermediate Jacobians $J(X_{t})$. This can be determined from $J^{2}%
H_{\mathrm{lim}}^{3}$. In Appendix \ref{app:LIMITS}, we compute that
\begin{equation}
J^{2}H_{\mathrm{lim}}^{3}\ \mathrm{is\ a\ }\left(  \mathbb{C}^{\ast}\right)
^{m-r}\ \mathrm{fibration\ over\ }J(\widetilde{X}),
\end{equation}
independent of the choice of smoothing $X_t$.
We further show in Appendix~A that the extension class of the
mixed Hodge structure $W_{3}H_{\mathrm{lim}}^{3}$ determines the global
structure of this $\left(  \mathbb{C}^{\ast}\right)  ^{m-r}$ fibration. Thus,
we conclude that the $\left(  \mathbb{C}^{\ast}\right)  ^{m-r}$ fiber over
$J(\widetilde{X})$ is the moduli space of the RR fields on $X$ associated with the singularities.

Near $X$, the hypermultiplet moduli space of $X$ fibers over the complex $m-r$
dimensional moduli space parametrizing the smoothings of the conifolds (the space whose
tangent space is the image of the rightmost map in (\ref{localtoglobal})).  The fibers
of this map in turn fiber over the
hypermultiplet moduli space of $\widetilde{X}$, with fibers $(\mathbb{C}^{\ast})^{m-r}$.
Locally, the two complex $m-r$ dimensional
spaces just described combine into $m-r$ quaternionic moduli.

If we do not assume that $X$ has a small K\"ahler resolution $\tilde{X}$,
we simply replace $J(\widetilde{X})$ with $J(\widehat{X})$, where $\widehat{X}$ is the
(non-Calabi-Yau) blowup of the conifolds of $X$.  Note that $J(\widehat{X})=
J(\widetilde{X})$ when $\widetilde{X}$ exists, since no new 3-cycles are created in
blowing up $\widetilde{X}$ to $\widehat{X}$.

\subsection{Isolated Curve of ADE\ Singularities \label{ssec:ISOLATED}}

We now turn to the first case of interest with non-abelian gauge symmetry, given by a
curve of ADE\ singularities in IIA\ string theory. Such singularities generate
the corresponding ADE\ gauge theory in the six-dimensional effective theory.
Non-simply laced gauge theories occur when there is an appropriate
monodromy action in the fiber \cite{BershadskyPLUS}.

In contrast to the previous case of the conifold, all of our examples from this point on have the property
that the limiting mixed Hodge structure analysis depends on the trajectory we take as $t \rightarrow 0$. Physically,
this is to be expected, because each choice of trajectory corresponds to a different identification of the
$W-$ and $Z-$ bosons of an $\frak{su}(2)$ factor. By sweeping over all possible trajectories,
however, we can extract the trajectory independent data associated with the
limiting mixed Hodge structures. In the physical theory, these different trajectories correspond to conjugation
by elements of the broken gauge symmetry.

To begin our geometric analysis, we recall the local presentation for the different ADE\ singularities:
\begin{align}
A_{N}  &  :xy=z^{N+1}+...\\
D_{N}  &  :y^{2}=x^{2}z+z^{N-1}+...\\
E_{6}  &  :y^{2}=x^{3}+z^{4}+...\\
E_{7}  &  :y^{2}=x^{3}+xz^{3}+...\\
E_{8}  &  :y^{2}=x^{3}+z^{5}+...
\end{align}
where the curve $C$ is defined by $x=y=z=0$ and $x,y,z$ lie in respective line
bundles $L_{x},L_{y},L_{z}$ on $C$. The Calabi-Yau condition together with
the adjunction formula and homogeneity of the above equations constrains
the bundle assignments for $L_{x}$, $L_{y}$, and $L_{z}$. For example, in the
$A_{N}$ singularity we have $L_{x}\otimes L_{y}\simeq K_{C}^{\otimes N{+1}}$ and
$L_{z}\simeq K_{C}$.
For additional explanation and the relation to gauge theory on a seven-brane,
see for example \cite{BHVI}.

The first order smoothing deformations of the singularities are captured by
$\mathcal{T}^{1}$ of equation (\ref{T1def}). In physical terms, we know that
this unfolding corresponds to activating a non-zero adjoint-valued Higgs
field, and explicit checks between the two descriptions have been carried out,
for example in \cite{BershadskyFOURD, Diaconescu:2005jw, Diaconescu:2006ry,
BHVI, DWI}. Here, $H^{0}(\mathcal{T}^{1})$ corresponds to the base of the
Hitchin fibration with gauge group corresponding to the singularity
type.\footnote{In \cite{Diaconescu:2005jw, Diaconescu:2006ry} only the local case was considered, but
$H^{0}(\mathcal{T}^{1})$ only depends on a neighborhood of $C$.}

In fact, even more is true. In \cite{Diaconescu:2005jw, Diaconescu:2006ry} it was observed
that away from the discriminant locus, the Hitchin system is identified with
the Calabi-Yau integrable system whose fibers are the local contributions to
the intermediate Jacobian.\footnote{There are subtle differences in comparing
families of complex tori between isomorphism and a torsorial relationship; we
ignore these issues here.} In this context, the discriminant locus
equivalently parametrizes singular Calabi-Yaus, or singular spectral covers.

Our assertion is that the local RR moduli space over $H^{0}(\mathcal{T}^{1})$
is completed over the discriminant locus by the Hitchin system. As with the
case of conifolds, this part of the moduli space can be realized globally
inside a moduli space associated with a limiting mixed Hodge structure.
Similar to the case of conifolds, the hypermultiplet moduli space can be related to
the hypermultiplet moduli space of the blowup $\widetilde{X}$ of $X$ by two
fibrations.

\subsubsection{Example: A Curve of $A_{N}$ Singularities}

Let us now illustrate these general considerations in the specific case of a curve of
$A_{N}$ singularities:
\begin{equation}
xy = z^{N+1}. \label{atcurvedef}%
\end{equation}
If we view this as the equation of a hypersurface in $\mathbb{C}^{4}$ with
coordinates $(w,x,y,z)$, the ring of first order deformations becomes
\begin{equation}
\mathbb{C}[w,x,y,z]/(x,y,z^N)\simeq\mathbb{C}[w,z]/(z^N),
\end{equation}
where $w$ is a local coordinate on $C$. Thus, to first order, deformations can
be written as
\begin{equation} \label{a1curvedef}
xy = z^{N+1}+\sum_{j=2}^{N+1}a_{j}(w)z^{N+1-j}.
\end{equation}
Since $z$ is a section of $K_{C}$ (via the Calabi-Yau condition), equation
(\ref{a1curvedef}) must now be interpreted as a section of $\mathcal{O}%
(K_{C}^{\otimes{N+1}})$, so we conclude that the $a_{j}(w)$ are to be
understood as sections of $\mathcal{O}(K_{C}^{\otimes{j}})$. We therefore see
that
\begin{equation}
\mathcal{T}^{1}=i_{\ast}\left(  \bigoplus_{j=2}^{n+1}\mathcal{O}%
(K_{C}^{\otimes{j}})\right)  ,
\end{equation}
where $i:C\hookrightarrow X$ is the inclusion.

Specializing further, we now verify the claim that in the case of an $A_{1}$
singularity, the remnant of the RR moduli is captured by the Hitchin system.
More precisely, our proposal is that the RR moduli of $X$ combined with the
moduli of global deformations of the complex structure
which preserve the singularity are fibered over the intermediate Jacobian of
the small resolution, with fiber the corresponding Prym in the fiber of the
Hitchin system. In other words, the Hitchin system describes the missing piece
of the RR moduli. A consequence of this proposal is that T-branes arise
naturally from Pryms of singular spectral covers. A similar but more involved
analysis holds for the case of $A_{N}$ singularities for all $N$.

Reverting to IIA language, we want to see the local contribution to the RR
moduli of $X$ when we have a curve $C$ of $A_{1}$ singularities. So we want to
see the fiber over the origin of the $SU(2)$ Hitchin system, i.e.\ the
nilpotent cone defined by $\mathrm{Tr}(\Phi^{2})=0$, which was worked out in detail in
\cite{del}. The nilpotent cone has many components; the component called
$N_{0}$ in \cite{del} corresponds to T-branes whose Higgs fields do not vanish
anywhere. One such example is given in \cite{HitchinSelf} by the bundle
$E=K_{C}^{1/2}\oplus K_{C}^{-1/2}$, with Higgs field
\begin{equation}
\Phi=\left[
\begin{array}
[c]{cc}%
0 & 1\\
0 & 0
\end{array}
\right]  ,
\end{equation}
where 1 is identified with a section of $\mathrm{Hom}(K_{C}^{1/2},K_{C}^{-1/2}\otimes
K_{C})$ in the natural way.

The general point of $N_{0}$ corresponds to a rank~2 bundle $E$ fitting into
an exact sequence:
\begin{equation}
0\rightarrow K_{C}^{-1/2}\rightarrow E\rightarrow K_{C}^{1/2}\rightarrow0.
\label{n0ext}%
\end{equation}
The corresponding Higgs field $\Phi$ is given by the composition
\begin{equation}
E\rightarrow K_{C}^{1/2}\simeq K_{C}^{-1/2}\otimes K_{C}\rightarrow E\otimes
K_{C}.
\end{equation}
The extensions (\ref{n0ext}) are parameterized by $\mathrm{Ext}^{1}(K_{C}^{1/2}%
,K_{C}^{-1/2})\simeq H^{1}(-K_{C})$, which is Serre dual to $H^{0}(2K_{C})$
and has dimension $3g-3$, as expected. Note that $\Phi^{2}=0$, although $\Phi$
can no longer be described in a block diagonal form, as $E$ need not be decomposable.

We now smooth the $A_{1}$ singularity and calculate the limiting mixed Hodge
structure. We have
\begin{equation}
H^{0}(\mathcal{T}^{1})\simeq H^{0}\left(  X,i_{*}\left(  \mathcal{O}%
(2K_{C})\right)  \right)  = H^{0}(C,\mathcal{O}(2K_{C})),
\end{equation}
the space of quadratic differentials, where $i:C\hookrightarrow X$ is the
inclusion. Then, a smoothing is given by $xy = z^2 + q$, with
$q\in H^0(C,\mathcal{O}(2K_C))$. For technical
reasons, we consider the degeneration
\begin{equation}
xy = z^{2}+t^{2}q
\label{a1curvedefcomp}
\end{equation}
as $t\rightarrow0$. We work out the limiting mixed Hodge structure in
Appendix \ref{app:LIMITS}. We will see, as already follows from
\cite{Diaconescu:2005jw, Diaconescu:2006ry}, that the limiting mixed Hodge
structure is actually a pure Hodge structure of weight~3. We compute that the
local contribution to $J^{2}H_{\mathrm{lim}}^{3}$ is the Prym variety of
$z^{2}+q=0$, the fiber of the Hitchin map over $Tr(\Phi^{2})=q$.

To define this Prym variety, let $Z\subset C$ be the zero set of $q$, assumed
to be $4g-4$ distinct points. We introduce the double cover:%
\begin{equation}
\pi:\widetilde{C}\rightarrow C,
\end{equation}
which is branched along $Z$, so that $\widetilde{C}$ has genus $4g-3$. Letting
$J_{g-1}(\widetilde{C})$ denote the space of line bundles of degree $g-1$ on
$\widetilde{C}$, then the Prym is given by:
\begin{equation}
\mathrm{Prym}(\widetilde{C}/C)=\left\{  L\in J_{2g-2}(\widetilde{C})\mid\iota^{\ast
}L\simeq \pi^*K_{C}\otimes L^{-1}\right\}
\end{equation}
where $i:\widetilde{C}\rightarrow\widetilde{C}$ is the involution which interchanges
the two sheets of the double cover. The space $\mathrm{Prym}(\widetilde{C}/C)$ can
be seen to have dimension $3g-3$.\footnote{The computation of the limiting
mixed Hodge structure gives more in fact: the limiting intermediate Jacobian
fibers over the intermediate Jacobian of the blowup of $X$, with fiber the
Prym variety discussed here.}

This situation is directly related to the $SU(2)$ Hitchin system as follows. First of
all, $q$ is naturally identified with a point of the Hitchin base, and
$\widetilde{C}$ is the associated spectral cover, as given by the equation
\begin{equation}
\widetilde{C}=\left\{  y\in\mathrm{Tot}(K_{C})\mid y^{2}=q\right\}.
\end{equation}
Then the Higgs bundle is reconstructed by the usual procedure, putting
$E=\pi_{\ast}(L)$ and deducing the Higgs field $E\rightarrow E\otimes K_{C}$
from the embedding $\widetilde{C}\subset\mathrm{Tot}(K_{C})$.

This computation makes clear that, in contrast to the conifold case, the
limiting mixed Hodge structure \emph{cannot\/} be the correct identification
of the RR moduli, since it depends nontrivially on the choice of $q$. As already mentioned,
this choice effectively dictates a trajectory for the smoothing to approach the singular limit.

However, we can also see that by sweeping over all different trajectories of smoothings, that
the calculation points directly to the Hitchin system. Note that the fiber of
the Hitchin system over $t^{2}q$ is independent of $t$ (as $z^{2}+q=0$ is seen
to be isomorphic to $z^{2}+t^{2}q=0$ by rescaling $z$), which explains why
naively taking the limit as $t\rightarrow0$ gives a spurious result.
Nevertheless, the Hitchin system itself provides a well-defined notion of RR
moduli, within which the $t\rightarrow0$ limit makes perfect sense and is
independent of $q$.

\subsubsection{Example: Unfolding an $A_{1}$ Factor}

In the previous example we focussed on recovering the remnant of T-brane data for
a curve of $A_{1}$ singularities. More generally, given $X$ a curve of
ADE\ singularities corresponding to a gauge theory of rank $r$, we can
blow up $r-1$ of the fiber $\mathbb{P}^{1}$'s to reach a still singular space
$\widetilde{X}_{\text{sing}}$. The remaining gauge symmetry in the Hitchin
system is $U(1)^{r-1}\times{SU}(2)$, and indeed, there is
a single smoothing deformation in this case as well. Since we are performing a
blowup on some factors, and a smoothing on others, we refer to this smoothing
as $\widetilde{X}_{t}$. As a consequence, the analysis of the previous section
easily extends to the unfolding patterns:%
\begin{align}
\label{aunfold} A_{N}  &  :xy=z^{N-1}(z^{2}+t^{2}q)\\
\label{dunfold} D_{N}  &  :y^{2}=x^{2}z+z^{N-2}(z+t^{2}q)=0,\\
E_{6}  &  :y^{2}=x^{3}+z^{2}\left(  z^{2}+t^{2}qx\right) \\
E_{7}  &  :y^{2}=x^{3}+xz(z^{2}+t^{2}qx)\\
\label{e8unfold} E_{8}  &  :y^{2}=x^{3}+z^{3}(z^{2}+t^{2}qx)
\end{align}
where in each case, $q$ corresponds to a quadratic differential for the
$\mathfrak{su}(2)$ valued Higgs field, i.e. $q = \mathrm{Tr}(\Phi^{2})$. Note that in all cases,
this quadratic differential is a section of $K_C^{2}$.

In the $A_N$ case, we see directly that the associated $SU(2)$
spectral curve $\widetilde{C}$ is
given by $z^{2}+t^{2}q=0$.  Since in the other cases there are
only cameral covers rather then spectral covers, we do not expect to see
the embedded $SU(2)$ spectral cover directly from the equation,
but it can be checked that in each case, the spectral cover is indeed
$z^{2}+t^{2}q=0$.\footnote{We caution the reader that here, $z$ is valued
in $\cO(K_C)$, so differs from the coordinate $z$ appearing in
(\ref{dunfold})---(\ref{e8unfold}).}  The main point is that after a change of
variables, the deformation after the partial blowup can be rewritten in the
form $xy=z^2+t^2q$ in each case \cite{KatzMorrison}.
It is now clear that we construct Prym$(\widetilde
{C}/C)$ in the same way as before. Taking the limit $t\rightarrow0$, we again
see that the limiting behavior of the intermediate Jacobian
$J(\widetilde{X}_{t})$ again points back to the T-brane data of the Hitchin system.

A similar analysis also allows us to cover some situations where the Hitchin system
couples to defect modes. Indeed, we can consider the deformation $X_{t}$
associated with the partial smoothing of the singularity. Following the
general procedure of \cite{KatzVafa, BershadskyPLUS}, in all of
these cases there are matter fields localized at the intersections of
the component curves.
Initiating a further unfolding pattern, we can
now in principle track the RR\ moduli by a further unfolding.

Let us give a concrete example of this type. We derive the moduli space in the
special case of breaking $SU(3)$, i.e. a curve of $A_{2}$
singularities which we unfold via:%
\begin{equation}
xy=\left(  z-\omega\right)  ^{2}\left(  z+2\omega\right)  ,
\end{equation}
where $\omega$ is a section of $K_{C}$. There are now two factors, which we
denote by $\widetilde{C}^{\prime}$ given by $z=-2\omega$, and the multiplicity~2 curve
$\widetilde{C}^{\prime\prime}$ given by $z=\omega$.

In this new system, there is localized matter trapped at the zeroes of
$\omega$. So, we get an $S(U(2)\times U(1))$ Hitchin system coupled to defects
which transform as bifundamentals of the algebra $\mathfrak{su}(2)\times
\mathfrak{u}(1)$. The $\mathfrak{su}(2)$ factor of the algebra is localized on
the component $z=\omega$, while the $\mathfrak{u}(1)$ factor is spread over
both $z = \omega$ and $z=-2\omega$.

The vevs of the trapped matter can generate localized T-brane data in the
original $SU(3)$ Hitchin system. Alternatively, one can view this as \textquotedblleft
gluing data\textquotedblright\ in the sense of \cite{glueI, glueII}. To see
this, observe that the Hitchin base for the $SU(3)$ system is $H^{0}%
(2K_{C})\oplus H^{0}(3K_{C})$, and the associated spectral curve $\widetilde{C}$
has equation:%
\begin{equation}
\left(  z-\omega\right)  ^{2}\left(  z+2\omega\right)  = 0.
\end{equation}
We pick a line bundle $L$ on $\widetilde{C}$ such that $\pi_{\ast}L$ is an $SU(3)$ bundle,
where $\pi$ is the projection to the base curve. Let $Z$ be the intersection
$\widetilde{C}^{\prime}\cap \widetilde{C}^{\prime\prime}$, which is a length~2 scheme at each of the
$2g-2$ points of intersection, corresponding to the zeroes of $\omega$.
Turning next to the classification of line bundles on $\widetilde{C}$, the exact sequence
\begin{equation}
0\rightarrow\mathcal{O}_{\widetilde{C}}^{\ast}\rightarrow\mathcal{O}_{\widetilde{C}^{\prime}}^{\ast
}\oplus\mathcal{O}_{\widetilde{C}^{\prime\prime}}^{\ast}\rightarrow\mathcal{O}_{Z}^{\ast
}\rightarrow0
\end{equation}
yields
\begin{equation}
0\to H^0\left(\mathcal{O}_{\widetilde{C}}^{\ast}\right)\rightarrow H^{0}\left(  \mathcal{O}_{\widetilde{C}^{\prime}}^{\ast
}\right)  \oplus H^{0}\left(  \mathcal{O}_{\widetilde{C}^{\prime\prime}}^{\ast
}\right)  \rightarrow H^{0}\left(  \mathcal{O}_{Z}^{\ast}\right)  \rightarrow
H^{1}\left(  \mathcal{O}_{\widetilde{C}}^{\ast}\right)  \rightarrow H^{1}\left(
\mathcal{O}_{\widetilde{C}^{\prime}}^{\ast}\right)  \oplus H^{1}\left(  \mathcal{O}%
_{\widetilde{C}^{\prime\prime}}^{\ast}\right)  \rightarrow0
\end{equation}
The first two terms are just constant functions, and the third has dimension
$2(2g-2)$, so the net contribution to the dimension of the group
$H^{1}(\mathcal{O}_{C}^{\ast})$ of line bundles $L$ on $C$ is
$4g-5$. Next we have $g+(4g-3)$ parameters from the next $H^{1}$'s (since they
are the fibers of a $U(1)$ and $U(2)$ Hitchin system, respectively).
Finally, we
have to reduce these moduli by
$g$ since we want $\pi_*L$ to be an $SU(3)$ bundle rather than just a
$U(3)$ bundle.  The conclusion is that we have $8g-8$ parameters in all,
and we have accounted for the entire fiber of the $SU(3)$ Hitchin system.

\subsection{More General Defects}
\label{sec:generaldefects}

In this subsection we extend our match of the limiting behavior of the
intermediate Jacobian to more general configurations where the Hitchin system supports
defect modes. In contrast to the analysis of the last subsection, we do not
assume the existence of a globally defined parent gauge theory which
Higgses down to the configuration of intersecting branes.

In field theory terms, we will now be interested in a Hitchin system on a
curve $C$ which is coupled to $k$ defects modes located at points
$p_{1},...,p_{k}\in C$. We saw in section \ref{sec:EFT}
that activating a vev for a localized matter field generates a pole for the
Higgs field, and a delta function concentrated flux. The mathematical object
which captures this pole data is a parabolic Higgs bundle. For an introduction
to this construction, see for example \cite{ParabolicHiggs}.

Here, we illustrate this for the case of $A_{N-1}$
singularities whose general sections at finitely many points are $A_{N}$
singularities. This is the situation which would arise in F-theory at
transverse collision points of $I_{N}$ and $I_{1}$ components of the
discriminant.

Let $P = p_{1}+...+p_{k}$ denote the divisor associated with the $k=|P|$ points
on $C$ where there is a defect. For ease of exposition, suppose that near any
$p_i$, the local equation of the singularity can be put in the form:%
\begin{equation}
xy = z^{N+1}+wz^{N},\label{enhanceda1}%
\end{equation}
where the curve $C$ is defined by $x=y=z=0$ and $w$ restricts to a local
coordinate on $C$ centered at $p_i$. We now study the deformation theory of this
singularity. Locally, the sheaf $\mathcal{T}^{1}$ is given by the Jacobian
ring, which in this case is
\begin{equation}
\mathbb{C}[x,y,z,w]/\left(  x,y,(N+1)z^{n}+Nwz^{N-1},z^{N}\right)
=\mathbb{C}[x,y,z,w]/\left(  x,y,z^{N},wz^{N-1}\right)  .
\end{equation}
Now, $\mathcal{T}^{1}$ is not a line bundle at $w=0$ due to the presence of
torsion: $z^{N-1}\in\mathcal{T}^{1}$ is a torsion element of $\mathcal{T}^{1}$
supported at $p_i$, as it is annihilated by the maximal ideal $(x,y,z,w)$ at
$p_i$. Modding out by this torsion element gives
\begin{equation}
\mathcal{T}^{1}/\left(  \mathbb{C}\cdot z^{N-1}\right)  \simeq\mathbb{C}%
[x,y,z,w]/(x,y,z^{N})\simeq\mathbb{C}[z,w]/z^{N}.\label{mattermodtors}%
\end{equation}
Since the quotient is a vector bundle as before, there can be no torsion element
other than scalar multiples of $z^{N-1}$.

Globally, the torsion subsheaf $\mathrm{Tors}\left(  \mathcal{T}^{1}\right)  $
of $\mathcal{T}^{1}$ is a skyscaper sheaf supported on $P$, one-dimensional
over each $p_i\in P$, and the quotient $\mathcal{T}^{1}/\mathrm{Tors}\left(
\mathcal{T}^{1}\right)  $ is a vector bundle on $C$. We can \ infer the global
structure of this bundle from the term $wz^{N}$ in (\ref{enhanceda1}), which
is a section of $\mathcal{O}(NK_{C})$ times a local equation for $p_i\in P$.
Globally, this is a section of $\mathcal{O}(NK_{C}+P)$.

The deformations of (\ref{enhanceda1}) corresponding to the local generators
$(z^{N-1},z^{N-2},\ldots,z,1)$ of $\mathcal{T}^{1}/(\mathrm{Tors}%
(\mathcal{T}^{1}))$ as an $\mathcal{O}_{C}$ bundle (cf \ref{mattermodtors})
can be written as
\begin{equation}
xy = z^{N+1}+wz^{N}+\sum_{j=2}^{N}a_{j}(w)z^{N-j}.
\end{equation}
Globally, the equation needs to hold in $\mathcal{O}(NK_{C}+P)$, and so can be
written as
\begin{equation}
xy = rz^{N+1}+sz^{N}+\sum_{j=2}^{N}a_{j}z^{N-j}, \label{defwithmatter}%
\end{equation}
where $z$ as before is in the bundle $\mathcal{O}(K_{C})$, $s$ is a global
section of $\mathcal{O}(P)$ vanishing at $P$, $r\in H^{0}(\mathcal{O}%
(P-K_{C}))$, and $a_{j}\in H^{0}(\mathcal{O}(jK_{C}+P))$.

This calculation shows that
\begin{equation}
\mathcal{T}^{1}/(\mathrm{Tors}(\mathcal{T}^{1}))\simeq i_{\ast}\left(
\mathcal{O}_{C}(2K_{C}+P)\oplus\mathcal{O}_{C}(3K_{C}+P)\oplus\cdots
\oplus\mathcal{O}_{C}(NK_{C}+P)\right)  .
\end{equation}
For later use, note that if we set $s=a_{i}=0$ (and add higher order terms),
we get an $SU(N+1)$ theory with matter localized at the zeros of $r$. Turning
on $s$ corresponds to partially Higgsing to $SU(N)$. If we use a parabolic
Hitchin system to describe the $SU(N+1)$ theory, then we will see later that
the complex structure deformation given by $s$ partners with RR moduli and
can collectively be described by a parabolic $SU(2)$ Hitchin system with
singularities at the zeros of $r$.

Returning to our main line of development, we therefore get an exact sequence
\begin{equation}
0\rightarrow\mathrm{Tors}\left(  \mathcal{T}^{1}\right)  \rightarrow
\mathcal{T}^{1}\rightarrow i_{\ast}\left(  \mathcal{O}_{C}(2K_{C}+P%
)\oplus\mathcal{O}_{C}(3K_{C}+P)\oplus\cdots\oplus\mathcal{O}_{C}(nK_{C}%
+P)\right)  \rightarrow0. \label{t1}%
\end{equation}
As a check, we have an $SU(N)$ gauge symmetry with $g$ adjoints and $|P|$
fundamentals. From our analysis in subsection \ref{ssec:SUNexamp}, we know
that we should find $(N^{2}-1)(g-1)+N|P|$ complex structure deformations. We
check this against $H^{0}(\mathcal{T}^{1})$. From (\ref{t1}) we get:%
\begin{equation}
0\rightarrow H^{0}(\mathrm{Tors}\left(  \mathcal{T}^{1}\right)  )\rightarrow
H^{0}(\mathcal{T}^{1})\rightarrow\oplus_{j=2}^{N}H^{0}(\mathcal{O}%
(jK_{C}+P))\rightarrow0. \label{cxmodulia}%
\end{equation}
But $h^{0}(\mathrm{Tors}\left(  \mathcal{T}^{1}\right)  )=|P|$, while
$h^{0}(\mathcal{O}(jK_{C}+P))=(2j-1)(g-1)+|P|$ by Riemann-Roch, and so
$h^{0}(\mathcal{T}^{1})=|P|+\sum_{j=2}^{N}((2j-1)(g-1)+|P|)=(N^{2}%
-1)(g-1)+N|P|,$ as expected.

Now we turn to the RR\ moduli for the smoothed system and track their behavior in the
limit where we switch off the smoothing parameters. We will show that a
parabolic Hitchin system emerges from the limiting mixed Hodge structure. For
each point $p\in P$ we associate a vector space $V_{p}$ with a
flag $0\subset W_{p}\subset V_{p}$ with $\dim W_{p}=1$. The parabolic Hitchin
system allows Higgs fields $\Phi$ which are meromorphic along $P$, and whose
residues are nilpotent with respect to the flag. The nilpotency condition
assures that the Casimirs of $\Phi$ are sections of $\mathcal{O}(j K_{C}+P)$,
which should be compared to (\ref{cxmodulia}).

In Appendix \ref{app:LIMITS} we present a special example where $N = 2$. We then
compute the limiting mixed Hodge structure of the degeneration
\begin{equation}
xy = rz^{3}+sz^{2}+t^{2}q,
\end{equation}
where $q\in H^{0}(\mathcal{O}(2K_{C}+P))$. As in the conifold case, the
limiting mixed Hodge structure has weights 2, 3, and 4. The weight 2 part is
$|P|$-dimensional. For the Jacobian of $H_{\mathrm{lim}}^{3}$, we get a
$(\mathbb{C}^{\ast})^{|P|}$ fibration over the Jacobian of the ordinary weight
3 Hodge structure $Gr_{3}H_{\mathrm{lim}}^{3}$. We also see that the
contribution of the singularity to $J^{2}(Gr_{3}H_{\mathrm{lim}}^{3})$ is
precisely the fiber over $q\in H^{0}(2K_{C} + P)$ of an $SU(2)$ parabolic Hitchin
system.

We therefore assert that the local part of the hypermultiplet moduli
space in which the location of the points $P$ are fixed
fibers over a parabolic Hitchin system with fibers the local RR moduli of complex
dimension $|P|$. The $|P|$-dimensional RR moduli are computed in
Appendix \ref{app:LIMITS}.  These RR moduli combine with the $|P|$ complex moduli corresponding
to complex structure deformations moving the points $P$ (i.e.\ to $H^{0}%
(\mathrm{Tors}(\mathcal{T}^{1}))$) to form $|P|$ quaternionic moduli.

\subsubsection{Example: A Local F-theory Model \label{ssec:E8pluspoles}}

To illustrate the considerations of the previous subsection, we now turn to a local F-theory model
involving a seven-brane wrapping a $\mathbb{P}^1$ with some number of punctures. Along these lines, we consider
the elliptic Calabi-Yau threefold with base the Hirzebruch surface $\mathbb{F}_{n}$:
\begin{equation}
y^2 = x^3 + g_{12 + n}(z') z^5 + f_{8}(z') x z^4 + g_{12}(z') z^6 + g_{12 - n}(z') z^7
\end{equation}
where $(z,z')$ are respectively coordinates in the fiber and base
of $\mathbb{F}_n$. To get a consistent F-theory model (i.e. not violate the Calabi-Yau condition after blowing up),
we need to take $-12 \leq n \leq 12$. For additional review,
see \cite{MorrisonVafaI, MorrisonVafaII, BershadskyPLUS} and our later discussion of
T-branes in global models in subsection \ref{het_f_review}.

For our present purposes, the main point is that there is an $E_8$ singularity located at $z = 0$ and another one at
$z = \infty$.\footnote{The $E_8$ at $z = \infty$ can be put in standard ADE form by switching to
a different affine coordinate patch.} These are associated with a seven-brane
wrapping a divisor given by a $\mathbb{P}^{1}$. So, without loss of generality, we shall focus
on the non-compact model defined by the equation:
\begin{equation}
y^2 = x^3 + g_{12 + n}(z') z^5.
\end{equation}
Following up on our previous discussion, we see that there is an associated parabolic Higgs bundle with
poles located at the zeroes of $g_{12 + n}(z')$. In contrast to previous cases where we could associate these poles with vevs of bifundamental
vevs, in the case of an $E_8$ gauge theory, here these poles are induced by condensing modes of a tensionless string theory
with degrees of freedom localized at these intersection points. This is simply because we cannot embed our $E_8$ gauge theory
in a unitary theory with a bigger simple gauge group. We can track the behavior of the intermediate Jacobian locally
by switching on a lower order deformation to an $E_7$ singularity:
\begin{equation}
y^2 = x^3 + g_{12 + n}(z') z^5 + f_{8 + n}(z') x z^3.
\end{equation}
Let us note that to activate this unfolding, we need to assume that $8 + n \geq 0 $. When this is not satisfied
there is no unfolding, i.e. Higgsing, available.

Assuming an unfolding to $E_7$ is possible, we can consider the $SU(2)$ parabolic Hitchin system in
the limit that we switch off the smoothing, i.e. send the coefficient $f_{8 + n}$ to zero. Working in a patch of the $\mathbb{P}^1$
which contains the zeroes of $f_{8 + n}$, we can describe the T-brane configuration as a Higgs field taking values in the $\frak{su}(2)$ factor of
the $\frak{e}_{7} \times \frak{su}(2) \subset \frak{e}_8$ subalgebra with matrix representative:
\begin{equation}
\Phi = \left[
\begin{array}
[c]{cc}%
0 & f_{8 + n}\\
\varepsilon & 0
\end{array}
\right]
\end{equation}
where the T-brane is reached by sending $\varepsilon \rightarrow 0$. The system therefore
contains $8 + n$ localized matter fields, corresponding to the number of
half hypermultiplets in the $\mathbf{56}$ of $\frak{e}_{7}$. To turn the
$\varepsilon \to 0$ limit into a globally well-defined
configuration on the compact $\mathbb{P}^{1}$ base wrapped by the seven-brane, we need to also mark the points
where $g_{12 + n}$ vanishes. Doing so, we can view the Higgs field as a map:
\begin{equation}
\Phi: E \rightarrow E \otimes K(g_{12 + n})
\end{equation}
where $K(g_{12 + n})$ is the bundle of differentials with poles on $g_{12 + n}$.

If $\Phi$ is nilpotent but nonzero, the methods of \cite{del} exhibit $E$ as
an extension of line bundles
\begin{equation}
0 \rightarrow L^{-1} \rightarrow E \rightarrow L \rightarrow 0.
\label{eext}%
\end{equation}
The Higgs field $\Phi$ is determined by the additional data of a nonzero section
$s\in H^0(\mathbb{P}^1,K(g_{12+n})\otimes L^{-2})$ as the composition
\begin{equation}
E\to L\to L^{-1}\otimes K(g_{12+n})\to E\otimes K(g_{12+n}).
\label{ehiggs}%
\end{equation}
In (\ref{ehiggs}), the second map comes from multiplication by $s$, and the
other maps are deduced from (\ref{eext}).

At this stage, it is convenient to further divide the analysis according
to whether $n$ is even or odd, since we are dealing with half
hypermultiplets. For $n$ even, there is a natural choice given by:
\begin{equation}
L = \mathcal{O}(  (10+n) / 2).
\end{equation}
In this case, the bundle $\mathcal{O}(K(g_{12+n}))\otimes L^{-2}$ is trivial,
so that $s$ and hence $\Phi$ may be taken to be nowhere vanishing.

For $n$ odd, we can see that there is a small subtlety, because there is no such line bundle.
Indeed, this is related to the fact that in the six-dimensional effective theory, one cannot
give a vev to a single half hypermultiplet. We can, however, still consider the moduli space
of rank two bundles consistent with our requirements on localized matter and poles.

Now, we can also see that there are $9 + n$ degrees of freedom for deforming these bundle (RR) moduli
preserving nilpotency. In the case of $n$ even, these are given by the space of extensions
\begin{equation}\label{RRextender}
0 \rightarrow \mathcal{O}( - (10 + n)/2) \rightarrow E' \rightarrow \mathcal{O}(  (10 + n) / 2 ) \to 0
\end{equation}
which is dual to $H^0(\mathbb{P}^1 , \mathcal{O}(8 + n))$. The latter characterization also works for $n$ odd. Now,
if we approach the $E_8$ locus by adding a term $\varepsilon f_{8 + n}(z') x z^3$ and let $\varepsilon$ go
to zero, that will constrain the RR moduli to Higgs fields that vanish along $f$ and for which $E'$ splits
as $\mathcal{O}(1) \oplus \mathcal{O}(-1)$ by the above description of extensions.

To summarize, we see that from the perspective of the parabolic Higgs bundle, we work over a $\mathbb{P}^{1}$ with $12 + n$ punctures. At
these punctures, we can specify additional non-normalizable boundary data for the Hitchin system, such as the asymptotic behavior of the Higgs field and gauge field
holonomies. We also have moduli from the positions of the localized
matter, i.e. $9+n$ such moduli associated with the choice of polynomial $f_{8+n}$. These are associated
with the breaking pattern $\frak{e}_{8} \rightarrow \frak{e}_7$, i.e. a choice of unfolding in the generic case.
We also see that the extensions in line (\ref{RRextender}) are the $9+n$ partners in the quaternionic Kahler moduli space for
the $h^{0}(\mathbb{P}^{1} , \mathcal{O}(8+n))$ deformations. Additionally, once we fix a choice of background Higgs field
and gauge bundle, we have matter fields localized at the zeroes of $f_{8+n}$.

Turning next to the characterization in the limiting mixed Hodge structure analysis, we can see a similar
split between the treatment of the moduli associated with $g_{12+n}$ and the moduli associated with $f_{8+n}$. The
moduli associated with the pole data, i.e. the zeroes of $g_{12+n}$ combine to form a $12+n$-dimensional quaternionic Kahler moduli space. Here,
these are associated with the complex structure moduli from moving around the zeroes of $g_{12+n}$ (which was the data $\mathcal{T}^1$ of the $SU(N)$ example
encountered previously) and the limiting behavior of the weight two part of the intermediate
Jacobian $J^{2}(H^{3}_{\mathrm{lim}})$. Additionally, we have the $9+n$ moduli associated with the choice of $f_{8+n}$. Altogether, these
combine to fill out a $21 + 2n$ dimensional quaternionic Kahler moduli space.

The distinction between the contributions to the moduli space from $g_{12+n}$ and $f_{8+n}$ reflected in the filtration of the limiting mixed Hodge structure
is naturally reflected in the mass hierarchies of the physical theory. Indeed, the background values at the zeroes of $g_{12 + n}$ specify a
UV cutoff for our local analysis (the background data of the parabolic Hitchin system), and when we unfold from $E_8$ to $E_7$, we are tilting the
stack of $E_8$ seven-branes. The moduli associated with this breaking pattern correspond to activating a choice of solution to the parabolic
Hitchin system, with boundary data at the zeroes of $g_{12 + n}$. For a given choice of background, the zeroes of $f_{8+n}$ lead to localized matter fields
in the $\mathbf{56}$ of $\frak{e}_{7}$. In our present discussion we have kept the vevs of the localized matter
switched off.

Once we recouple to gravity, however, we should really view all contributions to the moduli space on an equal footing.
The reason is that in this limit, the ``boundary data'' of the parabolic Higgs bundle is associated with a mass scale
which cannot be taken arbitrarily large, the upper bound being the Planck scale. So more generally, we expect the two
contributions to the quaternionic Kahler moduli space to combine into a single moduli space of quaternionic
dimension $21 + 2n$. In fact, we can see how this regrouping has to work by initiating a further
unfolding from $E_7$ down to $E_6$ by activating a vev for some combination of
the $\mathbf{56}$'s of $\frak{e}_{7}$. In this case,
the local profile of the spectral equation for the Higgs field in
the non-compact geometry $\mathcal{O}(K_C) \rightarrow C$ takes the form:
\begin{equation}
g_{12 + n} s^3 + f_{8+n} s + q_{6 + n} = 0
\end{equation}
where $s$ is a normal bundle coordinate such that $s = 0$ is the curve $C$. This is simply the spectral equation for
an $SU(3)$ parabolic Hitchin system with pole data at the zeroes of $g_{12 + n}$.
At each zero of $q_{6 + n}$ we have a $\mathbf{27}$ of $\frak{e}_{6}$. In the limit where the vevs of the $\mathbf{56}$ of $\frak{e}_{7}$
are taken to be very large compared to fluctuations in the $\mathbf{27}$ of $\frak{e}_{6}$, we see that this amounts to additional ``boundary data''
for the system. Indeed, we have the complex structure moduli associated with deformations of the polynomials $g_{12 + n}$ and $f_{8+n}$, and the remnants of the three-form
potential moduli are now packaged together in the limiting behavior of the weight two part of
$J^{2}(H^{3}_{\mathrm{lim}})$.

Anticipating our later discussion in section \ref{sec:COMPACT}, we will later show that the case $n = 12$
corresponds to the F-theory dual of the standard embedding of the spin connection in an $E_8$ factor
for heterotic strings on a K3 surface. Our treatment in this paper should therefore also be viewed as
generalizing as well as clarifying the heuristic treatment provided in \cite{glueII}.

\section{Defining Data in Six Dimensions \label{sec:PRESCRIPTION}}

Having identified the geometric remnant of T-brane data, we can now give a
more precise statement on the defining data of an F-theory compactification.
For the most part, this agrees with the operative definition used in the
literature, though in singular limits, our analysis demonstrates that
additional care is necessary. In fact, part of the point of this work is that the local
gauge theory provides a clean definition of various singular limits of the
closed string moduli space.

At a practical level, it is typically challenging to start from
a fully smooth $X_{\text{smth}}$ and then track various degeneration limits. Indeed,
in much of the literature on F-theory compactification, the logical
order is actually reversed: One starts from a singular threefold $X$ and either \textquotedblleft
unfolds\textquotedblright\ or \textquotedblleft blows up\textquotedblright\ the
singularities of $X$. There is a fly in the ointment, though, because
T-branes can obstruct such blowups! To properly define an
F-theory compactification on a singular space $X$, it is therefore necessary
to supplement this geometry by additional physical data.

We propose that in addition to the singular space $X$, we must also specify
some T-brane data. In the Hitchin system, this is associated with the
\textquotedblleft discrete flux data\textquotedblright\ and choice of a flat connection. In the lift
to a IIA / M-theory compactification, this will be captured by a four-form flux and three-form potential
moduli. For \textit{smooth} $X$, this data is captured by the Deligne cohomology of $X$ \cite{Donagi:1998vw}.\footnote{
This defining data is well-known for M-theory compactified on a smooth
eight manifold \cite{Becker:1996gj, Witten:1996md}.}
Motivated by our earlier analysis, we shall in fact propose the local Hitchin
system as the limiting behavior of the Deligne cohomology for \textit{singular} $X$.

The rest of this section is organized as follows. First, we show that the
standard operation of \textquotedblleft blowing up in the
fiber\textquotedblright\ is actually obstructed by T-branes.
Then, we turn to our proposal for how to unambiguously specify an F-theory
vacuum on a singular threefold $X$.

\subsection{Obstruction to a Blowup \label{ssec:OBSTRUCT}}

A standard way to understand the gauge symmetry and matter content of F-theory on
a singular threefold $X$ is to consider the dual M-theory compactification and perform a sequence
of blowup operations to a smooth Calabi-Yau $\widetilde{X}$. In this
section we show that the presence of a T-brane can obstruct this
sort of blowup operation.\footnote{Observations of a similar spirit have been made in the context of Green-Schwarz
anomalous $U(1)$'s in F-theory \cite{Grimm:2010ez,Grimm:2011tb}.}

We can see this obstruction by considering the physical origin of the
blowup modes and the T-brane moduli. It is simplest to consider F-theory compactified on
$S^{1}\times X$. At low energies this yields
a five-dimensional theory with eight real supercharges. The blowup parameters
then correspond to the vevs of the real scalars in the 5D vector multiplets.
Activating a vev for the vector multiplets moves the theory onto the Coulomb
branch, while activating a vev for the hypermultiplets moves the theory onto
the Higgs branch.

Now, the key point is that once we move onto the Higgs branch, we have given a
mass to the vector multiplet. As a consequence, we cannot give a vev to the
scalar in the vector multiplet, and so we cannot activate a blowup mode. Said differently,
activating a T-brane breaks the gauge symmetry, so the vector
multiplet has a mass.

It is instructive to illustrate this for the four-dimensional effective field theory
defined by wrapping a seven-brane over $T^{2}\times C$. Then, the
superpotential for this system is:%
\begin{equation}
W=\underset{i=1}{\overset{g}{\sum}}\text{Tr}_{\frak{g}}\left(  \varphi \cdot \left[
X_{i},\widetilde{X}_{i}\right]  \right)  +\underset{p}{\sum}\widetilde
{q}_{(p)}\cdot\varphi\cdot q_{(p)},
\end{equation}
so once we activate a vev for one of the hypermultiplets, we induce a mass
term for some components of $\varphi$, the vector multiplet complex scalar.
We can also track the fate of the previously massless gauge bosons of the vector multiplet.
In the dimensional reduction in the IIA / M-theory descriptions, these gauge bosons come from integrating the three-form
potential along the fiber $\mathbb{P}^{1}$ for the resolution. The equation of motion for the three-form
potential is:
\begin{equation}
(\Delta_{4D}+\Delta_{\mathrm{internal}})C_{(3)}=0,
\end{equation}
with $\Delta_{4D}$ the 4D Laplacian, and $\Delta_{\mathrm{internal}}$ the Laplacian in the internal directions. So, upon
expanding $C_{3} = A_{4D} \wedge \omega_{\mathrm{internal}}$, with $A_{4D}$ the 4D gauge field and $\omega_{\mathrm{internal}}$
a two-form dual to the fiber $\mathbb{P}^{1}$, we see that once the gauge boson has
picked up a mass, $\omega_{\mathrm{internal}}$ is no longer a harmonic two-form.

Similar considerations hold in the purely geometric context. Indeed, in the
limiting mixed Hodge structure analysis, we can see that at a generic point of
complex structure moduli, there is no singularity to speak of, and so there is
no blowup to perform.

\subsection{Disambiguation}

We now give a proposal for what additional data needs to be
attached to a singular threefold $X$ to unambiguously define the corresponding
effective field theory derived from F-theory.

We motivate our proposal by first summarizing the evidence accumulated so far.
In the local models defined by a Hitchin system, the extra data needed to
specify the theory is associated with the remnants of the three-form potential moduli, and the
choice of an abelian flux. In compactifications to four dimensions, such seven-brane fluxes can be lifted to
a corresponding four-form flux $\mathcal{G}$ in the resolved geometry. Indeed, taking
$T^{2}\times X$ with no T-brane data switched on, we can consider the blowup
of some singular ADE\ fiber to reach the geometry $T^{2}\times\widetilde{X}$.
Then, we can consider the difference of two four-form fluxes, $\mathcal{G}_{1} - \mathcal{G}_{2}$ which
by quantization needs to be valued in $ H^{4}(T^{2}%
\times\widetilde{X},\mathbb{Z})$. The main requirement is that integrating the difference $\mathcal{G}_1 - \mathcal{G}_2$
over the two-cycle in the fiber of the ADE singularity descends to the corresponding abelian flux
in the Hitchin system. Observe that for a rank $r$ gauge group, there
are precisely $r$ linearly independent combinations of $U(1)$ generators, and
there $r$ homologically distinct $\mathbb{P}^{1}$'s in the ADE fiber.

Now, there are a few potential issues with defining an F-theory
compactification in this way. First of all, since we have already
argued that T-branes can obstruct a K\"{a}hler resolution, how
can we speak of the geometry $T^{2}\times\widetilde{X}$? The point is
that although there is no K\"{a}hler parameter which allows us to perform a
blowdown $\widetilde{X}$ $\rightarrow X$, we can still construct the
topological space $T^{2}\times\widetilde{X}$, even when a T-brane is switched
on. The construction is simply to delete the singular ADE\ fibers and replace
them with the blown up fibers. Since we only need the integrality condition
for the difference of two G-fluxes, this is sufficient for our purposes.\footnote{For a recent approach to
reading off matter from geometry from a deformation-theoretic perspective, see \cite{Grassi:2013kha}.}

The second potential issue is that at least when $X$ is a \textit{smooth} Calabi-Yau threefold, it
is well-known that the supergravity equations of motion forbid the presence of
any four-form flux. The reason this is bypassed when T-branes are switched on
is that we are working with a \textit{singular} geometry. Indeed, in the Hitchin system
equations of motion, the flux often balances against the nilpotent Higgs
field. Turning the discussion around, specifying a G-flux in this way gives a
necessary and sufficient condition to recognize the appearance of a T-brane in
the geometry.

Finally, in the context of compactifications to four dimensions, there are
tight global consistency conditions related to tadpole cancellation for
D3-brane charge. So, we might expect similar constraints in the six-dimensional case.
In fact, we can see that the fluxes we are activating always descend from a non-abelian gauge group.
This means, for example, that the net amount of five-brane charge induced on a
seven-brane is zero, since $\mathrm{Tr}_{\frak{g}}(F)=0$. On the other hand, we do not know the
full list of consistent G-fluxes which can be switched on. What we can assert
is that when we can take a local degeneration limit $X_{t}\rightarrow X$, it
is possible to consistently switch on such a G-flux. In fact, we can see that
this flux should loosely be thought of as a $(2,2)$-form, since it must
descend to a $(1,1)$-form flux in the Hitchin system.

Putting these remarks together, we propose that in the geometry $\widetilde
{X}$, the remnant of T-brane data is a four-form G-flux, while on the smoothed
side $X_{t}$, the remnant of T-brane data is the three-form potential moduli valued in the
intermediate Jacobian $J(X_{t})$. Giving this data is then enough to specify the effective
six-dimensional theory of an F-theory compactification.

Now, as noted in \cite{Donagi:1998vw}, in the case of a \textit{smooth} Calabi-Yau $X_{\text{smth}}$, the $(2,2)$-flux data and
intermediate Jacobian can actually be packaged in terms of a \textit{single}
mathematical object known as the Deligne cohomology $H_{\mathcal{D}}^{4}(X_{\text{smth}%
},\mathbb{Z(}2))$ (see Appendix \ref{briefDeligne} for a brief introduction). For our current purposes,
the main point is that it fits into the short exact sequence:
\begin{equation}
0\rightarrow J^{2}(X_{\text{smth}})\rightarrow H_{\mathcal{D}}^{4}(X_{\text{smth}%
},\mathbb{Z}(2))\rightarrow H_{%
\mathbb{Z}
}^{2,2}(X_{\text{smth}})\rightarrow0.
\end{equation}
So we see that $H_{\mathcal{D}}^{4}(X_{\text{smth}},\mathbb{Z}(2))$ captures
both the discrete flux data $H_{%
\mathbb{Z}
}^{2,2}(X_{\text{smth}})$, as well as the data of the intermediate Jacobian
$J(X_{\text{smth}})$. Actually, the more precise statement is that to account
for the possibility of half integer shifts in the quantization condition we
should only require that the difference of this flux data across a domain wall
is an element of the Deligne cohomology.

But in the case of a \textit{singular} manifold $X$, the usual definition of
Deligne cohomology breaks down. This motivates a conjecture: We propose
that the Hitchin system provides the correct definition of Deligne cohomology in certain
singular limits. To see this, we observe that the Hitchin system data captures
both the discrete flux data of the blowdown $\widetilde{X}\rightarrow X$ as well as the limiting behavior of the
RR moduli valued in $J(X_{t})$ in the limit $t\rightarrow0$. So in the singular limit,
the Hitchin system unifies these two contributions.

\section{Global Models and Heterotic Duals \label{sec:COMPACT}}

In this section we present some global examples of T-brane phenomena. Our strategy
will be to show how to go from the globally defined geometry to a local limit. In
particular, we will explain how to isolate the relevant contributions to the
three-form potential moduli of the dual M-theory / IIA description in taking the
local limit of the geometry.

Now, in the global setting we face some additional complications, because it is typically
not possible to globally smooth away all singularities. So, while we can still unfold some
component of the discriminant locus, other components may intersect it. A common circumstance in
F-theory is the collision with additional $I_1$ components, which
generically contains various cusps. To deal with these cases, we shall therefore need to
show how to isolate the relevant contributions to the three-form potential moduli in taking the local limit
of the geometry. Though more technically involved, in principle we can simply carry over our
previous discussion of the theory of limiting mixed Hodge structures to this case as well.

To give some concrete examples of global models with T-branes, we focus on some simplifying cases where a heterotic
dual description exists. In technical terms, the reason this leads to a simplification is that the
task of computing the relevant contributions to the intermediate Jacobian
$J(X)$ reduces to the calculation of the Jacobian $J(\widetilde{C})$ of an algebraic curve $\widetilde{C}$. This
curve $\widetilde{C}$ is nothing other than the spectral curve which figures in both the heterotic string
construction of the vector bundle, as well as in the local behavior of intersecting seven-branes
on the F-theory side.

Another benefit of this analysis is that we will be able to see what T-brane data turns into in the
dual heterotic description. As one might expect, this corresponds to situations where the spectral
curve used in the spectral cover construction of bundles on K3 has become singular.

More striking is that this occurs in a host of rather ordinary and well-known cases! For example, the
standard embedding of the spin connection in an $E_8$ vector bundle is a perfectly smooth vector
bundle which nevertheless has a singular spectral curve \cite{Aspinwall:1998he} (see also \cite{glueII}).
We provide compact examples along these lines which illustrate various features of T-brane data
and their manifestation in both the F-theory and heterotic descriptions.

The rest of this section is organized as follows. First, we give a brief review of heterotic / F-theory
duality in six dimensions, and show how the elements of the intermediate Jacobian relevant for T-branes
reduce to related questions in the associated spectral curve. After this, we turn to some examples of
T-branes and their heterotic duals.

\subsection{Review of 6D Heterotic F-theory Duality \label{het_f_review}}

Since we shall be making heavy use of it later, in this subsection we briefly review
some aspects of the duality in six-dimensions between heterotic
strings compactified on a K3 surface and F-theory compactified on a Hirzebruch $\mathbb{F}_{n}$ base.
Most checks of this duality have been performed at \textit{generic} points in the moduli space.
Indeed, we will see that certain ambiguities crop up in singular limits, and need to be treated with
additional care.

Loosely speaking, the six-dimensional duality is obtained by applying fiberwise the eight-dimensional duality
between the $E_8 \times E_8$ heterotic string on $T^2$ and F-theory on an elliptic K3 surface. Fibering over a common $\mathbb{P}^{1}$,
we arrive at a K3 surface on the heterotic side, and on the F-theory side, an elliptically fibered Calabi-Yau threefold
with Hirzebruch surface as the base. We recall that the Hirzebruch surfaces are given for $n \in \mathbb{Z}$ by a
fibration $\mathbb{P}^{1}_{\mathrm{fiber}} \rightarrow \mathbb{F}_{n} \rightarrow \mathbb{P}^{1}_{\mathrm{base}}$ defined by projectivizing the rank~2 bundle
$\cO_{\mathbb{P}^1}\oplus\cO_{\mathbb{P}^1}(n)$. For physics applications,
we need to restrict to $-12 \leq n \leq 12$. In the heterotic dual description, the parameter $n$ indicates the number of instantons
$(12+n , 12-n)$ in the $E_8 \times E_8$ bundle. For additional
details on the proposed duality and numerous previous checks, see e.g. \cite{MorrisonVafaI, MorrisonVafaII, BershadskyPLUS}.

The F-theory threefold, $\pi: X \to B$ can be described in minimal Weierstrass form as
\begin{equation}\label{fweier}
y^2=x^3 + f x + g
\end{equation}
where $f \in H^0(B, K_{B}^{-4})$, $g \in H^0(\mathbb{F}_n, K_{B}^{-6})$, $K_{B}$ is the canonical
bundle of the base $B = \mathbb{F}_n$. Introducing affine coordinates $(z, z')$ for the fiber and base, respectively,
we can perform a further expansion:
\begin{equation} \label{expansion}
f = \sum_{i} f_{8 + n(4 - i)}(z')z^{i} \,\,\,\, \text{and} \,\,\,\, g = \sum_{j} g_{12 + n(6-j)}(z')z^{j}
\end{equation}
where the sum on $i$ and $j$ is over all non-negative degree terms. Here,
$z=0$ specifies the base $\mathbb{P}^{1}_{\mathrm{base}}$ of the Hirzberuch surface degree.
The discriminant locus:
\begin{equation}\label{delta_again}
\Delta=4 f^{3}+27 g^2
\end{equation}
defines a divisor in the Hirzebruch surface, and components
of the discriminant locus indicate the locations of
seven-branes in the geometry.

For our present purposes, the primary advantage of this class of examples is the existence of a
globally defined K3 fibration.\footnote{For a recent proposed extension of heterotic / F-theory duality which
does not require a global K3 fibration, see \cite{Heckman:2013sfa}.} In a suitable limit of moduli for the Calabi-Yau metric, the K3
fiber can be split into two ``1/2 K3's'', i.e. del Pezzo nine ($dP_9$)
surfaces where the fiber $\mathbb{P}^{1}_{\mathrm{fiber}}$ asymptotes to a cylinder, i.e a sphere with two punctures. In
the stable degeneration limit, $X$ splits up into the components
$X = \mathcal{X}_{1} \cup \mathcal{X}_{2}$, where each $\mathcal{X}_{i}$ is given by a $dP_9$ fibration over the base $\mathbb{P}^{1}_{\mathrm{base}}$.
In the dual heterotic description, these two factors are associated with the two $E_8$ vector bundles.

In the dual heterotic M-theory description \cite{VafaFTHEORY, MorrisonVafaI, MorrisonVafaII}, the heterotic dilaton is
given to leading order by the expression:
\begin{equation}
\exp(-2 \phi_{\mathrm{het}}) = \frac{\mathrm{Vol}(\mathbb{P}^{1}_{\mathrm{base}})}{\mathrm{Vol}(\mathbb{P}^{1}_{\mathrm{fiber}})}
\end{equation}
Moreover, we can also recognize the dual heterotic K3 surface. In the middle region of the cylinder, we can locally approximate the Calabi-Yau threefold as:
\begin{equation}
y^2=x^3 + f_{8}(z')z^{4} x + g_{12}(z')z^{6}
\end{equation}
which for $z$ held fixed defines an elliptically fibered K3 surface in the variables $x,y,z^{\prime}$. In fact, these middle coefficients should be viewed as defining the moduli of the
K3 surface in the dual heterotic string description:\footnote{There is a slight subtlety with this statement owing to the non-trivial fibration over the base $\mathbb{P}^{1}_{\mathrm{base}}$. We have $9 + 13$ complex coefficients, but we have fixed the locations $z = 0$ and $z = \infty$, where the two stacks of branes are located. This means that the total number of moduli is given by $22 - 2 = 20$. This accounts for the $18$ complex structure moduli of the heterotic K3, as well as the two complexified K\"ahler moduli of the elliptic K3 in the heterotic dual description.}
\begin{equation}\label{hetK3}
v^2 = u^{3} + f_{8}(z') u + g_{12}(z'),
\end{equation}
where to make the context clear, we reserve $(x,y,z,z^{\prime})$ for coordinates in
the F-theory geometry, and $(u,v,z^{\prime})$ for coordinates in the dual
heterotic K3.

The other complex structure moduli of equation (\ref{fweier}) translate to deformations of the vector bundles $\mathcal{V}_1, \mathcal{V}_2$ of the
two $E_8$ factors. The terms in $f$ and $g$ of respective degrees less than four and six (i.e. concentrated near $z = 0$) correspond
to moduli for $\mathcal{V}_1$, while the terms of higher degree (i.e. concentrated near $z = \infty$) correspond to moduli for $\mathcal{V}_2$. A quick way
to see this is to start from the singular model:
\begin{equation}\label{singWeier}
y^2 = x^3 + g_{12 + n}(z') z^5 + f_{8}(z') x z^4 + g_{12}{z'} z^6 + g_{12 - n}(z') z^7,
\end{equation}
and to then unfold by switching on deformations of the singularity near $z = 0$ and $z = \infty$. In \cite{MorrisonVafaI, MorrisonVafaII}, the
further unfolding to a lower singularity was interpreted as dissolving the small instantons located at the zeroes of
$g_{12 + n}(z')$ and $g_{12 - n}(z')$ back into smooth vector bundle moduli.\footnote{In slightly more detail, the idea is that
when all unfolding parameters have been switched off, we should perform a blowup at each intersection of $g_{12 + n}(z') = 0$ with $z = 0$. These blowups
corresponding to pulling an M5-brane off the end of the world nine-brane in the heterotic M-theory description. Similar considerations hold for the
blowups at the intersections of $g_{12 - n}(z') = 0$ with $z = \infty$.} As the astute reader will have no doubt noticed,
there is an ambiguity in just specifying the physical system by the singular geometry of equation (\ref{singWeier}), since T-brane data could be hiding
in the discriminant. We return to this in subsection \ref{ssec:SuperSTD}.

\subsubsection{The Spectral Cover Construction}

Some of the most detailed checks of heterotic / F-theory duality have been performed for vector bundles
produced via the spectral cover construction. Here we shall be interested in the case of heterotic strings
compactified on a K3 surface. As we shall
repeatedly stress, sometimes a singular spectral cover description can correspond to
a perfectly smooth vector bundle.

We are mainly interested in vector bundles on the K3 surface, so for now we simply reference
a vector bundle $\mathcal{V}$ with structure group $G \subset E_8$. For
ease of exposition, we shall take $G = SU(N)$, though nothing depends on this restriction. Now,
for an elliptically fibered K3 surface, the defining data of the spectral
cover consists of a pair $(\widetilde{C}, L_{\widetilde C})$ where $\widetilde{C}$ is a compact curve
inside of K3, given by an $r$-sheeted cover of the base $\mathbb{P}^1$,
and $L_{{\widetilde C}}$ is a rank $1$ sheaf defined over the curve ${\widetilde C}$ (See Appendix \ref{app:StabbingDegenerates}
and \cite{Friedman:1997yq} for a review). We turn this into a vector bundle on the elliptic K3 by
applying a fiberwise T-duality, i.e. Fourier-Mukai transform (see e.g. \cite{Friedman:1997ih,Curio:1998bva}). The
number $r = rk(\mathcal{V})$ then corresponds to the rank of the vector bundle.
One can also go the other way, starting from a general vector bundle $\mathcal{V}$ and
by applying a Fourier-Mukai transform producing some pair $(\widetilde{C}, L_{\widetilde C})$, though this
may lead to a singular spectral curve $\widetilde{C}$.

For an $SU(N)$ vector bundle, we can write the associated spectral curve $\widetilde{C}$
as the zero set \cite{Friedman:1997yq}:
\begin{equation}\label{spec_cov}
\widetilde{C} = \{ a_0 w^N +a_2 u w^{N-2}+a_3 v w^{N-3}+ \ldots = 0 \}
\end{equation}
ending in $a_N u ^{N/2}$ for $N$ even and $a_N u^{(N-3)/2}v$ for $N$ odd. Here, the
$a_i$ are given as sections:
\begin{equation}
a_i \in H^0(\mathbb{P}^1, K_{\mathbb{P}^1}^{\otimes i} \otimes \mathcal{O}(12 + n)).
\end{equation}
Anticipating our discussion of the F-theory geometry, we note that $a_i(z')$ specifies
a degree $12 + n - 2i$ polynomial in the variable $z'$. The class of the curve $\widetilde{C}$ is
$[\widetilde{C}]=N \sigma_{0} + (12+n) f$ where $\sigma_{0}$ is the class of the zero-section of K3 and $f$ is the fiber class. The two
classes satisfy $\sigma_{0} \cdot f = 1$, $f \cdot f = 0$ and $\sigma_{0} \cdot \sigma_{0} = -2 $. The
genus of the spectral curve $\widetilde{C}$ follows from an application of intersection theory:
\begin{equation} \label{genusSmooth}
2 g(\widetilde{C}) = -2 N^2 + 2N(12 + n) + 2 = -2 (N^2 - 1) + 2 N c_2(V) =  h^{1}(K3 , \mathrm{End}_0(V))
\end{equation}
where the right hand side follows from an index computation. This match is not an accident, and reflects the fact that
the vector bundle moduli space of the ${\cal N}=2$ theory appears as a pairing in the degrees of freedom
parameterized by deformations of the curve $\widetilde{C}$, $Def(\widetilde{C})$ with those in its Jacobian, $J(\widetilde {C})$.
The space of local deformations is counted by $h^{0}(\widetilde{C} , \mathcal{O}(K_{\widetilde{C}})) = g(\widetilde{C})$, while the
Jacobian is simply the set of all locally free sheaves (i.e. line bundles)
on ${\widetilde C}$ and $\mathrm{dim}_{\mathbb{C}}J({\widetilde C})=g(\widetilde{C})$.

Now, we have already encountered a rather similar pairing in F-theory between complex deformations of the geometry, and the
associated intermediate Jacobian of the Calabi-Yau threefold. As explained in more detail in Appendix \ref{app:StabbingDegenerates}, we can again
track the behavior of this structure in the stable degeneration limit, where the relevant correspondence now involves
the complex deformations and intermediate Jacobian of $\mathcal{X}$, one of the factors in $X = \mathcal{X}_{1} \cup \mathcal{X}_{2}$.
We summarize the correspondence between the relevant heterotic and F-theory structures in table 1.

{\small \begin{table}[t!]
\begin{center}
\begin{tabular}{|c|c|}\hline
Het/Spectral Cover  &  F-theory/Global \\\hline\hline
$Def({\widetilde C})$   &  $Def({\cal X})$ \\ \hline
$J({\widetilde C})$   &  $J(\mathcal{X}) $ \\ \hline
\end{tabular}
\caption{\it\small Summary of six-dimensional heterotic F-theory duality and moduli matching for smooth spectral covers. On the heterotic side, we have the
deformations of the spectral curve $Def({\widetilde C})$, and the Jacobian of the curve, $J({\widetilde C})$. On the F-theory side, we have
the space $Def({\cal X})$, which refers to the space of complex structure deformations of (the resolution of) $\mathcal{X}$,
and $J(\mathcal{X})$ refers to the intermediate Jacobian. Implicit in the definition and the correspondence is the assumption
that all quantities are non-singular.}
\end{center}\label{mod_table}
\end{table}}

\begin{figure}
[t!]
\begin{center}
\includegraphics[
height=3.5068in,
width=4.6311in
]%
{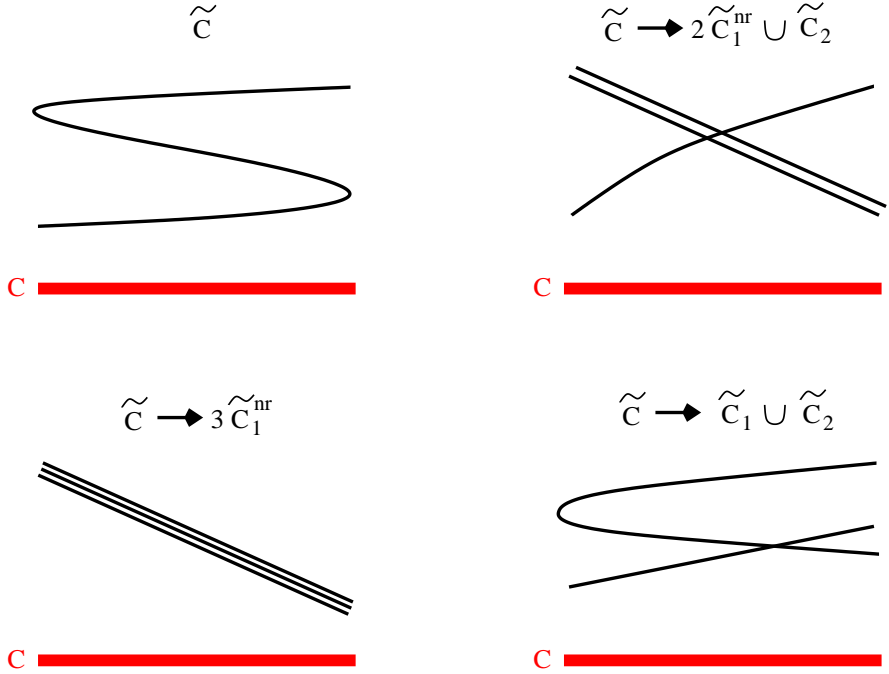}%
\caption{Depiction of the spectral curve $\widetilde{C}$ and some possible
degenerations. In the figure, we illustrate in the case of a three-sheeted
cover of a curve $C$. Possible degenerations include a non-reduced scheme of
length three, as indicated by $\widetilde{C}\rightarrow3\widetilde{C}%
_{1}^{\text{nr}}$, by a factorization into two smooth reducible components, as
indicated by $\widetilde{C}\rightarrow\widetilde{C}_{1} \cup \widetilde{C}_{2}$,
and a factorization into two reducible components, one of which is a
non-reduced scheme, as indicated by $\widetilde{C}\rightarrow2\widetilde
{C}_{1}^{\text{nr}} \cup \widetilde{C}_{2}$.}%
\label{spectralcurve}%
\end{center}
\end{figure}

Of course, in many cases of interest for physics we cannot expect to remain at a completely generic point of the moduli space. In these limits,
we need to extend the correspondence of table 1 to cover singular limits as well. From our analysis in earlier sections, we know that such singular limits
arise when we turn off smoothing parameters. In the dual heterotic description, this translates to limiting behavior for the spectral curve, and there are two basic ways
this can occur:
\begin{itemize}

\item $\widetilde{C}$ becomes reducible, i.e. splits into a collection of factors $\widetilde{C} \to \widetilde{C}_1 \cup \widetilde{C}_2 \cup \ldots$

\item $\widetilde{C}$ becomes a non-reduced scheme, i.e. it contains a factor $w^n=0$ for some function $w$.

\end{itemize}
In principle, both types of behavior could be present for a given spectral curve. See figure \ref{spectralcurve}
for an illustration of such degenerations.

But in spite of this apparent singular behavior, the resulting heterotic vector bundle \textit{could still be perfectly smooth} \cite{Aspinwall:1998he}.
Indeed, it is worth noting that there are a wide range of smooth heterotic bundles on K3 surfaces and Calabi-Yau threefolds
that give rise to singular spectral covers. For example, according to \cite{Bershadsky:1997zv}, most bundles built via the monad construction (see also \cite{Anderson:2009mh,Anderson:2008uw,Anderson:2007nc}) will yield a degenerate spectral cover. An important
example of this type is the tangent bundle on K3 \cite{Aspinwall:1997ye, glueII}. The extra data that makes it possible to
obtain a smooth bundle under the Fourier-Mukai transform is that for such
singular curves, the Jacobian of $\widetilde{C}$ is no longer a smooth torus and can now contain more exotic rank $1$ sheaves. The
new types of rank $1$ sheaves that can be defined over such singular curves have been well-studied in the mathematics
literature (see e.g. \cite{Markman, Donagi:1995am, del, SimpsonMartin}). Of course, we should expect something
like this to occur in the heterotic description, because on the F-theory side of the duality, these
singular limits are correlated with the possibility of T-brane data.

\subsection{From $J(X)$ to $J(\widetilde{C})$}

Having reviewed some basic elements of heterotic / F-theory duality, in this subsection we now show how
T-branes fit into this correspondence. One pragmatic reason for carrying out this detailed match is that in practice
it is often far simpler to calculate the limiting behavior of the Jacobian of a spectral curve compared with
the limiting behavior of the intermediate Jacobian of a Calabi-Yau threefold.

To this end, we first explain how to embed the compact spectral curve $\widetilde{C}$ in  $\mathcal{X}$,
one of the $dP_9$ fibered components appearing in the stable degeneration limit. Using this, we show
how the data of the intermediate Jacobian filters down to the spectral curve, and moreover, how to track the singular behavior of the
spectral curve. We should note that for \textit{non-singular} heterotic / F-theory pairs, this correspondence is by now well established \cite{MorrisonVafaI,MorrisonVafaII,Friedman:1997yq,Friedman:1997ih,Curio:1998bva} and continues to
yield new insights  \cite{Hayashi:2009ge, WatariTATARHETF, DWIII}. Our aim is to track the behavior of this correspondence
in various \textit{singular} limits.

\subsubsection{Projectivization \label{ssec:PROJ}}

In order to facilitate our analysis, it will prove convenient to switch to a different presentation of an elliptic curve.
As is standard in much of the F-theory literature, up to now we have presented the elliptic fiber on both the
heterotic and F-theory sides of the correspondence as a weight six hypersurface in the weighted projective space
$\mathbb{P}^{2}_{1,2,3}$. To avoid unnecessary complications from the orbifold singularities we now consider a presentation of
the elliptic fiber in a $\mathbb{P}^{2}$ bundle. At the expense of introducing spurious orbifold singularities,
we can of course switch back and forth between these presentations.

Let us now turn to the parametrization of the elliptic fibration, viewed as a $\mathbb{P}^{2}$ bundle
over a base curve $C = \mathbb{P}^{1}_{\mathrm{base}}$ on the heterotic side, and over the base Hirzebruch surface
$B = \mathbb{F}_{n}$ on the F-theory side. Along these lines, we introduce the $\mathbb{P}^{2}$ bundles:
\begin{equation}\label{projFIBER}
\mathcal{P} = \mathbb{P}(\mathcal{O}_{C} \oplus \mathcal{O}_{C}(4) \oplus \mathcal{O}_{C}(6))
\end{equation}
for the heterotic configuration, and:
\begin{equation}\label{projFIBER}
\mathcal{Y} = \mathbb{P}(\mathcal{O}_{B} \oplus \mathcal{O}_{B}(-4 \widetilde{K}_B) \oplus \mathcal{O}_{B}(-6 \widetilde{K}_B))
\end{equation}
for the F-theory configuration.  In (\ref{projFIBER}), $\widetilde{K}_B$
is the log canonical divisor $K_B+ D_{\mathrm{glue}}$ of the log pair $(B, D_{\mathrm{glue}})$,
where $D_{\mathrm{glue}}$ is the divisor along which the other component of the stable degeneration
is glued. Then, we can present the heterotic K3 surface as:
\begin{equation}\label{k3}
V^2 W = U^3 + f_8(z') U W^2 + g_{12}(z') W^3
\end{equation}
where $(W,U,V)$ are fiber coordinates:
\begin{equation}\label{hetweights}
W \in H^0(\mathcal{P} , \mathcal{O}_{\mathcal{P}}(\xi),\,\,\, U \in H^0(\mathcal{P}, \mathcal{O}_{\mathcal{P}}(\xi + 4h)),\,\,\, V \in H^0(\mathcal{P}, \mathcal{O}_{\mathcal{P}}(\xi +6h)),
\end{equation}
where $h$ is a fiber class and $\xi = c_1(\mathcal{O}_{\mathcal{P}}(1))$. Similar
considerations hold for the F-theory geometry $\mathcal{X}$, where by abuse of notation, we use the fiber coordinates $(\lambda , x , y)$.
The context will be clear when we are using the $\mathbb{P}^{2}$ bundle versus $\mathbb{P}^{2}_{1,2,3}$ bundle fiber coordinates. For the
$\mathbb{P}^{2}$ fiber bundle coordinates, we have sections:
\begin{equation}
\lambda\in H^0\left(\mathcal{Y},\cO_{\mathcal{Y}}(\xi)\right),\
x\in H^0\left(\mathcal{Y},\cO_{\mathcal{Y}}(\xi + 2\sigma_{b} + \left(2n+4\right) \sigma_{f})\right),\
y\in H^0\left(\mathcal{Y},\cO_{\mathcal{Y}}(\xi + 3\sigma_{b} + \left(3n+6\right) \sigma_{f})\right).\
\label{fweights}
\end{equation}
and $\xi = c_1(\mathcal{O}_{\mathcal{Y}}(1))$.
Here, $\sigma_{b}$ is a section of the Hirzebruch surface with $\sigma_{b} \cdot \sigma_{b} = -n$, which is
where the other component of the stable degeneration will be glued.  The
$E_7$ singularity is located at a disjoint section, whose class is
$\sigma_{b} + n \sigma_{f}$. Introducing homogenous coordinates $(z_0,z_1)$ on the fibers of $\mathbb{F}_n$ with $z_1$
vanishing on $\sigma$ and $z_0$ vanishing on the $E_7$ locus, the
F-theory model on $\mathcal{X}$ is:
\begin{equation}
y^2 \lambda = x^3+\left(f_{8 + n}(z'){z_0}^3z_1+f_{8}(z'){z_0}^4\right)x \lambda^2+\left(g_{12 + n}(z'){z_0}^5z_1+g_{12}(z')
{z_0}^6\right) \lambda^3.
\label{fe7}
\end{equation}

In this presentation, the spectral curve also appears somewhat differently. For example,
for an $SU(N)$ vector bundle, we can write, for $k = \lceil N / 3\rceil $ the associated spectral curve $\widetilde{C}$ as:\footnote{For $k\ge3$,
this representation is not unique, as a multiple of the Weierstrass equation can be freely added.}
\begin{equation}\label{projspec_cov}
\widetilde{C} = \{ \alpha_{k,0,0} W^k + \alpha_{k-1,1,0} W^{k-1} U + \ldots + \alpha_{0,0,k} V^k = 0 \},
\end{equation}
i.e. a general homogeneous polynomial of degree $k$ in $(W,U,V)$. Here, the
coefficients $\alpha_{a,b,c}$ are sections:
\begin{equation}
\alpha_{a,b,c} \in H^{0}(\mathbb{P}^{1} , \mathcal{O}(12 + n) \otimes \mathcal{O}(-4b - 6c))
\end{equation}
where $a + b + c = k$, as required by homogeneity. Indeed, the general form is
deduced by requiring overall homogeneity of (\ref{spec_cov}) after
assigning $W,U,V$ the respective weights $0,4,6$.

In this presentation, the geometric characterization depends on the value of $N$ mod 3. For $N = 3k$, there are
no further restrictions, while for $N = 3k - 1$, the same form as equation (\ref{spec_cov}) holds
with $ \alpha_{0,0,k} = 0$, but an explanatory point is needed. Since the zero section is given by $w=u=0$, the spectral equation vanishes
on the zero section when $\alpha_{0,0,k} = 0$.  The $SU(N)$ spectral cover is obtained by
removing the zero section from the solution curve of (\ref{spec_cov}). Finally, if $N = 3k - 2$,
we still have the equation (\ref{spec_cov}), but we require the zero section to
be a solution with multiplicity $2$, and then
remove two copies of the zero section.  This requires the vanishing of $\alpha_{0,0,k}$
as well as the coefficient of $U V^{k-1}$, i.e. $\alpha_{0,1,k-1}$.

\subsubsection{Embedding $\widetilde{C}$ in $\mathcal{X}$}

Having set up our notation, we now proceed to the embedding of the spectral curve $\widetilde{C}$ directly in the
F-theory geometry. Using this correspondence, we will be able track the limiting behavior of the intermediate
Jacobian for $\mathcal{X}$ in terms of the limiting behavior of $J(\widetilde{C})$.

Let $\rho:S\to \mathbb{P}^1$ be an elliptically fibered K3, and $\mathcal{V}$ be an
$E_8$ bundle on $S$.  Then $\mathcal{V}$ restricts to an $E_8$ bundle on the elliptic
fiber $E_p$ over any $p \in \mathbb{P}^1$, and hence gives rise to $D_p$, a $dP_9$ which
contains $E_p$ as a fiber.  Varying $p$ gives rise to a $dP_9$
fibration ${\cal X}\to\mathbb{P}^1$.  The threefold ${\cal X}$ can be viewed as $dP_9$ fibered
over $\mathbb{P}^1$ or elliptically fibered over a Hirzebruch surface $\mathbb{F}_n$,
where $c_2(\mathcal{V})=12 + n$.

Now suppose that the structure group of $\mathcal{V}$ can be reduced to $SU(N)\subset
E_8$.\footnote{The method employed here does not require this restriction
and applies in principle. Details are left for future work.}
Then for each $p\in \mathbb{P}^1$, we have that $\mathcal{V}|_{E_p}$ is an
$SU(N)$ bundle on $E_p$, which can be represented by points $p_i=p_i(p)\in E_p$ satisfying
$\sum_{i=1}^N p_i(p)=p_0(p)$. For additional details, see Appendix \ref{app:StabbingDegenerates}.
Varying $p$, the points $p_i(s)$ sweep out the $SU(N)$ spectral cover
\begin{equation}
\begin{array}{ccc}
\widetilde{C}&\subset&S\\
&&\downarrow\\
&&\mathbb{P}^1
\end{array}
\end{equation}

On the F-theory side, to each point $p_i(p)$ of $\widetilde{C}$,
we get a section $E_i(p)$ of the $dP_9$ $D(p)$, the fiber of ${\cal X}\to \mathbb{P}^1$
over $p$.  Thus $\widetilde{C}$ parameterizes a family of curves in
$\mathcal{X}$.\footnote{This is the same family of curves found by Aspinwall in an
explicit situation using the Mordell-Weil group \cite{Aspinwall:1998bw}. The analysis here holds more generally.}
The coincidence of the points $p_{N+1}(p)=\cdots=p_8(p)=p_0(p)$
leads to the conclusion there there is a curve of singularities in ${\cal X}$,
located along the zero section $S_0\subset S\subset {\cal X}$, with $S_0\simeq\mathbb{P}^1$.

For clarity of exposition, let us first perform a small resolution of the singularity of ${\cal X}$
to get a smooth $\widetilde{{\cal X}}$.  Then $\widetilde{C}$ parameterizes
a family of curves in $\widetilde{{\cal X}}$, the pullback of the family of curves in ${\cal X}$
to a family of curves in $\widetilde{{\cal X}}$ (recall that the fibers of $\widetilde{{\cal X}}\to
{\cal X}$ are curves).  So we get an Abel-Jacobi mapping
\begin{equation}
J({\widetilde C})\to J(\widetilde{{\cal X}})
\end{equation}
whose image is an explicit description of the RR moduli associated with
the $SU(N)$ part of the geometry of ${\widetilde{\cal X}}$.

But now we can think of elements of $J(\widetilde{C})$ as line bundles on
$\widetilde{C}$, which is precisely the additional data needed to reconstruct
$\mathcal{V}$ from the $\{\mathcal{V}|_{E_p}\}$, which is all that is determined by the complex
structure of ${\cal X}$.  So the Abel-Jacobi mapping gives an explicit description
of heterotic / F-theory duality in this context.

Now, blowing up to get $\widetilde{{\cal X}}$ was really just a crutch, at least as
far as the physics is concerned.  In the spirit of the Appendices, we can define
an Abel-Jacobi mapping
\begin{equation}
J(\widetilde{C})\to J({\cal X}),
\end{equation}
where $J({\cal X})$ denotes the three-form potential moduli of ${\cal X}$, which
are in principle determined by a limiting mixed Hodge structure analysis as in the Appendices.

This last point can be made more precise \textit{if} the family of curves in $\mathcal{X}$
parameterized by $\widetilde{C}$ is disjoint from the singular locus $S_0$.  However, this is typically
not the case, since whenever $\widetilde{C}$ meets $S_0$ at a point $p_0(p)$, i.e. when
$p_i(p)=p_0(p)$ for some $i\le N$, the corresponding curve $E_i(p)\subset {\cal X}$
intersects $S_0\subset {\cal X}$. In fact, this apparent complication can be
turned into a virtue. Let
\begin{equation}\label{poles}
Z=\rho\left(\widetilde{C}\cap S_0\right)\subset \mathbb{P}^1.
\end{equation}
Now, as we saw in the context of the local models of section \ref{sec:REMNANT} and the Appendices, the collection of
points $Z$ correspond to the poles of the $SU(N)$ parabolic Higgs bundle. In this context, the spectral curve
$\widetilde{C}$ should be viewed as an abstract cover of the locus $C$ wrapped by the physical seven-brane. What we have just seen
is that $\mathrm{Pic}(\widetilde{C})$ should then be identified with the corresponding fiber
of this parabolic Hitchin system.

Let us now show how this construction works for the case of $SU(2)$ spectral covers. Recall that in
section \ref{sec:REMNANT}, we showed how to see the emergence of a parabolic Higgs bundle from the geometry of a local curve of
singularities. The main idea there was to isolate an $\frak{su}(2)$ subalgebra, and perform the corresponding unfolding. Our
aim here will be to show how to recover this description in the compact setting.

Consider the $SU(2)$ spectral cover. Recall that we can describe
$SU(2)$ bundles on an elliptic curve $E$ in terms of points $\kappa_1,\ \kappa_2\in E$
such that $\kappa_1+\kappa_2=0$ in the group law for $E$.  If $E$ is a Weierstrass fiber,
the origin of the group law is $\kappa_0=(W,U,V)=(0,0,1)$, and $\kappa_1+\kappa_2=0$
if and only if $\kappa_0,\ \kappa_1$, and $\kappa_2$ are collinear.
The coordinates of $\kappa_0$ show that this line must be of the form:
\begin{equation}
a U + b W = 0.
\label{su2eqn}
\end{equation}
Now varying the fibers, we identify $a$ and $b$ with homogeneous
polynomials on $C$, and then equation (\ref{k3}) shows that the degrees
of $a$ and $b$ must differ by $4$.  In fact, if the $SU(2)$ bundle $\mathcal{V}$
has $c_2(\mathcal{V})=12 + n$, then globally the equation defining the spectral cover is
of the form:\footnote{We are not claiming that the spectral cover
is a complete intersection of (\ref{k3}) and (\ref{su2spec}).  In fact
both of these equations vanish on the zero section $U = W = 0$.  The complete
intersection has two components, one of which is the zero section,
the  other one being the spectral cover.}
\begin{equation}
f_{8 + n} U + g_{12 + n} W = 0.
\label{su2spec}
\end{equation}
The spectral curve $\widetilde{C}$ meets $S_0$ (given by $X = z =0$) over the
points in $C = \mathbb{P}_{\mathrm{base}}^1$ where $f_{8+ n}=0$.
Let $Z\subset\mathbb{P}_{\mathrm{base}}^1$ be the set of points given by the zeros of $g_{12+n}$.

Let us note that the base of an $SU(2)$ parabolic Hitchin system with poles on $Z$ is
\begin{equation}\label{base_para}
H^0(\mathbb{P}_{\mathrm{base}}^1,2K+Z)\simeq H^0(\mathbb{P}_{\mathrm{base}}^1,\cO(-4+n+12))=H^0(\cO(n+8)),
\end{equation}
corresponding precisely to the moduli of $f_{8 + n}$ in (\ref{su2spec}).   The spectral
cover (\ref{su2spec}) is isomorphic to the Hitchin spectral cover as can
be seen by comparing branch points. The moduli
of $g_{n+12}$ are precisely the complex structure part of the quaternionic
moduli in the hypermultiplet space which are the fibers over the
parabolic Hitchin system. We will return to this explicitly below.

We now describe the F-theory dual corresponding to unfolding $E_8$ to $E_7$.
Returning to our general discussion near equation (\ref{projFIBER}), we consider
projective fiber coordinates so that the associated F-theory model on $\mathcal{X}$ is:
\begin{equation}
y^2 \lambda = x^3+\left(f_{8 + n}(z'){z_0}^3z_1+f_{8}(z'){z_0}^4\right)x \lambda^2+\left(g_{12 + n}(z'){z_0}^5z_1+g_{12}(z')
{z_0}^6\right) \lambda^3
\label{fe7}
\end{equation}
We now consider a slight modification of equation (\ref{su2spec}):
\begin{equation}
f_{8 + n}(z')x + g_{12 + n}(z') \lambda z_0^2=0.
\label{su2fcurves}
\end{equation}
The equations (\ref{fe7}) and (\ref{su2fcurves}) both contain the zero
section.  As with the spectral cover, the complete intersection of
(\ref{fe7}) and (\ref{su2fcurves}) contains the zero section, and another
component, which we call $S$.

We claim that the surface $S$ is ruled by a family of (rational) curves which
are sections of $\mathcal{X}$ over the fibers of $\mathbb{F}_n$, and that the parameter space for
these curves is precisely the heterotic spectral cover.\footnote{This assertion
may be viewed as a concrete verification of a general claim made
in Appendix~\ref{spec_cov_review}.}  To see this,
let $(\lambda_0,x_0,y_0)$ be the coordinates of a
point $p$ of the spectral cover lying over a point $p \in C$, i.e. satisfying
(\ref{k3}) and (\ref{su2spec}).  Then the point
$(\lambda_0,x_0{z_0}^2,y_0{z_0}^3)$ satisfies (\ref{fe7}) and (\ref{su2fcurves}).  We
verify this by rewriting (\ref{fe7}) as
\begin{equation}
y^2 \lambda = \left(x^3+f_8{z_0}^4 x \lambda^2+g_{12}{z_0}^6 \lambda^3\right)+
{z_0}^3{z_1} \lambda^2\left(f_{n+8} x + g_{n+12} \lambda {z_0}^2\right).
\end{equation}
Then (\ref{su2fcurves}) vanishes at $(\lambda_0,x_0{z_0}^2,y_0{z_0}^3)$ because
(\ref{su2spec}) vanishes at $(\lambda_0,x_0,y_0)$, and the indicated regrouping shows
(\ref{fe7}) vanishes at $(\lambda_0,x_0{z_0}^2,y_0{z_0}^3)$ because of the vanishing
of (\ref{k3}) and (\ref{su2spec}) at $(\lambda_0,x_0,y_0)$. Now
$z_0\mapsto (\lambda_0,x_0{z_0}^2,y_0{z_0}^3)$ parameterizes a curve $C_p$, which is
visibly a section of $\mathcal{X}$ over the Hirzebruch fiber
over $p$ (which is parameterized by ${z_0}$). This
completes the proof of the claim.

\subsection{Examples}

Having illustrated the general map from the Jacobian of the spectral curve
to the intermediate Jacobian of the F-theory geometry, we now
turn to several examples which illustrate different types of T-brane phenomena.
First, we present some examples based on a spectral curve which is a
non-reduced scheme, and we then present an example where the spectral curve is reducible.

\subsubsection{Standard Embedding versus Small Instantons \label{ssec:SuperSTD}}

A rather striking example of T-brane phenomena in the heterotic string
already occurs for $\mathcal{V} = T_{K3}$, i.e. the standard embedding of the spin connection in
one of the $E_8$ factors \cite{Aspinwall:1997ye,Aspinwall:1998he}. This corresponds to a rank two vector bundle with $c_2(\mathcal{V}) = 24$. A
helpful feature of the standard embedding is that the heterotic anomaly constraint is automatically satisfied.

In the general spirit of our approach, we work at a generic point in the moduli space, and then show how T-brane data arises at some singular loci in the spectral
cover construction. To this end, first consider heterotic strings on K3, in the presence of a rank two vector bundle with instanton number $24$ which breaks
one of the $E_8$ factors to the commutant subgroup $E_7$. The hypermultiplet matter is completely fixed by the
topological condition $c_2(\mathcal{V}) = 24$ and consists of the following quaternionic degrees of freedom:
\begin{align}
&H^1(T_{K3})=20~\text{(singlets corresponding to the deformations of the K3 surface)} \\
&H^1(\mathcal{V})=20~\text{(charged-matter fields: 1/2 hypers in the}~{\bf 56}~\text{of}~E_7) \\
&H^1(\mathrm{End}_0(\mathcal{V}))=90~SU(2)~\text{(bundle moduli: 1/2 hypers which are singlets of}~E_7)
\end{align}
A generic $SU(2)$ bundle of this type can be described via the two-sheeted spectral cover:
\begin{equation}
g_{24}(z')w^2 + f_{20}(z')u=0
\label{su2c24}
\end{equation}
where we view the heterotic K3 as embedded in a $\mathbb{P}^{2}_{1,2,3}$-bundle over the base curve $C$. Viewing the
heterotic K3 as embedded in a $\mathbb{P}^{2}$-bundle over $C$, we can also write:
\begin{equation}
g_{24}(z') W + f_{20}(z') U = 0
\end{equation}
where we recall our convention that the
spectral cover is obtained by removing the zero section.  For later use, note that
the Weierstrass equation shows that $W = 0$ defines the zero section with
multiplicity~3, since substituting $W = 0$ into the Weierstrass
equation forces $U^3 = 0$.

A surprise arises, however, in the case of the standard embedding. Recall that this is given by the
tangent bundle of K3, i.e. we embed the spin connection of K3 in one of the $E_8$ factors. As
was demonstrated in \cite{Aspinwall:1998he}, performing a Fourier-Mukai transform
on the tangent bundle of the K3 surface, the spectral equation takes a very particular
and singular form:
\begin{equation}\label{tan_spec}
g_{24}(z')  w^{2} = 0.
\end{equation}
Here, $g_{24} \in H^{0}(\mathbb{P}^{1}_{\mathrm{base}} , \mathcal{O}(24))$, it is the coefficient $a_0$ appearing in equation (\ref{spec_cov}),
or equivalently the coefficient $\alpha_{k,0,0}$ appearing in equation (\ref{projspec_cov}).

This defines a spectral curve which is both a non-reduced scheme, and reducible. It is reducible since it is
the union of the base $\mathbb{P}^1$ (given by $w=0$ in K3), and $24$ vertical components
given by the zeros of $g_{24}$. In fact, for the tangent bundle the 24 fibers appearing
in the spectral cover are the $24$ singular $I_1$ fibers of the K3 \cite{Friedman:1997ih,Aspinwall:1998he}. It is
non-reduced because $w^2 = 0$. Equivalently, since $W=0$ contains the zero section
with multiplicity~3 and we have removed one copy of the zero section, we
see that  $g_{24}(z')  W = 0$ contains the zero section with multiplicity~2.

The spectral data in (\ref{tan_spec}) appears degenerate and indeed, is exactly what would
be expected from one very singular point in bundle moduli space: the limit in which the
$SU(2)$ bundle is dissolved into $24$ point-like instantons on K3 \cite{Aspinwall:1998he, Aspinwall:1997ye}.
Indeed, if the location of the instantons is given by the zeros of $g_{24}$, the Fourier-Mukai
transform of $\bigoplus_{r=1}^{24} {\cal I}_{r}$ yields exactly the vertical ($I_1$ fiber)
components of (\ref{tan_spec}).

But how is it that one spectral cover can describe two very different gauge configurations,
one of which corresponds to highly non-perturbative physics and tensionless strings? A
resolution to this puzzle was proposed in \cite{Aspinwall:1998he} by considering the
values of the three-form potential moduli. More precisely, non-zero vevs for the three-form
potential moduli prevent the effective theory of the standard embedding
from approaching the singular small-instanton limit.

As we can see from table 1, the remnants of the three-form potential moduli in the dual M-theory description
arise in the heterotic geometry via the rank $1$ sheaves, $L_{\widetilde C}$ over the spectral cover. In the case of a non-reduced
spectral cover, the Jacobian of this degenerate curve can have multiple, disjoint components.
It is this extra geometric data in $J(\widetilde{C})$ which guarantees that the singular spectral curve
still gives rise to a smooth bundle (in this case $T_{K3}$) under the Fourier-Mukai transform.

More explicitly, it was pointed out in \cite{del} that the Jacobian $J(C)$ of a non-reduced curve could contain rank $1$ sheaves with exotic origins. In particular, if $C$ is a smooth curve of genus $g>2$ and $\mathcal{V}_C$ a vector bundle of rank $n$ and degree $k +(n^2-n)(1-g)$ on $C$, then $\mathcal{V}_C$ may also be viewed as {\it coherent, rank $1$ sheaf}, $\mathcal{V}_C=L_{nC}$, on the non-reduced curve $nC$. As we will see below, it is exactly such rank 1 sheaves (the remnants of higher rank bundles on the base curve) which appear in non-reduced heterotic spectral covers and can explain mysteries like the one above.\footnote{For other work on Jacobians of non-reduced curves (frequently called ``ribbons'') see for example \cite{0853.14016}.}

For the tangent bundle, the spectral sheaf, $L_{\widetilde C}$ is characterized by the fact that $
T_{K3}|_E$, the restriction to the elliptic fiber, generically does not split as a sum of two line
bundles ${\cal O}_E \oplus {\cal O}_E$, but instead the non-trivial extension
\begin{equation}
0 \to {\cal O}_E \to T_{K3} |_E \to {\cal O}_E \to 0
\end{equation}
while the $24$ point-like instantons trivially decompose.

Likewise, for the instantons, the spectral sheaf trivializes (see Section 3.3 of \cite{Aspinwall:1998he}).
In contrast, for the tangent bundle, the sheaf over the spectral cover is indecomposable. As explained in \cite{glueII},
the rank one sheaf ${\cal L}_C$ on the non-reduced and reducible curve can be defined via the following short exact sequences:
\begin{align} \label{tansheaf}
0 \to i_{*}{T\mathbb{P}^1}^{\vee} \to L_{\widetilde C} \to {\cal K} \to 0 \\
0 \to \bigoplus_{i=1}^{24} {\cal O}_{E_i} \to {\cal K} \to i_{*}T\mathbb{P}^1 \to 0
\end{align}
where $i$ imbeds the zero section into K3. This rank $1$ sheaf then is defined via line bundles
over each of the $24$ vertical fibers as well as a rank $1$ sheaf over the double cover of $\mathbb{P}^1$.
At each of the $24$ points of intersection we must define non-trivial extensions of the form given in
(\ref{tansheaf}). This ``gluing data" identifies the sheaves at each point of intersection. As was shown
in \cite{glueII}, the gluing data at the $24$ points of intersection are not all independent. In fact,
only $21$ degrees of freedom are sufficient to specify the non-trivial extensions defining the spectral
sheaf, $L_{\widetilde C}$. These gluing vevs, together with the choice of $24$ line bundles on the singular
fibers give rise to the $45$ degrees of freedom in the Jacobian of the singular curve in (\ref{tan_spec}).

Having given a detailed discussion of the heterotic string side of the story, let us now turn to the dual F-theory description.
Here, we have the Hirzebruch base $\mathbb{F}_{n}$, where $n = 12$ as required by the value of the instanton number
in the heterotic description. The associated Weierstrass model is:
\begin{equation} \label{smallK3}
y^2 = x^3 + g_{24}(z^{'}) z^5 + f_{8}(z') x z^4 + g_{12}(z') z^6 + g_{0} z^7.
\end{equation}
In this case, the coefficient multiplying $z^7$ is a constant, reflecting the fact
that one of the $E_8$ factors is frozen to a ``non-Higgsable'' $E_8$ factor. On the
other $E_8$ factor, however, we see a high degree coefficient, reflecting the fact
that there are many moduli available to unfold the singularity.

Now, the important point for our present discussion is that equation (\ref{smallK3}) could
refer to two \textit{completely different vacua}. In the heterotic dual description, these two
vacua are the standard embedding and the limit of small instantons.

To illustrate, consider first the small instanton limit. As proposed
in \cite{MorrisonVafaI, MorrisonVafaII} one can consider blowing up in
the base the $24$ points where the zeroes of $g_{24}(z')$ and $z = 0$
intersect. In the dual heterotic M-theory dual description, these blowups
correspond to pulling the M5-branes off of the $E_8$ wall, and moving them
into the bulk. The distance of each M5-brane from the wall corresponds to
the size of the K\"ahler resolution parameter. After performing these blowups in the base, we
reach a new F-theory model, with base $\widetilde{B}$ such that $h^{1,1}(\widetilde{B}) = h^{1,1}(B) + 24$. These
blowup modes correspond to additional tensor multiplets in the six-dimensional supergravity theory.
After blowing up the $E_8$ singularities, we reach the ``Coulomb branch'' of the theory.

Now, we can go back to the origin of the Coulomb branch by blowing down all the K\"ahler classes, which in the
dual heterotic M-theory description includes moving all the M5-branes back to the $E_8$ wall. At
this point, we can dissolve the small instantons back into smooth vector bundles, moving onto
the Higgs branch. However, only \textit{some} of these moduli will show up as complex deformations
of equation (\ref{smallK3}), with the rest captured by the three-form potential moduli.

Returning to our general discussion in subsection \ref{ssec:E8pluspoles}, we saw there that
locally, the dynamics of the gauge theory are controlled by a parabolic Higgs bundle with pole
data localized at the zeroes of $g_{24}$. These poles should be viewed as specifying background vevs for
modes in a tensionless string theory with degrees of freedom charged under the $E_8$ gauge theory. We can next
perform a smoothing to an $E_7$ singularity:
\begin{equation} \label{smallK3_smoothed}
y^2 = x^3 + f_{20}(z') x z^3 + g_{24}(z^{'}) z^5 + f_{8}(z') x z^4 + g_{12}(z') z^6 + g_0 z^7.
\end{equation}
Then as expected, the T-brane moduli are captured by the $SU(2)$ parabolic Higgs bundle, with
poles at the locus $Q = \{ g_{24} = 0 \} \cap \{ z = 0 \}$. The $21$-dimensional space of
non-trivial extensions:
\begin{equation}
0 \to {\cal O}(-11) \to E \to {\cal O}(11) \to 0
\end{equation}
which pair with $H^0(\mathbb{P}^1, O(20))$ to form a hypermultiplet are dual to the heterotic degrees
of freedom in the Jacobian given in (\ref{tansheaf}).

To summarize our discussion, we can see that in the correspondence between the heterotic / F-theory pairs, the deformations of the spectral
curve $\widetilde{C}$ agree with the appropriate deformations of $\mathcal{X}$. However, we can also see that the behavior of the
Jacobian $J(\widetilde{C})$ exhibits more structure in the singular limit, and that this is matched by the degeneration of the limiting
behavior of the intermediate Jacobian $J(\mathcal{X})$. Indeed, the $24$ vertical components of the spectral curve associated with small instantons
correspond in the F-theory description to the location of the poles of the Higgs field, which is the point set $Z$ of (\ref{poles}).  In the limiting mixed Hodge structure
analysis, this corresponds to the weight two part of the mixed Hodge structure, associated with $J^{2}(H^{3}_{\mathrm{lim}})$. The moduli of the intermediate Jacobian
correspond to choices of line bundles on the nonreduced scheme of multiplicity~2 which arises as the spectral cover associated to the origin in the base of
the corresponding parabolic Hitchin system. Moreover, the $20$ locations of
gluing data in the spectral cover simply translate to the limiting behavior
of the three-form potential moduli, just as we already observed in the local
configurations of subsection \ref{ssec:E8pluspoles}.

It is also instructive to compare this with the structure of the hypermultiplet moduli space in the heterotic string description. There, all of the $45$
moduli are on an equal footing, even though in the parabolic Higgs bundle there is a natural split between the ``boundary data'' fixed at the zeroes of
$g_{24}$, and the dynamical moduli associated with $f_{20}$. As we observed at the end of section \ref{ssec:E8pluspoles}, once we move away from the decoupling limit
of the seven-brane gauge theory by recoupling to gravity, these moduli get put back
on an equal footing.

There are some clear generalizations of our analysis. For example, we can see that similar
phenomena will arise for other choices of instanton numbers, i.e. choices of $n$.
Additionally, it is worth pointing out that in making the heterotic / F-theory comparison above,
we have focused on rank $2$ heterotic bundles. It is equally possible to consider higher rank
bundles with degenerate spectral covers. For example, if we include another copy of the zero
section, $W = U = 0$ in the definition of the spectral cover given in (\ref{tan_spec}) and
compare the F-theory geometry given in (\ref{smallK3}) with a heterotic spectral cover of the form
\begin{equation}
g_{24}(z')W=0,
\end{equation}
we have the union of the zero section with multiplicity~3 with the fibers
located over the zeros of $g_{24}$. In the equivalent presentation with fiber coordinates $(w,u,v)$ for $\mathbb{P}^{2}_{1,2,3}$, this
would read as:
\begin{equation}
g_{24}(z') w^3 = 0.
\end{equation}
In this case, once again it is possible for the degenerate spectral cover to be have a smooth $SU(3)$ bundle as its Fourier-Mukai transform. The Jacobian of the non-reduced curve would lead to an additional $19$ degrees of freedom, corresponding in the dual F-theory geometry to the three-form potential moduli partners of the smoothing parameters $q_{18}z^2x$ in the Weierstrass equation. See Appendix \ref{App:Su3} for further details of such $SU(3)$ examples.

Similar degenerate spectral covers can correspond
to smooth, higher rank vector bundles. For example, given an
$\mathfrak{su}(N) \subset\mathfrak{e}_{8}$ subalgebra, we can produce an $N$-sheeted
spectral cover with degenerate spectral equation:
\begin{equation}
g_{12 + n}w^N = 0
\end{equation}
This would likewise give rise to a rich possible structure of rank 1 sheaves on ${\widetilde C}$. In this case, some choice of
smoothing, connected with a choice of limiting behavior for the three-form potential moduli would
again differentiate between small instantons and specific smooth $SU(N)$ bundles.

The above example clearly illustrates one important lesson in heterotic / F-theory duality: In
considering a degenerate spectral cover and its F-theory dual, we must first decide what F-theory
smoothing deformation, or equivalently, heterotic bundle moduli space, we want to compare. We have
already encountered this ambiguity in Section \ref{sec:REMNANT} and
Appendix \ref{hitchinmhs}. This freedom is precisely the choice of smoothing that must be included
to fix the limiting mixed Hodge structures and define the emergent parabolic Hitchin system.

\subsubsection{Reducible Spectral Covers}

T-brane data can also arise when the spectral curve becomes reducible, i.e.
factors into two or more components. Here we focus on the case where the
spectral curve factors into two components $\widetilde{C}=\widetilde{C}%
_{1}\cup\widetilde{C}_{2}$. Now, the naive expectation would be that since the
spectral cover splits into two components, so does the vector bundle,
$V\rightarrow \mathcal{V}_{1}\oplus \mathcal{V}_{2}$. From the point of view of the moduli space
of properly slope-stable sheaves \cite{huybrechts2010geometry}, a poly-stable bundle $\mathcal{V}_{1}\oplus
\mathcal{V}_{2}$ would correspond to a singular point, at which we would expect an
enhancement of symmetry in the heterotic effective theory, and possibly
additional light states (see e.g. \cite{Lukas:1999nh,Anderson:2009nt,Louis:2011hp}).

However, such a splitting of the bundle only occurs if the \textit{full data
of the spectral cover is reducible}. That is, it is not enough to consider
just the split form of the spectral cover $\widetilde{C}$. To illustrate,
consider an $SU(N)$ bundle, $\mathcal{V}$, and its characterization as a spectral cover.
This will be captured by a pair $(\widetilde{C},L_{\widetilde{C}})$. For this
to become a direct sum $\mathcal{V}_{1}\oplus \mathcal{V}_{2}$ of stable bundles with $c_{1}%
(\mathcal{V}_{1})=-c_{1}(\mathcal{V}_{2})$ and $rk(\mathcal{V}_{1})+rk(\mathcal{V}_{2})=N$, the spectral cover must
become reduced, i.e. we need to have two independent pairs $(\widetilde{C}%
_{1},L_{\widetilde{C}_{1}})$ and $(\widetilde{C}_{2},L_{\widetilde{C}_{2}})$.
So in other words, in addition to the condition that $\widetilde{C}$
degenerates to two components $\widetilde{C}_{1}\cup\widetilde{C}_{2}$, to get
a reducible bundle we must also demand that the rank one sheaf $L_{\widetilde
{C}}$ splits up as a direct sum. More generally, all we can say is that we
have a pair $(\widetilde{C}_{1}\cup\widetilde{C}_{2},L_{\widetilde{C}_{1}%
\cup\widetilde{C}_{2}})$ where $L_{\widetilde{C}_{1}\cup\widetilde{C}_{2}}$ is
some rank one sheaf with support on $\widetilde{C}_{1}\cup\widetilde{C}_{2}$.

As should now be clear, it is possible to have a reducible spectral curve
$\widetilde{C}\rightarrow\widetilde{C}_{1}\cup\widetilde{C}_{2}$, but an
irreducible vector bundle $\mathcal{V}$. This will occur whenever the rank one sheaf
$L_{\widetilde{C}_{1}\cup\widetilde{C}_{2}}$ does not trivialize at the
intersection points $\{p_{i}\}=\widetilde{C}_{1}\cap\widetilde{C}_{2}$.
Indeed, to define $L_{\widetilde{C}_{1}\cup\widetilde{C}_{2}}$ we must, for
each point $p\in\widetilde{C}_{1}\cap\widetilde{C}_{2}$ choose an isomorphism
$\phi:L_{\widetilde{C}_{1}}|_{p}\rightarrow L_{\widetilde{C}_{2}}|_{p}$. The
map $\phi$ is referred to as the \textquotedblleft gluing data" and the
triple, $(L_{\widetilde{C}_{1}},L_{\widetilde{C}_{2}},\phi)$ determines the
line bundle $L_{\widetilde{C}}$.

Schematically, we can describe the bundle $\mathcal{V}$ over the whole K3 by considering
$\mathit{{D}}=$ $\mathit{\pi^{-1}(\pi(\widetilde{C}_{1}\cap\widetilde{C}_{2}))}$ (the
collection of points viewed now as points in K3), and the short exact
sequence
\begin{equation}
0\rightarrow \mathcal{V}\rightarrow \mathcal{V}_{1}\oplus \mathcal{V}_{2}\rightarrow\mathcal{T}\rightarrow0
\end{equation}
where $\mathcal{T}$ is a torsion sheaf supported at the points $p\in D$. In
the case that $\mathcal{T}=(\mathcal{V}_{1})|_{D}=(\mathcal{V}_{1})|_{D}$ the bundle $\mathcal{V}$ splits as
$\mathcal{V}\rightarrow \mathcal{V}_{1}\oplus \mathcal{V}_{2}(-D)$ (where the sum has now been modified by
the familiar dualizing sheaf). Thus, in order for the bundle to decompose as a
direct sum globally, we must not only describe the factorization of the
spectral cover, but also the gluing data and the decomposition of the line
bundles over $\widetilde{C}_{1}$ and $\widetilde{C}_{2}$. Returning to our
discussion near equation (\ref{genusSmooth}), we can see that these extra
moduli are accounted for by the intersection points. Indeed, applying the
intersection theoretic genus formula, we learn:%
\begin{equation}
g(\widetilde{C})=g(\widetilde{C}_{1})+g(\widetilde{C}_{2})+\widetilde{C}%
_{1}\cdot\widetilde{C}_{2}+1.
\end{equation}
Since the genus tracks the complex dimension of the Jacobian, we see that
roughly speaking, the intersection points make up the additional degrees of
freedom for the gluing data.

We now turn to a specific example. On the heterotic side, we consider
an $SU(4)$ vector bundle which initiates a breaking pattern in the second
factor of $\mathfrak{so}(10)\times\mathfrak{su}(4)\subset\mathfrak{e}_{8}$.
For generic bundle valued moduli, this will leave us with an $\mathfrak{so}%
(10)$ gauge symmetry factor. By tuning the vector bundle moduli, we can reach
a spectral curve which factorizes, and so would naively correspond to a
breaking pattern $\mathfrak{so}(12)\times\mathfrak{so}(4)\subset
\mathfrak{e}_{8}$. However, once we include the gluing data, we will be able
to see that in spite of the reducible form of the spectral curve this could
still be a perfectly smooth $SU(4)$ vector bundle. In other words, further
tuning must be specified to get a reducible bundle. We then turn to the
F-theory description of these models, where the appearance of the gluing data
is reflected in the choice of a T-brane configuration.

Consider first the heterotic side of the correspondence. We construct a
smooth vector bundle with structure group $SU(4)$ via the
four-sheeted spectral cover of $\mathbb{CP}^{1}$. In this and more involved cases, it proves convenient
to adopt the presentation with fiber coordinates $(W,U,V)$ for a $\mathbb{P}^{2}$ bundle. The spectral curve can then be written as:
\begin{equation}
g_{12+n}W^{2}+f_{8+n} UW +q_{6+n}V W +h_{4+n}U^2 = 0, \label{smooth}%
\end{equation}
where the subscripts on $g_{12+n}$, $f_{8+n}$, $q_{6+n}$ and $h_{4+n}$
indicate their degrees viewed as polynomials in the coordinate $z^{\prime}$ of
$\mathbb{P}_{\text{base}}^{1}$. Recalling our discussion near the end of subsection \ref{ssec:PROJ}, since $N = 3 k - 2$ with $k = 2$,
here we have removed two copies of the zero section which in turn requires the vanishing of $V^2$ and $UV$. For the
sake of comparison, we also give the spectral curve in a presentation with
fiber coordinates $(w,u,v)$ for a $\mathbb{P}^{2}_{1,2,3}$ bundle:
\begin{equation}
g_{12+n} w^{4}+f_{8+n} u w^2 +q_{6+n}v w +h_{4+n}u^2 = 0, \label{smooth}
\end{equation}
In what follows, however, we shall stick to the $\mathbb{P}^{2}$-bundle presentation.

Applying Fourier-Mukai to the pair
$(\widetilde{C},L_{\widetilde{C}})$, we get a rank $4$ bundle of instanton
number $c_{2}(\mathcal{V})=12+n$, leaving us with an unbroken $\mathfrak{so}%
(10)\subset\mathfrak{e}_{8}$ subalgebra. Now, we next consider a specific
limit where the curve factorizes. We arrange this by switching off $q_{6+n}$,
and setting $g_{12+n}=\alpha_{4+r}\gamma_{8+n-r}$, $f_{8+n}=\beta_{r}%
\gamma_{8+n-r}+\alpha_{4+r}\delta_{4+n-r}$ and $h_{4+n}=\beta_{r}%
\delta_{4+n-r}$, for some $r$. The equation for the spectral curve then
factors as:%
\begin{equation}
(\alpha_{4+r} W + \beta_{r} U)(\gamma_{8+n-r}W + \delta_{4+n-r}U)=0
\label{tuned}%
\end{equation}
to two components $\widetilde{C}=\widetilde{C}_{1}\cup\widetilde{C}_{2}$.

We might be tempted to say from this reducible curve alone that the resulting
heterotic theory has gauge symmetry $SO(12)$ (see \cite{BershadskyPLUS} for such a
heterotic / F-theory pair). However, by careful choices of line bundle
$L_{\widetilde{C}}$ on the spectral cover it is possible for the bundle $V$ to
be a smooth, indecomposable bundle with either an $SU(4)$ or $SO(5)$ structure
group, leading to $SO(10)$ or $SO(11)$ theories.

Next, let us explain how this shows up in the case of the dual F-theory
description. Following the general outline in earlier sections, we first give
the Hitchin-like system description, and then turn to the global model. Our
interest is in the unfolding of the $\mathfrak{su}(4)$ factor in the
decomposition $\mathfrak{so}(10)\times\mathfrak{su}(4)\subset\mathfrak{e}_{8}%
$. It is therefore enough to focus on an $\mathfrak{su}(4)$ Hitchin system on
the curve $C=\mathbb{P}_{\text{base}}^{1}$ with punctures. The punctures are
specified by the zero set $g_{12+n}(z^{\prime})=0$ on the curve, and indicate
the existence of possible poles and delta function supported fluxes.

We first take a pair of independent $SU(2)$ Higgs bundles, and then explain how to glue
them back together. Along these lines, we introduce two $SU(2)$ bundles
$E_{1}$ and $E_{2}$, and two Higgs fields:%
\begin{align}
\Phi_{1} &  :E_{1}\rightarrow E_{1}\otimes K\left(  \alpha_{4+r}\right)  \\
\Phi_{2} &  :E_{2}\rightarrow E_{2}\otimes K\left(  \gamma_{8+n-r}\right)  ,
\end{align}
where $K\left(  \alpha_{4+r}\right)  $ corresponds to the sheaf of
differentials with a pole along $\alpha_{4+r}=0$, with similar notation for
$K\left(  \gamma_{8+n-r}\right)  $. We can now start to construct much more
general $SU(4)$ Higgs bundles:%
\begin{equation}
\Phi:E_{SU(4)}\rightarrow E_{SU(4)}\otimes K(g_{12+n})
\end{equation}
by taking extensions of the parabolic Higgs bundles for the $SU(2)$ case. This
corresponds to gluing the two factors back together.

Our primary interest is in those cases where this gluing retains the original
form of the reducible spectral curve. To do this, we first begin with
the construction of the bundles $E_{i}$, viewed as appropriate extensions:%
\begin{align}
0  & \rightarrow L_{1}^{-1}\rightarrow E_{1}\rightarrow L_{1}\rightarrow0\\
0  & \rightarrow L_{2}^{-1}\rightarrow E_{2}\rightarrow L_{2}\rightarrow0,
\end{align}
where the line bundles $L_{i}$ are chosen so as to satisfy the analogue of the
degree requirements seen on the heterotic side for equation (\ref{tuned}),
i.e.:%
\begin{equation}
L_{1}=K_{C}^{-1/2}\otimes\mathcal{O}(r/2)\text{ \ \ and \ \ }L_{2}%
=K_{C}^{-1/2}\otimes\mathcal{O}((4+n-r)/2).
\end{equation}
Here, we assume that all degrees are integers, though the count of the number
of moduli associated with these extensions holds more generally.

To produce a more general $SU(4)$ bundle, we can then perform a further gluing
by activating some choice of vevs for the localized matter trapped at the
zeroes of $\beta_{r}$ and $\delta_{4+n-r}$. So in other words, we can pick a
non-trivial extension in  either $\mathrm{Ext}^{1}(E_{1},E_{2})$ or alternatively in
$\mathrm{Ext}^{1}(E_{2},E_{1})$. In these cases, we can retain the condition of T-brane
data which remains hidden from the spectral equation. To give an example of
this type, we could consider, in some local patch of $C$, the Higgs
field:%
\begin{equation}
\Phi =\left[
\begin{array}
[c]{cccc}%
0 & \beta_{r}/\alpha_{4+r} & T_{10+n}/g_{12+n} & 0\\
\varepsilon_{1} & 0 & 0 & -T_{10+n}/g_{12+n}\\
0 & \varepsilon_{2} & 0 & \delta_{4+n-r}/\gamma_{8+n-r}\\
0 & 0 & \varepsilon_{3} & 0
\end{array}
\right]  .
\end{equation}
where notation is as in equation (\ref{tuned}), and $T_{10+n}\in
H^{0}(\mathbb{P}^{1},\mathcal{O}(10+n))$. The degree is fixed by the condition
of homogeneity in the spectral equation:%
\begin{equation}
g_{12+n}\det\left(  s - \Phi \right)  =(\alpha_{4+r}%
s^{2}-\varepsilon_{1}\beta_{r})(\gamma_{8+n-r}s^{2}-\varepsilon_{3}%
\delta_{4+n-r})+\varepsilon_{2}\left(  \varepsilon_{3}-\varepsilon_{1}\right)
sT_{10+n}=0.
\end{equation}
When $\varepsilon_{2}=0$, the moduli associated with $T_{10+n}$ do not appear
at all. Of course, following our previous discussion, we can track these
moduli by first working at generic values of $\varepsilon_{2}$, and then
passing to a singular limit.

Now we turn to the F-theory realization of these configurations. We first
consider the case of the smooth spectral cover, as described by
equation (\ref{smooth}). Using the fact that we can embed $\widetilde{C}$ in
$\mathcal{X}$, we can write the threefold $\mathcal{X}$ as:%
\begin{equation}
y^{2}\lambda=\left(  x^{3}+f_{8}xz^{4}\lambda^{2}+g_{12}z^{6}\lambda
^{3}\right)  +z\lambda\left(  g_{12+n}z^{4}\lambda^{2}+f_{8+n}xz^{2}%
\lambda+q_{6+n}zy\lambda+h_{4+n}x^{2}\right)
\end{equation}
where $(\lambda , x, y)$ are fiber coordinates for a $\mathbb{P}^{2}$ bundle.
So, we see as in earlier cases that we get a surface in the F-theory geometry
ruled by a family of (rational)\ curves which are sections of $\mathcal{X}$
over the fibers of $\mathbb{F}_{n}$. In this presentation, we have grouped the
terms into two contributions to reflect their different roles in the duality.
If the second term had been switched off, we would be describing the equation
for the heterotic K3 surface. With the second term switched on, we can clearly
identify the contribution from the spectral equation. To put this in minimal
Weierstrass form, we would need to complete the square in $x$.

Next, consider the case where we tune the moduli of the spectral equation, as in equation
(\ref{tuned}). In this case, the model for $\mathcal{X}$ becomes:%
\begin{equation}
y^{2}\lambda=\left(  x^{3}+f_{8}xz^{4}\lambda^{2}+g_{12}z^{6}\lambda
^{3}\right)  +z\lambda(\alpha_{4+r}z^{2}\lambda+\beta_{r}x)(\gamma
_{8+n-r}z^{2}\lambda+\delta_{4+n-4}x).
\end{equation}
To track the behavior the behavior of the intermediate Jacobian in this limit,
we can now blow up the $SO(10)$ factor, and see what happens as $q_{6+n}%
\rightarrow0$ and the other components of the spectral cover factorize. Since
we have essentially reduced the problem to a more involved example along the
same lines considered in section \ref{sec:REMNANT}, we see that the remnants of the
three-form potential moduli descend to T-brane data of the local $SU(4)$ Hitchin system on the
curve $C$ with punctures at the zeroes of $g_{12 + n}$.

\section{Conclusions} \label{sec:CONC}

In F-theory, the interplay between open and closed string degrees of freedom
provides a vast generalization of perturbative IIB\ vacua. T-branes
correspond to a class of non-abelian bound states which are
straightforward to construct in the open string description, but are
surprisingly subtle to identify in the closed string moduli. In
this paper we have argued that the geometric remnants of
T-brane data are, in the dual M-theory description, associated
with periods of the M-theory three-form potential. To track this data
in a singular limit, we have applied the theory of limiting
mixed Hodge structures, showing in particular how it directly points to the emergence
of an associated Hitchin system coupled to defects.

While the structure of the hypermultiplet moduli space (pairing complex structure
degrees of freedom with RR-moduli ) has long been understood in six-dimensional
compactifications of F-theory in the case of smooth resolutions of Calabi-Yau
threefolds, the degrees of freedom in singular limits of the geometry have remained
largely unexplored. In this paper we have investigated for the first time the definition
and structure of this moduli space in the \textit{singular} limit, including singular
Calabi-Yau geometries which admit no K\"ahler resolution.

We have found that a generalization of the results of \cite{Diaconescu:2006ry} to the
compact setting provides a remarkable new tool in this study. The periods of the
three-form potential valued in the intermediate Jacobian of $X_{\mathrm{smth}}$ can be
tracked in singular limits using the theory of limiting mixed Hodge structures and
lead to an emergent Hitchin-like system coupled to defects on the discriminant locus
of $X_{sing}$. Our results concretely link the degrees of freedom in local/global
F-theory compactifications and can also be non-trivially verified in the case of
theories with heterotic duals (corresponding to singular/degenerate spectral
covers). This work provides not only a (singular) geometric description of the
notion of T-branes \cite{TBRANES,glueI,glueII}, but also resolves a number of
outstanding puzzles in heterotic/F-theory duality \cite{Aspinwall:1998he}.

To unambiguously characterize an F-theory compactification on a singular threefold $X$, we have found that
it is necessary to include additional discrete ``flux'' and continuous
``three-form potential moduli''. This data is a natural generalization of Deligne
cohomology to the case of a singular threefold. We have provided some explicit
checks of our proposal in the case of compact models, finding agreement
with the heterotic dual description, when it exists. In the remainder of
this section we discuss some potential avenues of future investigation.

Our strategy in this paper has been to study deformations of the closed string sector to track the
remnant of T-branes in geometry. In this sense, all of the branches of T-brane vacua discussed in this paper
fit within known (albeit singular limits of) geometric phases of F-theory, in accord with the
classification results given in \cite{Morrison:2012np}. From the perspective of the six-dimensional effective
field theory, however, this is not strictly necessary. It would be interesting to study possible
T-brane solutions which do not have a geometric remnant. Such vacua would
correspond to new, non-geometric phases of F-theory.

Clearly it would be important to extend our considerations to four-dimensional supersymmetric F-theory vacua.
In this setting, we have a theory with four real supercharges, so we should not expect the intermediate Jacobian to fill out
half the degrees of freedom of a quaternionic Kahler moduli space, as would happen in a theory with eight real supercharges.
Even so, the intermediate Jacobian will still fiber over the space of complex deformations, so the
general theory of limiting mixed Hodge structures still enables us to study
the limiting behavior of the three-form potential moduli in this setting.
Based on this, it is reasonable to propose that the geometric remnant of
T-brane data in a smooth four-fold is captured by the Deligne cohomology, and
that in the limit where we degenerate to a singular component of the
discriminant locus, the Vafa-Witten theory
(see e.g. \cite{VafaWitten, BershadskyFOURD, BHVI}) coupled to defects
and pointlike Yukawas fills in the remnants of the three-form potential moduli and flux.
What this means is that the results of this paper should persist
to four-dimensional vacua, but will need to be further restricted
by compatability conditions associated with superpotential obstructions. It would be
quite interesting to explore these details in future work.

\section*{Acknowledgements}

We thank C. Vafa for many helpful discussions and collaboration at an early
stage of this work. We also thank A. Collinucci, C. C\'{o}rdova, R. Donagi, D.R. Morrison, T.
Nevins, R. Plesser, S. Sch\"afer-Nameki, and W. Taylor for helpful discussions. The work of LBA was
supported by the Fundamental Laws of Nature Initiative of the Center for
Fundamental Laws of Nature, Harvard University. The work of JJH is supported
by NSF grant PHY-1067976. The work of SK is supported by NSF grants
DMS-05-55678 and DMS-12-01089, and SK further acknowledges NSF grant
PHYS-1066293 and the hospitality of the Aspen Center for Physics in supporting
work on this project.


\appendix

\section{Limits of the Intermediate Jacobian \label{app:LIMITS}}

In this Appendix we review some basic features of the intermediate Jacobian for $X$ a Calabi-Yau threefold. We shall
be particularly interested in how to make sense of this structure in limits where $X$ develops singularities, since this
is where T-brane data can hide.

To begin, let $X$ be a \textit{smooth} compact K\"ahler threefold. Then the real dimension
of $H^{3,0}(X)+H^{2,1}(X)$ is the same as the rank of $H_3(X)$. The map
\begin{equation}
\phi:H_3(X,\mathbb{Z})\to \left(H^{3,0}(X)+H^{2,1}(X)\right)^*,\qquad
\phi(\gamma): \omega\mapsto\int_\gamma\omega
\end{equation}
is easily seen to map $H_3(X,\mathbb{Z})$ to a full rank lattice in
$(H^{3,0}(X)+H^{2,1}(X))^*$.  The quotient
\begin{equation}
J(X) = \left(H^{3,0}(X)+H^{2,1}(X)\right)^*/H^3(X,\mathbb{Z}),
\label{intjacdef}
\end{equation}
is a complex torus, the \textit{intermediate Jacobian} of $X$.
The intermediate Jacobian is not an Abelian variety if $H^{3,0}(X)\ne0$, so
in particular, it is not an abelian variety
when $X$ is Calabi-Yau. The intermediate Jacobian varies holomorphically
in $X$.  When we are only concerned with the real structure of $J(X)$, we
will sometimes write it more simply as $J(X) = H^3(X,\mathbb{R}) / H^3(X,\mathbb{Z})$.

If $X$ becomes singular, the intermediate Jacobian need not be defined.
We can, however, still sometimes make sense of a limit of intermediate Jacobians.
Our plan in the remainder of this Appendix will be to explain how to track this behavior. To this end,
we first review some background on mixed Hodge structures. Then, we show how to apply this in some simple
examples. In the cases of primary interest for applications to F-theory, this prescription becomes incomplete,
and is supplemented by the emergence of a Hitchin-like system.

\subsection{Background on Mixed Hodge Structures}

In preparation for our later discussion, we now give some background details on mixed
Hodge structures. See also \cite{Donagi:2012ts} for additional discussion on
applications to F-theory.

We recall that a Hodge Structure of weight $k$ consists of the following data:
\begin{itemize}
\item A finitely generated abelian group $H_\mathbb{Z}$
\item A decreasing filtration $F^\bullet$ on $H_\mathbb{C} \equiv H_\mathbb{Z} \otimes\mathbb{C}$
\end{itemize}
The filtration is required to satisfy
\begin{equation}
F^p\cap\overline{F^{k-p+1}}=0.
\end{equation}
The filtration $F^\bullet$ is called the Hodge filtration.
The prototypical example is $H^k(X)$, where $X$ is a compact K\"ahler manifold.
We put $H_\mathbb{Z} = H^k(X,\mathbb{Z})$ and
\begin{equation}
F^pH^k(X,\mathbb{C})=\bigoplus_{p'\ge p}H^{p',k-p'}(X).
\end{equation}
The Hodge decomposition of $X$ can then be recovered from the Hodge filtration by
\begin{equation}
H^{p,q}(X)=F^p\cap\overline{F^q}.
\end{equation}
The Hodge filtration varies holomorphically in families, while the Hodge
decomposition does not.  An element of $H^{p,q}(X)$ is said to have {\em Hodge type\/} $(p,q)$.

There is a natural space of Hodge structures, which is not compact.  In order to study limits of Hodge structures,
we will need a more general notion. A {\em mixed Hodge structure\/} consists of
\begin{itemize}
\item A finitely generated abelian group $H_\mathbb{Z}$
\item An increasing filtration $W_\bullet$ on $H_\mathbb{Q}=H_\mathbb{Z}\otimes\mathbb{Q}$, called the weight filtration
\item A decreasing filtration $F^\bullet$ on $H_\mathbb{C} \equiv H_\mathbb{C}\otimes\mathbb{C}$, called the Hodge filtration
\end{itemize}
Let $\Gr^W_\ell = W_\ell/W_{\ell-1}$, which is obtained from the finitely generated abelian group
$(\Gr^W_\ell)_\mathbb{Z} = (H_\mathbb{Z} \cap W_\ell)/(H_\mathbb{Z} \cap W_{\ell-1})$ by tensoring with $\mathbb{Q}$.
Furthermore, $F^\bullet$ induces decreasing filtrations $F^\bullet(\Gr^W_\ell\otimes\mathbb{C})$. The data
$(\Gr^W_\ell)_\mathbb{C}, F^\bullet(\Gr^W_\ell\otimes\mathbb{C})$ is required to be a Hodge structure of
weight $\ell$ for each $\ell$.

Now, we can associate generalized Jacobians to any mixed Hodge structure:
\begin{equation}
J^p(H_\mathbb{Z},F^\bullet,W_\bullet) = F^p\backslash \left(H_\mathbb{Z} \otimes \mathbb{C} \right)/H_\mathbb{Z}.
\end{equation}
This generalized Jacobian does not depend on the weight filtration.
These generalized Jacobians are functorial in the sense that we can
define natural categories of mixed Hodge structures and of
generalized Jacobians. Observe that if $X$ is a compact
K\"ahler threefold, then $J^2$ of the usual Hodge structure on $H^3(X)$ is just the usual
intermediate Jacobian $J(X)$:
\begin{equation}
J^2=\left(H^{3,0}(X)+H^{2,1}(X)\right)\backslash H^3(X,\mathbb{C})/H^3(X,\mathbb{Z}).
\end{equation}
Note that by Poincar\'e duality, we can write
$(H^{3,0}(X)+H^{2,1}(X))\backslash H^3(X,\mathbb{C})\simeq(H^{3,0}(X)\oplus H^{2,1}(X))^*$.

\subsection{Limiting Mixed Hodge Structure}
\label{limmhs}
We review the limiting Mixed Hodge Structure of a semistable degeneration $\pi: \mathcal{X} \to \Delta$, where $\mathcal{X}$ and
$X_t \equiv \pi^{-1}(t)$ are smooth for $t\ne0$, while $X_0=Y_1\cup\ldots\cup Y_c$ is a union of smooth normal crossings
divisors $Y_i$.

The cohomology groups $H^k(X_t,A)$ for $t\ne0$ form the fibers of local systems $\cH^k_A$ over $\Delta^*$
for $A=\mathbb{Z},\mathbb{Q}$, or $\mathbb{C}$.  The Hodge filtrations on $H^k(X_t,\mathbb{C})$ give rise to a decreasing filtration $\cF^\bullet$
of the vector bundle $\cH^k_\mathbb{C}\otimes\mathcal{O}_{\Delta^*}$ on $\Delta^*$ by subbundles.  The local system gives rise
to a flat connection $\nabla$ on $\cH^k_\mathbb{C}\otimes\mathcal{O}_{\Delta^*}$, which obeys Griffiths transversality:
$\nabla\cF^p\subset \cF^{p-1}\otimes\Omega^1_{\Delta^*}$.  This is the prototypical example of a variation of
Hodge structure.

We would like to take the limit of the Hodge filtration as $t\to0$, but monodromy can prevents us from doing so.
Instead, we pull back to the universal cover where we can disentangle the monodromy.  Let $\mathbb{H}$ be the upper
half plane, which we realize as the universal cover of $\Delta^*$ via $u:\mathbb{H}\to\Delta^*$, where
$t=u( \tau )=\exp(2\pi i \tau)$.  The pullback $u^*\cH^k_\mathbb{C}\otimes\mathcal{O}_{\mathbb{H}}$ of the vector bundle to the universal cover
becomes trivial, so we can and will trivialize the bundle as $H^k(X_t,\mathbb{C})\otimes\mathcal{O}_{\mathbb{H}}$ for some fixed
$t\ne0$.  The flat connection need not be the trivial one, but the covering map $\tau \mapsto \tau + 1$ leads to a
monodromy transformation $T:H^k(X,\mathbb{Z})\to H^k(X,\mathbb{Z})$ which naturally extends over $\mathbb{Q}$ or $\mathbb{C}$.  For each
$\tau \in \mathbb{H}$, the fiber $\cF^\bullet_\tau$ of $\cF^\bullet$ at $\tau$ together with $H^3(X_t,\mathbb{Z})$
defines a Hodge structure of weight $k$, which undergoes monodromy as $\tau \mapsto \tau + 1$.

In the situation of a semistable degeneration, $T$ is known to be unipotent, $(T-I)^{n+1}=0$
for some $n\ge 0$ \cite{landman}. For a unipotent transformation $T$, we can define its logarithm as
\begin{equation}
N=\log(T)=(T-I)-\frac{(T-I)^2}2+\frac{(T-I)^3}3-\ldots,
\end{equation}
which is a finite sum. The transformation $N$ is nilpotent. We can then define a filtration
\begin{equation}
\exp\left(- \tau N\right)\cF^\bullet
\end{equation}
which by construction is invariant under $\tau \mapsto \tau + 1$.  It can be shown that the limit
\begin{equation}
F^\bullet_{\lim}=\lim_{\tau \mapsto i\infty}\exp\left(- \tau N \right)\cF^\bullet
\end{equation}
exists.

This filtration does not define a Hodge structure, but is the Hodge filtration of a mixed
Hodge structure with respect to a suitably defined weight filtration $W_\bullet$ on $H^k(X_t,\mathbb{Q})$.  The weight filtration
is uniquely characterized by
\begin{equation}
N(W_\ell)\subset W_{\ell-2},\ N^\ell:\Gr^W_{k+\ell}\simeq\Gr^W_{k-\ell}.
\end{equation}
The mixed Hodge structure $(H^k(X_t,\mathbb{Z}), F^\bullet{\lim}, W_{\bullet})$
is called the {\em limiting mixed Hodge structure\/} and will be denoted $H^k_{\lim}$ for
brevity. For example, if $N^2=0$, we have
\begin{equation}
\begin{array}{ccc}
W_{k+1}&=&H^k(X_t,\mathbb{Q})\\
W_k &=& \ker N\\
W_{k-1}&=&\im N\\
W_{k-2}&=&0
\end{array}
\end{equation}
so that $\Gr^W_{k+1}\simeq \mathrm{coker} (N),\ \Gr^W_k\simeq\ker N/\im N,\ \Gr^W_{k-1}\simeq\im N$ and all other graded
pieces of the weight filtration vanish.

The limiting mixed Hodge structure is related to the mixed Hodge structure
on $H^*(X_0)$ by the Clemens-Schmid sequence.  We sketch this connection in
the case of interest, where there are two components, $X_0=Y_1\cup Y_2$.

Let $Y_{12}=Y_1\cap Y_2$.  Then for any $k$,
$H^k(X_0)$ has a mixed Hodge structure whose
only nonvanishing graded pieces are of  weights $k-1$ and $k$:
\begin{equation}
Gr_kH^k(X_0)=\ker\left(H^k(Y_1)\oplus H^k(Y_2)\to H^k(Y_{12})\right)
\end{equation}
\begin{equation}
Gr_{k-1}H^k(X_0)=\mathrm{coker}\left(H^{k-1}(Y_1)\oplus H^{k-1}(Y_2)\to H^{k-1}(Y_{12})\right).
\end{equation}

The mixed Hodge structure on $H^k(X_0)$ is related to the mixed Hodge structure
on $H^k_{lim}$ by the Clemens-Schmid sequence.  Letting $n=\dim(X_0)=
\dim(X_t)$, the Clemens-Schmid sequence reads
\begin{equation}
\cdots\to H_{2n+2-k}\stackrel{\alpha}{\to}H^k(X_0)\stackrel{i^*}{\to}
H^k_{lim}\stackrel{N}{\to}H^k_{lim}\stackrel{\beta}{\to}
H_{2n-k}(X_0)\stackrel{\alpha}{\to}H^{k+2}(X_0)\to\cdots
\label{clemensschmidgeneral}
\end{equation}
In (\ref{clemensschmidgeneral}), the map $i^*$ is deduced from the inclusion
of $X_0$ into the total space of the family, while $\alpha$ and $\beta$
are deduced from Poincar\'e duality.  All maps are morphisms of mixed Hodge
structures, which means that they shift the Hodge types by a fixed amount,
called the {\em Hodge type\/} of the morphism.  The maps $\alpha,\ i^*,\ N,$ and
$\beta$ have respective Hodge types $(n+1,n+1),\ (0,0),\ (-1,-1)$, and
$(-n,-n)$ respectively.  See \cite{morrisoncs} for more details.

\subsection{The Conifold Transition \label{limmhscon}}

Let us illustrate the computation of limiting mixed Hodge structure in the case of a degeneration of a smooth threefold to a conifold
singularity.  We refine this later to the case of a conifold transition. We assume that $X_t$ is
Calabi-Yau which defines a smoothing of $X_0$, the conifold. While $X_t$ is not
a semistable degeneration, one arrives at a semistable degeneration by blowing up the conifold point of $X_0$.  This
blowup does not affect the $X_t$ for $t\ne0$ which is all that the limiting mixed Hodge structure depends on.  So we
can still construct a limiting mixed Hodge structure as above.

It is well known that $X_0$ can be obtained from $X_t$ by contracting an $S^3$.  By Poincar\'e duality, this
corresponds to a vanishing cycle $A_0\in H^3(X_t,\mathbb{Z})$.  The monodromy is given by
\begin{equation}
T(\gamma)=\gamma+(\gamma\cdot A_0)A_0
\end{equation}
Sometimes the monodromy transformation is written by extending $A_0$ to a symplectic basis $(A_i,B_i)$ for
$H^3(X_t,\ZZ)$ with $0\le i\le h^{2,1}(X_t)$.\footnote{This is where we first use the Calabi-Yau
condition.} We then have
\begin{equation}
\begin{array}{ccc}
T(A_i)&=&A_i\\
T(B_0)&=&-A_0+B_0\\
T(B_i)&=&B_i,\ i\ne0.
\end{array}
\label{conmon}
\end{equation}
Since $(T-I)^2=0$, we have $N=T-I$ and $N^2=0$.  From (\ref{conmon}) we get for the log of monodromy
\begin{equation}
\begin{array}{ccc}
N(A_i)&=&0\\
N(B_0)&=&-A_0\\
N(B_i)&=&0,\ i\ne0.
\end{array}
\label{conlogmon}
\end{equation}
We immediately get for the weight filtration
\begin{equation}
\begin{array}{ccl}
W_4&=&H^3(X_t,\mathbb{Q})\\
W_3&=&\mathrm{span}_{\mathbb{Q}}\left\{A_0,\ldots,A_{h^{21}},B_1,\ldots,B_{h^{21}}\right\}\\
W_2&=&\mathbb{Q}\cdot A_0\\
W_1&=&0,
\end{array}
\end{equation}
where $h^{21}=h^{21}(X)$.
We immediately get
\begin{equation}
\begin{array}{|c|c|c|}\hline
\ell&\dim\Gr^W_\ell&{\rm basis}\\ \hline
4&1&\left\{B_0\right\}\\ \hline
3&2h^{21}&\left\{A_1,\ldots,A_{h^{21}},B_1,\ldots,B_{h^{21}}\right\}\\ \hline
2&1&\left\{A_0\right\}\\ \hline
\end{array}
\end{equation}
Since $\Gr^W_4$ and $\Gr^W_2$ are each 1-dimensional and defined over the rationals (in particular
they are real), they must have Hodge type $(2,2)$ and $(1,1)$, respectively.

\smallskip
For the limiting Hodge filtration we have
\begin{equation}
\begin{array}{|c|c|}\hline
p & \dim F^p_{\lim}\\ \hline
0 & 2h^{21}+2\\ \hline
1 & 2h^{21}+1\\ \hline
2 & h^{21}+1\\ \hline
3 & 1\\ \hline
\end{array}
\end{equation}

Note that $\Gr^W_3$ is an ordinary Hodge structure of weight 3 and dimension $2h^{21}$.  So its Jacobian
$J^2(\Gr^W_3)$ is an ordinary complex torus of dimension $h^{21}$.\footnote{The
notion of a  {\em polarization\/} of a mixed Hodge structure is needed to confirm this assertion.  We omit further discussion of polarizations for brevity.} A straightforward
calculation using the Clemens-Schmid sequence shows that $J^2(\Gr^W_3)$
is isomorphic to the intermediate Jacobian
$J(\widetilde{X_0})$ of
the (non-Calabi-Yau) blow up of $X_0$ at the conifold point.

\smallskip\noindent
{\bf Claim.} $J^2(H^3_{\lim})$ is a $\mathbb{C}^*$ fibration over $J(\widetilde{X_0})$.

\smallskip
This follows from two claims.
\begin{itemize}
\item $J^2(H^3_{\lim})=J^2(W_3)$, the second generalized Jacobian of the sub-mixed Hodge
structure $W_3$.\footnote{In fact, by Clemens-Schmid, the mixed Hodge structure $W_3$
is precisely the mixed Hodge structure on $X_0$ and so is independent of the
chosen degeneration of $X_t$ to $X_0$.}
\item $J^2(W_3)$ is a $\mathbb{C}^*$-bundle over $J^2(\Gr^W_3)$.
\end{itemize}

To see the first subclaim,
we first note
that since $\Gr^W_4$ has Hodge type $(2,2)$, we have for the induced Hodge filtration on $\Gr^W_4$ that
$F^2\Gr^W_4=\Gr^W_4$.  In terms of the limiting Hodge filtration this means:
\begin{equation}
H^3(X_t,\mathbb{Z})=F^2+W_3.
\label{f2w3}
\end{equation}
For brevity, in (\ref{f2w3}) and the sequel, $F^2$ denotes $F^2H^3{\lim}$.

The inclusion $W_3\subset H^3(X_t,\mathbb{Z})$ induces a map
\[
\left(F^2\cap W_3\right)\backslash W_3/\left(W_3\cap H^3(X_t,\mathbb{Z})\right)=J^2(W_3)\to J^2(H^3_{\lim})
= F^2\backslash H^3(X_t,\mathbb{C})/H^3(X_t,\mathbb{Z})
\]
which we now show is an isomorphism.

First of all, surjectivity follows immediately from (\ref{f2w3}). Consider next injectivity.
Suppose $\omega\in W_3\cap(F^2+H^3(X_t,\mathbb{Z}))$.  We must show that $\omega\in (F^2\cap W_3)+(H^3(X_t,\mathbb{Z})\cap W_3)$.
Now consider $F^2\cap\overline{F^2}\cap H^3(X_t,\mathbb{Z})$.  Since $\Gr^W_4$ has type $(2,2)$ and is spanned by $B_0$, we have
a element $\widetilde{B_0}\in F^2\cap\overline{F^2}\cap H^3(X_t,\mathbb{Z})$ which is equivalent to $B_0$ mod $W_3$.  We have
$N\widetilde{B_0}=NB_0=-A_0$.

Write
\begin{equation}
\omega=\phi+\eta,\qquad \phi\in F^2,\  \eta\in H_\mathbb{Z}
\label{inkernel}
\end{equation}
Applying $N$ to (\ref{inkernel}), we have
$0=N\omega=N\phi+N\eta$,
where the first equality holds since $\omega\in W_3$.  Thus $N\phi=-N\eta$ hence is an element
of $W_2\cap H_\mathbb{Z}$, and we conclude that
\begin{equation}
N\phi=kA_0
\end{equation}
for some integer $k$.  Then we can rewrite (\ref{inkernel}) as
\begin{equation}
\omega = \left(\phi+k\widetilde{B_0}\right)+\left(-k\widetilde{B_0}+\eta\right)
\label{iszero}
\end{equation}
Each of the two terms in parentheses in (\ref{iszero}) is in $W_3$ since they were constructed to be contained in
the kernel of $N$.  Also $\phi+k\widetilde{B_0}\in F^2$ since it is a sum of terms in $F^2$, and $-k\widetilde{B_0}+\eta\in
H^3(X_t,\mathbb{Z})$ since it is a sum of integral terms.

For the second subclaim, we consider the natural map
\begin{equation}
\left(F^2\cap W_3\right)\backslash W_3/\left(W_3\cap H^3(X_t,\mathbb{Z})\right)=J^2(W_3)\to J^2(\Gr^W_3)
\end{equation}
which is surjective by the surjectivity of $W_3\to \Gr^W_3$. Further, we have
\begin{equation}
J^2(\Gr^W_3)=
\left(\left(F^2\cap W_3\right)/\left(F^2\cap W_2\right)\right)\backslash\Gr^W_3/\left(\left(W_3\cap H^3(X_t,\mathbb{Z})\right)/\left(W_2\cap H^3(X_t,\mathbb{Z})\right)\right)
\end{equation}
We let
$\phi\in W_3$ be such that
\begin{equation}
\phi\in \left(F^2\cap W_3\right)+\left(H^3(X_t,\mathbb{Z})\cap W_3\right)+W_2
\end{equation}
and we need to classify the possible $\phi$ modulo $\left(F^2\cap W_3\right)+\left(H^3(X_t,\mathbb{Z})\cap W_3\right)$.
Clearly the only flexibility is to add an element of $W_2$ to $\phi$, i.e.\ a complex multiple of $A_0$, modulo
an integer multiple of $A_0$, which is a $\mathbb{C}^*$, as claimed.

It is now straightforward to extend this analysis to the type IIA conifold transition in the situation
where $m$ hypermultiplets charged under $U(1)^r$ get Higgsed. Geometrically, we have a
smooth Calabi-Yau threefold $\widetilde{X}$ containing $m$ curves $C_i\simeq\mathbb{P}^1$s whose
homology classes span an $r$-dimensional subspace of $H_2(\widetilde{X})$.  We simultaneously contract these curves to obtain
a threefold $X_0$ which has $m$ conifold singularities and is smooth elsewhere. We then
deform $X_0$ to a smooth $X_t$ for $t\ne0$.

For $t$ near 0, we get for each conifold point $p_i\in X_0$ a vanishing
cycle $A_i\in H^3(X_t,\mathbb{Z})$.
The monodromy is given by the Picard-Lefschetz transformation
\begin{equation}
T(\gamma)=\gamma+\sum_i(\gamma\cdot A_i)A_i.
\end{equation}
The vanishing cycles satisfy $r$ independent relations in cohomology
\begin{equation}
\sum_{i=1}^mc_{ij}A_i=0,\qquad j=1,\ldots,r,
\end{equation}
so that the $A_i$ span an $m-r$-dimensional space of vanishing cycles
$\cV\subset H^3(X_t,\mathbb{Z})$.

Repeating the previous analysis, we get a weight filtration with $W_2=\Gr^W_2=\cV$, $\Gr^W_4\simeq\Gr^W_2$ (via
$N$) and rank $\Gr^W_3=2h^{21}+2-2(m-r)$.  In this case, $J^2(\Gr^W_3)$ is the ordinary intermediate
Jacobian of $\widetilde{X}$.  Our result is that:
\begin{equation}
J^2(H^3_{\lim}) \,\,\, \text{is a} \,\,\, (\mathbb{C}^*)^{m-r} \,\,\, \text{fibration over} \,\,\, J(\widetilde{X}).
\end{equation}
Finally, we have to explain the global structure of this
$(\mathbb{C}^*)^{m-r}$ fibration in order to give a complete picture of the RR
moduli.  This is again given by the limiting Hodge structure, specifically
its extension class.

Recall the result of \cite{carlsonext} in a special case adapted to our
purposes. Given a mixed Hodge structure $H$, consider the exact sequence
\begin{equation}
0\to Gr_k^W H\to H\to H/Gr_k^WH\to 0
\label{mhsext}
\end{equation}
This exhibits $H$ as an extension of the mixed Hodge structure
$H/Gr_k^WH\to 0$ by the mixed Hodge structure $Gr_k^WH$. It is shown
in \cite{carlsonext} that such extensions are classified by
the generalized complex torus $J^0\mathrm{Hom}\left(H/Gr_k^WH,Gr_k^W H\right)$
defined as
\begin{equation}
\mathrm{Hom}\left(\left(H/Gr_k^WH\right),Gr_k^W H\right)/
\left(F^0\mathrm{Hom}\left(\left(H/Gr_k^WH\right),Gr_k^W H\right)+
\mathrm{Hom}_\mathbb{Z}
\right),
\label{mhsextclass}
\end{equation}
where $F^0\mathrm{Hom}\left(H/Gr_k^WH,Gr_k^W H\right)$ is defined as
\begin{equation}
\left\{
\phi\in \mathrm{Hom}\left(\left(H/Gr_k^WH\right),
Gr_k^W H\right)\mid \phi\left(F^p
H/Gr_k^WH\right)\subset F^pGr_k^W H\ \forall p\right\}
\end{equation}
and $\mathrm{Hom}_\mathbb{Z}$ denotes the homomorphisms preserving the integral
lattices.

Letting $H$ be $W_3$ of
the limiting mixed Hodge structure above, and $k=2$, then (\ref{mhsext})
becomes
\begin{equation}
0\to \cV_{\mathbb{C}}\to W_3H^3_{\mathrm{lim}}\to H^3(\widetilde{X})\to 0.
\end{equation}
Passing to Jacobians, we see that the extension class (\ref{mhsextclass}),
combined with an arbitrary character of $(\mathbb{C}^*)^{m-r}=\cV_{\mathbb{C}}
/\cV_{\mathbb{Z}}$ produces an element of
$\mathrm{Hom}(H^3_{\mathrm{lim}}/F^2,\mathbb{C})/H^3(\widetilde{X},\mathbb{Z})$, a complex torus
dual to $J(\widetilde{X})$, with points which parameterize line bundles on $J(X)$.
This completely characterizes the bundle structure. Finally, note that the weight $3$ part only
depends on $X_0$ and not on the choice of the smoothing $X_t$, so we have
intrinsically described a space that we assert are the RR moduli.

\subsection{Emergent $SU(2)$ Hitchin-like system \label{hitchinmhs}}

In this Appendix we provide some additional technical details on the analysis of
limiting mixed Hodge structures for a curves of ADE singularities given in section \ref{sec:REMNANT}.
These methods are borrowed from techniques used in a related collaboration
of the third author with D.~Morrison and R.~Plesser \cite{kmptoappear}.  The main
point is that in contrast to the case of a conifold, for a curve of ADE singularities, the
classical intermediate Jacobian does not have a canonical limit. However, the theory of limiting mixed Hodge structures
directly points to the appearance of a Hitchin-like system.

Along these lines, we first present some additional details on the case of an isolated curve of $A_1$ singularities discussed in
section \ref{sec:REMNANT}. This analysis is
performed from another perspective in \cite{Diaconescu:2005jw, Diaconescu:2006ry}. The primary novelty here is that in preparation
for our application F-theory, we will couch this analysis in the formalism of limiting mixed Hodge structures. Additionally, we shall
extend the analysis of \cite{Diaconescu:2005jw, Diaconescu:2006ry} to cover Hitchin-like systems coupled to defect modes.

To begin, we return to a Calabi-Yau threefold $X$ containing a curve $C$ of $A_1$
singularities and no enhancements. We deform it using a quadratic differential
$q$ as in (\ref{a1curvedefcomp}).  We make the simplifying assumption
that $q$ has simple zeros, so the zero set $Z$ of $q$ has $|Z|=4g-4$. We study the limiting mixed Hodge structure
and identify it with the fiber of the $SU(2)$ Hitchin system over the point
$q$ of the Hitchin base.  It is known that this fiber can be identified with
a generalized Prym variety of the associated spectral cover $z^2=q$.

To achieve a normal crossings situation, we make the substitution
$\epsilon=t^2$ to get
\begin{equation}
xy+z^2=t^2q
\label{ssdegen}
\end{equation}
and blow up the singularity along $C\times\{0\}$ defined by $x=y=z=t=0$.
The central fiber $X_0$ has two smooth components: $\widetilde{X}$, the blowup of
$X$ along $C$, in which $C$ is replaced by a family of $\mathbb{P}^1$s parametrized
by $C$, and the exceptional divisor $E$, which can conveniently be thought of
as being defined by the same equation (\ref{ssdegen}), with $(x,y,z,t)$ now
interpreted as homogeneous coordinates on the projective bundle
$\mathbb{P}(L_x\oplus L_y\oplus L_z\oplus \mathcal{O}_C)$ over $C$, with line bundle assignments
as in subsection \ref{ssec:ISOLATED}. So the threefold
$E$ is fibered over $C$, with generic fiber isomorphic to a smooth quadric
surface in $\mathbb{P}^3$.  But the fibers over $Z$ are singular quadrics, isomorphic to the quadric cone
$xy+z^2=0$ in $\mathbb{P}^3$.  The intersection $F=\widetilde{X}\cap E$ is just the
exceptional divisor of the blowup of $X$ (without the product with $t\in \mathbb{C}$)
and is a $\mathbb{P}^1$-bundle over $C$.  Thus we have achieved normal crossings.

We now compute the mixed Hodge structure of $H^3(X_0)$.  We have
\begin{equation}
\begin{array}{ccl}
Gr_3 H^3(X_0)&=&\ker\left(H^3(\widetilde{X})\oplus H^3(E)\to H^3(F)\right)\\
Gr_2 H^3(X_0)&=&\mathrm{coker}
\left(H^2(\widetilde{X})\oplus H^2(E)\to H^2(F)\right)
\end{array}
\end{equation}
From this it follows by direct computation that $Gr_2 H^3(X_0)=0$
and the ``local'' part of $Gr_3(H^3(X_0)$ is isomorphic to $H^3(E)$ (with its
Hodge structure).  In a little more detail, the map $H^3(\widetilde{X})\to H^3(F)$
is surjective, and its kernel is the ``non-local part'' of $H^3(\widetilde{X})$.

The description of $E$ together with a Mayer-Vietoris calculation gives
$8g-6$ for the 3rd Betti number of $E$.
It follows that the 3rd (local) Betti number of $X_0$ is $8g-6$.  So the Jacobian
associated to this Hodge structure is just the intermediate Jacobian $J(E)$,
a compact complex torus of dimension $4g-3$.  In fact, this is an abelian
variety since $h^{3,0}(E)=0$.  We show that this abelian
variety is isogenous to
the Jacobian of the spectral cover $C_q$
with equation $z^2=q$.

To see this, we rewrite the spectral cover in projective
coordinates as $z^2 = t^2 q$, where the homogeneous coordinates $(z,t)$
live in $K_C\oplus\mathcal{O}_C$.   Then, for each point of the spectral cover,
we get two $\mathbb{P}^1$s in $E$: substituting $z^2=t^2q$ in (\ref{ssdegen})
gives $xy=0$.  Either of these families of $\mathbb{P}^1$s gives an Abel-Jacobi map
\begin{equation}
J(C_q)\to \mathcal{}J(E),
\label{aje}
\end{equation}
which can be shown to be surjective by the surjectivity of the related map
$H^1(C_q)\to H^3(E)$.  It follows that (\ref{aje}) is an isogeny, being a
surjective map of abelian varieties of the same dimension.

To identify the limiting mixed Hodge structure of (\ref{ssdegen}), we use
the Clemens-Schmid sequence, which includes the terms
\begin{equation}
H_5(X_0)\to H^3(X_0)\to H^3_{\mathrm{lim}}\stackrel{N}{\to} H^3_{\mathrm{lim}}.
\label{clemensschmid}
\end{equation}
Looking at graded pieces gives
\begin{equation}
Gr_{-6}H_5(X_0)\to Gr_2 H^3(X_0)\to Gr_2H^3 \to 0
\end{equation}
and
\begin{equation}
Gr_{-5}H_5(X_0)\to Gr_3H^3(X_0)\to Gr_3H^3_{\mathrm{lim}}\to 0
\end{equation}
Another Meyer-Vietoris calculation gives $Gr_{-6}H_5(X_0)=0$
and $Gr_{-5}H_5(X_0)\simeq H_1(C)$. The conclusion is that the local part of
$H^3_{\lim}$ is a quotient of $H^3(E)$ by the image of a map
$H^1(C)\to H^3(E)$.  On Jacobians, we get a map
$J(C)\to J(X)$.

Since $J(X)$ is isogenous to $J(C_q)$, the local part of the
the limiting mixed Hodge structure is a quotient of $J(C_q)$ by $J(C)$ up to
finite order.

The generalized Prym of $C_q$ is the fiber of the Hitchin map over $q$, and
up to a finite group is the quotient of $J(C_q)$ by $J(C)$ (the mapping
$J(C)\to J(C_q)$ being given by pullback after a shift).  So, it is natural
to expect the local part of $H^3_{\lim}$ to be the generalized Prym of $C_q$
and local part of the full hypermultiplet moduli space to be equal to the
Hitchin system.  In fact, this has already been verified in
\cite{Diaconescu:2006ry} away from the discriminant locus.

Having covered the case of an isolated curve of $A_1$ singularities, we now extend our analysis
to the case where the Hitchin-like system is coupled to defect modes. To this end, we now consider the
case of $X$ a non-compact threefold given by a curve $C$ of $A_1$ singularities. Let $P\subset C$ be the location of the localized
matter associated with the zeroes of the section $\beta \in \mathcal{O}_{C}(P)$. The geometry we consider is:
\begin{equation}
xy=\alpha z^3+ \beta z^2.
\label{globala1a2}
\end{equation}

Now, we saw in section \ref{sec:generaldefects}
that there are two complex structure moduli
for each point of $P$, one arising from torsion in $\cT^1$ and the other
comes from replacing $2K_C$ with $2K_C+P$ in deforming \ref{globala1a2}.  So
the count of complex structure moduli is $3g-3+2k$ where $k=|P|$,
as one would expect from Higgsing $g$ SU(2) adjoints and $k$ fundamentals.

We now describe the most general complex structure deformation.  From the
analysis in section \ref{sec:generaldefects}, the parameters are:
\begin{itemize}
\item $q$: a section of $2K_C + P$
\item $\gamma$: a section of $K_C + P$.  Only $\gamma|_P$ is a true modulus,
but we need to fix a choice of $\gamma$ for a concrete model.
These deformations correspond to the torsion in $\cT^1$.
\end{itemize}
The equation of the deformation can be taken to be
\begin{equation}
xy=\alpha z^3+ \beta z^2+t\gamma z+t^2q.
\end{equation}
We compute the limiting mixed Hodge structure as $t\to0$.  To achieve a
semistable degeneration, we blow up $x=y=z=t=0$ inside the
fourfold which is the total space of the deformation.  There are now two
components over $t=0$:
The blowup $\widetilde{X}$ of $X$, and the
exceptional divisor $E$ of the blowup.  $E$ is fibered over
$C$, with fiber having equation
\begin{equation}
xy=\beta z^2+t\gamma z+t^2q.
\label{eeqn}
\end{equation}
In (\ref{eeqn}), $x,y,z,t$ are now coordinates in a {\em projective\/}
bundle over $C$, so that $E$ is a bundle of quadric surfaces over $C$.
The generic fibers are smooth quadrics, but the fibers are singular over
the discriminant locus $\Delta$ with
equation $\beta q - 4\gamma^2=0$.  As this is a section
of $2K+2P$, there are $4g-4+2k$ singular fibers, where $k=|P|$.  In
the generic situation, all singularities are quadric cones.

Recall that $H^2$ of a smooth quadric is two-dimensional, as there are
two families of $\mathbb{P}^1$s on a smooth quadric.   Also $H^2$ of a quadric
cone is one-dimensional, corresponding to the unique family of lines (which
all pass through the vertex of the cone) on the quadric cone. Next,
letting $F=E\cap\widetilde{X}$, we see that $F$ is just the exceptional divisor of
$\widetilde{X}$, which is a $\mathbb{P}^1$ bundle over $C$ which degenerates to a union
of two $\mathbb{P}^1$s over $P$.

We now compute the limiting mixed Hodge structure. The relevant pieces are
already visible in the mixed Hodge structure on the central fiber
$X_0=\widetilde{X}\cup E$. The weight three part is
\begin{equation}
W_3H^3(X_0)=\mathrm{ker}\left(
H^3(\widetilde{X})\oplus H^3(E)\to H^3(F)
\right)
\end{equation}
and the weight two part is
\begin{equation}
W_2H^3(X_0)=\mathrm{coker}\left(
H^2(\widetilde{X})\oplus H^2(E)\to H^2(F)
\right)
\end{equation}

We now specialize to the case $C=\mathbb{P}^1$, the case that is relevant for
heterotic / F-theory duality.
For $W_3$, note that $H^3(F)=0$, so $W_3=H^3(\widetilde{X})\oplus H^3(E)$.
The ``local'' part comes from $H^3(E)$, so we compute this.  From the Leray
spectral sequence, we look at the part $H^1(R^2\pi_*\mathbb{Z})$, where $\pi:E\to
C$ is the projection.  The contribution we need is from the locus where the
quadrics are smooth.  These have a pair of $H^2$ classes in the fiber,
corresponding to the lines $\ell',\ell''$
of the respective rulings on the quadrics.  The class $\ell'-\ell''$ is
odd under monodromy around points of $\Delta$
(where the two rulings come together:
a generic singular quadric surface is a cone with a singular ruling give by
the lines through the singular point).  Let $\widetilde{C}$ be the double
over of $C$ branched along $\Delta$, and let $\widetilde{\Delta}\subset \widetilde{C}$
be the ramification locus.  Let $H^1(\widetilde{C}-\widetilde{\Delta})^-$ be the
part of cohomology which is odd under monodromy.  Then the
relevant classes in $W_3H^3$ are given by $p\otimes ([\ell']-[\ell''])$.
These classes correspond to the Prym and will relate to the Hitchin system.

Let us now count the degrees of freedom associated with our system:
\begin{itemize}
\item $\mathrm{euler}(C-\Delta)=(2-2g)-(4g-4+2k)=6-6g-2k$
\item $\mathrm{euler}(\widetilde{C}-\widetilde{\Delta})=2(6-6g-2k)
=12-12g-4k$
\item $h^1(C-\Delta)=6g+2k-7;\ h^1(\widetilde{C}-\widetilde{\Delta})=12g+4k-13$
\item $\dim H^1(\widetilde{C}-\widetilde{\Delta})^- = (12g+4k-13)-(6g+2k-7)
=6g-6+2k$
\item The corresponding part of the intermediate Jacobian of
the limiting MHS has dimension half of that: $3g-3+k$.  This can be matched
to the fibers of a parabolic Hitchin system.
\end{itemize}

The missing $k$ moduli come from $W_2H^3$.  For each point $p \in P$, consider
the difference of the pair of lines over $p$ in $F$.  This is a class in
$H^2(F)$ which survives in the cokernel.  Like the conifold calculation, in the
corresponding intermediate Jacobian, we got a $\mathbb{C}^*$ factor. So in conclusion, the
net count of RR moduli is $3g-3+2k$, matching the complex struture moduli, as expected in the physical theory.

To summarize then, in this section we have seen that the Hitchin System is contained in the local part of the Calabi-Yau integrable system. This result can be succinctly summarized by the following diagram:
\begin{equation}
\begin{array}{ccc}
&&M\\
&\nearrow&\downarrow\\
\pi^*H&\to&\widetilde{M}_{\mathrm{cplx}}\\
\downarrow&&\phantom{\pi}\downarrow\pi\\
H&\to&M_{\mathrm{loc}}
\end{array}
\end{equation}
where $H$ is the Hitchin moduli space, $\widetilde{M}_{\mathrm{cplx}}$ the complex structure moduli space of
the resolved geometry, and the maps are defined such that the bottom map is the Hitchin fibration
and the top map is an inclusion.

\section{Brief Introduction to Deligne Cohomology \label{briefDeligne}}

In the context of the Hitchin-like system and its lift to a global F-theory geometry, there is a natural sense in which
one can unify the space of flat connections with flux data. Here we briefly this unified description for a smooth analytic
variety $X$. As we now explain, the relevant mathematical object is the Deligne cohomology of $X$. For a
review of Deligne cohomology, see for example \cite{DeligneReview}.

The starting point for our considerations is a \textit{smooth} analytic variety $X$. We then form the Deligne
complex $\mathbb{Z}(p)_{\mathcal{D}}$:
\begin{equation}
0 \rightarrow \mathbb{Z}(p) \rightarrow \Omega^{0}_{X} \rightarrow \Omega^{1}_{X} \cdots \rightarrow \Omega^{p-1}_{X}
\end{equation}
where $\mathbb{Z}(p) = (2 \pi i )^p \mathbb{Z}$, and $\Omega^{j}_{X}$ refers to the sheaf of
holomorphic $j$-differentials on $X$. For each value of $p$, we can define an associated cohomology theory, which we label as:
\begin{equation}
H^{q}_{\mathcal{D}} (X , \mathbb{Z}(p))\equiv H^{q}(X , \mathbb{Z}(p)_{\mathcal{D}})
\end{equation}
i.e. the Deligne cohomology is defined by the hypercohomology of the complex.

For the purposes of this paper, the key feature of Deligne cohomology is that it provides a unified perspective on the discrete
data of fluxes, and three-form potential moduli. Indeed, for any $p$, we have the short exact sequence:
\begin{equation}
0 \rightarrow J^{p}(X) \rightarrow H_{\mathcal{D}}^{2p}(X , \mathbb{Z}(p)) \rightarrow H^{p,p}_{\mathbb{Z}}(X) \rightarrow 0.
\end{equation}
where $J^{p}(X)$ is the intermediate Jacobian in the sense of Griffiths \cite{GriffithsI, GriffithsII, GriffithsClemens}.

For details on the geometric interpretation of the various Deligne cohomology groups, see for example \cite{GAJER}.
At least for low values of $p$, there is a simple geometric interpretation of this data. For example, for $C$ an algebraic curve and $p = 1$, we
can see that the sequence reduces to:
\begin{equation}
0 \rightarrow J(C) \rightarrow H_{\mathcal{D}}^{2}(C , \mathcal{O}_{C}^{\ast}) \rightarrow H^{1,1}_{\mathbb{Z}}(C) \rightarrow 0,
\end{equation}
i.e. the Deligne cohomology captures the space of flat connections and discrete flux data as well. This provides a simple way to characterize
the gauge theoretic data of the Hitchin system. Similar though more involved considerations hold for higher Deligne cohomology groups.

Now, an important caveat in this discussion is that it works for \textit{smooth} varieties $X$. Of course, in the applications to F-theory,
we typically do not have this luxury. In fact, the analysis of this paper motivates the conjecture that
a Hitchin-like system provides a \textit{definition} of Deligne cohomology in certain singular limits of Calabi-Yau threefolds. A similar
proposal holds for singular Calabi-Yau fourfolds, where now we have the Vafa-Witten theory coupled to matter fields and Yukawas.

\section{Intermediate Jacobians and Stable Degeneration \label{app:StabbingDegenerates}}

In section \ref{sec:COMPACT} we considered a class of examples where the calculation of the
relevant components of the intermediate Jacobian $J(X)$ could be reduced to a simpler calculation of the
Jacobian of a spectral curve. In this Appendix we present some additional details on this limiting behavior of the intermediate Jacobian in the
stable degeneration limit. We shall explain how to apply this analysis in some specific examples of singular spectral curves.

First, consider the simplest case where $X$ is a smooth, generic
elliptically fibered Calabi-Yau threefold over a Hirzebruch base
$\mathbb{F}_n$.  We consider the stable degeneration limit so that it becomes a union
of two elliptic fibrations ${\cal X}_1,{\cal X}_2$ over $\mathbb{F}_n$, intersecting over
an elliptically fibered K3 surface, $S$.  We assume that the ${\cal X}_i$ and $S$
are all generic, and in particular smooth.  We want to understand what happens to $J(X)$ in this
process.

Letting $X_t$ be a smooth Calabi-Yau with $t\ne0$, and let $X_0={\cal X}_1\cup {\cal X}_2$
as above,then $H^3(X_t,\mathbb{Z})$ undergoes monodromy as $t$ goes around zero.
The monodromy $T$ is unipotent, so $N=\log(T)$ is nilpotent.  It can be
shown that $N^2=0$, so $\im N\subset \ker N$.

The transformation $N$ defines the {\em monodromy weight filtration\/}
$\cdots\subset W_i\subset W_{i+1}\subset\cdots H^3(X_t,\mathbb{C})$ by
\begin{equation}
\begin{array}{ccl}
W_4&=&H^3(X_t,\mathbb{C})\\
W_3&=&\ker N\\
W_2&=&\im N\\
W_1&=&0
\end{array}
\label{wtfilt}
\end{equation}
Next, letting $\Gr_i=W_i/W_{i-1}$ be the associated graded pieces of the weight
filtration, we have:
\begin{equation}
\begin{array}{ccl}
\Gr_4&\simeq& H^2(S,\mathbb{C})\\
\Gr_3&\simeq&H^3(\mathcal{X}_1,\mathbb{C})\oplus H^3(\mathcal{X}_2,\mathbb{C})\\
\Gr_2&\simeq& H^2(S,\mathbb{C})\\
\Gr_i&=&0,\qquad i\ne 2,3,4
\label{stablewtfilt}
\end{array}
\end{equation}
The Hodge numbers of $X_t$ can be related to these graded pieces, with
shifts on $\Gr_2$ and $\Gr_4$.  The computation is:
\begin{equation}
\begin{array}{ccccl}
H^2(S,\mathbb{C})&(1&20&1&0) \qquad({\rm from\ }\Gr_4)\\
H^3({\cal X}_1,\mathbb{C})&(0&111&111&0)\\
H^3({\cal X}_2,\mathbb{C})&(0&111&111&0)\\
H^2(S,\mathbb{C})&(0&1&20&1) \qquad({\rm from\ }\Gr_2)\\
\end{array}
\end{equation}
and the columns add up to the Hodge numbers $(1,243,243,1)$ of $X_t$,
as they must. More precisely, let
\begin{equation}
F^\bullet_{\mathrm{lim}}=\lim_{t\to0}\mathrm{exp}\left(-\frac1{2\pi i}\log(t)N\right).
\end{equation}
Then this limit exists and induces the familiar Hodge structures on each of
the graded pieces (\ref{stablewtfilt}). The usual Hodge structures can be
recovered asymptotically as $t\to0$ as
$\mathrm{exp}\left(\frac1{2\pi i}\log(t)N\right)F^\bullet_{\mathrm{lim}}$.

Next, suppose that $X_t$ has a singular limit $X_0$.  We assume that this
is a semistable degeneration (a mild requirement, often achievable
by a substitution $t\mapsto t^n$ and a blowup of $X_0$).  Then the Hodge
structure approaches a {\em limiting mixed Hodge structure\/}
$H^3_{\mathrm{lim}}$, a complex vector space of the same dimension as $H^3(X)$,
equipped with a Hodge filtration and a finite increasing weight filtration
$\cdots\subset W_i\subset W_{i+1}\subset\cdots H^3_{\mathrm{lim}}$.  We
put $\Gr_i=W_i/W_{i-1}$ (see \cite{Donagi:2012ts} for details for Calabi-Yau
fourfolds). For example, suppose $X_0={\cal X}_1\cup {\cal X}_2$ is the usual stable degeneration limit
to two $\mathrm{dP}_9$ fibrations.  Then we can return to line (\ref{stablewtfilt}). In
this and similar cases, $W_3$ represents the part of $H^3(X_t)$
that survives in the limit as $t\mapsto0$.

\subsection{Spectral Covers and $dP_9$ Fibrations}\label{spec_cov_review}

Having reduced the computation of the limiting mixed Hodge structure
to a calculation of the intermediate Jacobian of $\mathcal{X}_{1}$, one of the
components appearing in the stable degeneration, we can now specialize further to a discussion
of spectral covers and $dP_9$ fibrations.

To begin, we reviewing the dictionary between $E_8$ bundles on an elliptic
curve $E$ and $dP_9$'s containing $E$ as an elliptic fiber. For additional
discussion, see e.g. \cite{Friedman:1997ih, Friedman:1997yq}.
Let $G$ be a semisimple group and let $\cM_G(E)$ be the moduli space of
semistable $G$-bundles on $E$.  Let $T\subset G$ be a maximal torus.  Then
the structure group of any $G$ bundle can be reduced to $T$, leading to a
description
\begin{equation}
\cM_G(E)\simeq \cM_T(E)/W,
\label{mge}
\end{equation}
where $W$ is the Weyl group of $G$. Now, if $\chi$ is a character of $T$ and
$\mathcal{P}$ is a principal $T$-bundle on $E$,
we get an induced line bundle $\mathcal{V}_\chi$ on $E$, giving rise to a homomorphism
\begin{equation}
\Lambda\to \mathrm{Pic}^0(E),\qquad \chi\mapsto \mathcal{V}_\chi,
\end{equation}
where $\Lambda$ is the lattice of characters of $T$, which we identify with
the weight lattice $\Lambda_G$ of $G$. The degree zero constraint follows from the condition that our
flux has vanishing first Chern class. Then, we can rewrite (\ref{mge}) as
\begin{equation}
\cM_G(E)=\mathrm{Hom}\left(\Lambda_G,\mathrm{Pic}^0(E)\right)/W.
\end{equation}

Now let $D$ be a $dP_8$ containing $E$ as an anticanonical divisor.  The
anticanonical linear system on a $dP_8$ has a unique base point $p_0$, which
we require to be the origin of the group structure on $E$. We sometimes blow
up $p_0$ to get a $dP_9$ with an elliptic fibration
containing $E$ as a fiber. Next, we define
\begin{equation}
H^2(D,\mathbb{Z})^\perp \equiv \left\{
\gamma\in H^2(D,\mathbb{Z})\mid \gamma\cdot E =0
\right\}.
\end{equation}
Then $H^2(D,\mathbb{Z})^\perp$ is isomorphic to $\Lambda_{E_8}$.  The isomorphism is
not canonical but is determined up to the action of $W(E_8)$ on $\Lambda_{E_8}$. Further, we see that
the pair $(D,E)$ determines an $E_8$ bundle on $E$ as follows.  Composing the
homomorphism
\begin{equation}
H^2(D,\mathbb{Z})\to \mathrm{Pic}^0(E),\qquad \gamma\mapsto \mathcal{O}_D(\gamma)|_E
\end{equation}
with the isomorphism $\Lambda_{E_8}\simeq H^2(D,\mathbb{Z})$, we get a homorphism
$\Lambda_{E_8}\to \mathrm{Pic}^0(E)$, which determines an $E_8$ bundle.  If we
modify the choice of isomorphism by an element of $W(E_8)$, we still get the
same $E_8$ bundle.

Conversely, given an $E_8$ bundle $\mathcal{V}$ on $E$, we can represent $\mathcal{V}$
by a $T$-bundle $\oplus_{i=1}^8\mathcal{O}_E(p_i-p_0)$.  We construct a $dP_8$ containing
$E$ as an anti-canonical section as follows.  First, embed
$E\hookrightarrow\mathbb{P}^2$ as a Weierstrass cubic using the linear system $|3p_0|$,
i.e.\ write $E$ in Weierstrass form with $p_0$ corresponding to $(x,y,z)=
(0,1,0)\in\mathbb{P}^2$.  Then blow up the points $p_1,\ldots,p_8$ inside $\mathbb{P}^2$
to obtain the desired $dP_8$, or blowup $p_0,p_1,\ldots,p_8$ to obtain a
$dP_9$ containing $E$ as a fiber.

We can similarly describe $SU(N)$ bundles
on $E$ as
\begin{equation}
\bigoplus_{i=1}^N\mathcal{O}_E(p_i-p_0),\qquad \sum p_i=p_0,
\label{sunbundle}
\end{equation}
where in (\ref{sunbundle}) the sum $\sum p_i$ denotes addition in the group
structure of $E$. So, when $\sum_{i=1}^Np_i=p_0$ and $p_{N+1}=\cdots=p_8=p_0$, then the structure
group of the corresponding $E_8$ bundle can be reduced to $SU(N)$.  More
generally, any $T$-bundle which can be brought to this form by an element
of the Weyl group has an associated $E_8$ bundle whose structure group can be
reduced to $SU(N)$. These two constructions are inverse to each other and identify $E_8$
bundles on $E$ with $dP_9$ fibrations containing $E$ as a fiber.

\subsection{An $SU(3)$ Example}\label{App:Su3}

In section \ref{sec:COMPACT} we explained how to embed the spectral curve associated to an
$SU(2)$ spectral curve into the geometry of the elliptic fibration $\pi: \mathcal{X}_{i} \rightarrow \mathbb{F}_{n}$
associated with the stable degeneration limit of an elliptic Calabi-Yau threefold $X = \mathcal{X}_{1} \cup \mathcal{X}_{2}$
In this subsection we perform a similar analysis for an $SU(3)$ spectral curve. As per our discussion in subsection \ref{ssec:PROJ}, we use a presentation of the
minimal Weierstrass model embedded in a $\mathbb{P}^2$ bundle over an appropriate base space.

To begin, we start on the heterotic side of the duality, associated with compactification of the heterotic
string on a K3 surface in the presence of a vector bundle $\mathcal{V}$ with
structure group $SU(3)$. Now, an $SU(3)$ bundle on an elliptic curve $E$ is described in terms of three points
$p_1, p_2, p_3\in E$ such that $p_1 + p_2 + p_3 = 0$.  If $E$ is in Weierstrass form,
this just means that the points $p_i$ are collinear.  Comparing to the $SU(2)$
case, we just have to replace (\ref{su2eqn}) with
\begin{equation}
f_{8+n} U +q_{6+n} V + g_{12+n}W = 0,
\label{su3eqn}
\end{equation}
where the relationship between the degrees of $f,q,g$ is determined by
(\ref{hetweights}). Note that with $Z\subset\mathbb{P}^1$ as before, we have
\begin{equation}
\begin{array}{ccc}
H^0(\mathbb{P}^1,2K+Z)&\simeq&H^0(\mathcal{O}(n+8))\\
H^0(\mathbb{P}^1,3K+Z)&\simeq&H^0(\mathcal{O}(n+6)),
\end{array}
\end{equation}
perfectly matching the parameters $f_{8+n}$ and $q_{6+n}$ in (\ref{su3eqn})
as in the $SU(2)$ case.  The spectral
cover (\ref{su3eqn}) is isomorphic to the corresponding
Hitchin spectral cover as can
be seen by comparing branch points.

Let us turn next to the F-theory description. To begin, recall that the Weierstrass equation is
\begin{equation}
y^2 \lambda=x^3+\left(f_{8 + n}{z_0}^3{z_1}+f_{8}{z_0}^4\right)x \lambda^2+\left(q_{6 + n}^2{z_0}^4 {z_1}^2+
g_{12 + n}{z_0}^5{z_1}+g_{12}
{z_0}^6\right) \lambda^3.
\label{fe6}
\end{equation}
On the F-theory side, to get the homogeneity required by (\ref{fweights}),
we must modify (\ref{su3eqn}) to
\begin{equation}
f_{8+n}x+q_{6+n}y+g_{12+n}{z_0}^2 \lambda = 0.
\label{su3fcurves}
\end{equation}
We uncomplete the square in $y$ to rewrite the Weierstrass equation in new coordinates as
\begin{equation}
y^2 \lambda + 2y \lambda^2 q_{6 + n}{z_0}^2 {z_1}=x^3+\left(f_{8 + n}{z_0}^3{z_1}+f_{8}{z_0}^4\right)x \lambda^2+\left(
g_{12 + n}{z_0}^5{z_1}+g_{12}
{z_0}^6\right) \lambda^3,
\end{equation}
Redefining $q$ by rescaling to take care of the factor of 2 and the sign to match the spectral cover,
we rewrite this as
\begin{equation}
y^2 \lambda =x^3+f_{8}{z_0}^4x \lambda^2 + g_{12}
{z_0}^6 \lambda^3+\left(f_{8+n}x{z_0}+q_{6+n}y+g_{12+n}{z_0}^3 \lambda \right){z_0}^2{z_1}\lambda^2
\end{equation}
so as in the $SU(2)$ case, the spectral cover parameterizes a family of
curves in the F-theory model, giving an Abel-Jacobi mapping
from the Jacobian of the spectral curve to $J(\mathcal{X})$.

We now tune the coefficient $q$ to 0. This forces the spectral curve to become
reducible, with the zero section of the heterotic K3 as a component.
We want to understand the limit of the Abel-Jacobi mapping as
$q\to0$. The two components of the spectral curve
intersect at $8 + n$ pairs of points where $f_{8 + n}=0$, and the $E_6$ enhances
to $E_7$.  The corresponding
curve in the F-theory model is the zero section over the fiber of $\mathbb{F}_n$
corresponding to a point of $\mathbb{P}^1$ at which $f_{8 + n}=0$.
Note that this curve intersects the $E_7$ locus by our explicit
parametrization of this curve.

We resolve the $E_6$ singularity and see what we are left with.
This can be done explicitly by resolving the $E_6$ singularity,
but there is an easier way: the $E_6$ or $E_7$ singularities live at
$x= y = z_0 =0$, but the zero section is at $x=\lambda=0$. Hence the curve is
disjoint from the singularity. By explicit calculation, the Hilbert scheme of these curves
is supported on the reducible spectral curve, and also contains embedded points
at the intersections.

\newpage

\bibliographystyle{utphys}
\bibliography{GlobalTbranes}

\providecommand{\href}[2]{#2}\begingroup\raggedright\begin{thebibliography}{10%
0}

\bibitem{VafaFTHEORY}
C.~Vafa, ``{Evidence for F-Theory},''
  \href{http://dx.doi.org/10.1016/0550-3213(96)00172-1}{{\em Nucl. Phys.}
  {\bfseries B469} (1996) 403--418},
\href{http://arxiv.org/abs/hep-th/9602022}{{\ttfamily arXiv:hep-th/9602022}}.

\bibitem{BHVI}
C.~Beasley, J.~J. Heckman, and C.~Vafa, ``{GUTs and Exceptional Branes in
  F-theory -- I},'' \href{http://dx.doi.org/10.1088/1126-6708/2009/01/058}{{\em
  JHEP} {\bfseries 01} (2009) 058},
\href{http://arxiv.org/abs/0802.3391}{{\ttfamily arXiv:0802.3391 [hep-th]}}.

\bibitem{BHVII}
C.~Beasley, J.~J. Heckman, and C.~Vafa, ``{GUTs and Exceptional Branes in
  F-theory -- II: Experimental Predictions},''
  \href{http://dx.doi.org/10.1088/1126-6708/2009/01/059}{{\em JHEP} {\bfseries
  01} (2009) 059},
\href{http://arxiv.org/abs/0806.0102}{{\ttfamily arXiv:0806.0102 [hep-th]}}.

\bibitem{DWI}
R.~Donagi and M.~Wijnholt, ``{Model Building with F-Theory},'' {\em Adv. Theor.
  Math. Phys.} {\bfseries 15} (2011) 1237--1318,
\href{http://arxiv.org/abs/0802.2969}{{\ttfamily arXiv:0802.2969 [hep-th]}}.

\bibitem{DWII}
R.~Donagi and M.~Wijnholt, ``{Breaking GUT Groups in F-Theory},'' {\em Adv.
  Theor. Math. Phys.} {\bfseries 15} (2011) 1523--1604,
\href{http://arxiv.org/abs/0808.2223}{{\ttfamily arXiv:0808.2223 [hep-th]}}.

\bibitem{HVCKM}
J.~J. Heckman and C.~Vafa, ``{Flavor Hierarchy From F-theory},''
  \href{http://dx.doi.org/10.1016/j.nuclphysb.2010.05.009}{{\em Nucl. Phys.}
  {\bfseries B837} (2010) 137--151},
\href{http://arxiv.org/abs/0811.2417}{{\ttfamily arXiv:0811.2417 [hep-th]}}.

\bibitem{Font:2008id}
A.~Font and L.~Ibanez, ``{Yukawa Structure from $U(1)$ Fluxes in F-theory Grand
  Unification},'' \href{http://dx.doi.org/10.1088/1126-6708/2009/02/016}{{\em
  JHEP} {\bfseries 0902} (2009) 016},
\href{http://arxiv.org/abs/0811.2157}{{\ttfamily arXiv:0811.2157 [hep-th]}}.

\bibitem{BHSV}
V.~Bouchard, J.~J. Heckman, J.~Seo, and C.~Vafa, ``{F-theory and Neutrinos:
  Kaluza-Klein Dilution of Flavor Hierarchy},'' {\em JHEP} {\bfseries 01}
  (2010) 061,
\href{http://arxiv.org/abs/0904.1419}{{\ttfamily arXiv:0904.1419 [hep-ph]}}.

\bibitem{Randall:2009dw}
L.~Randall and D.~Simmons-Duffin, ``{Quark and Lepton Flavor Physics from
  F-Theory},''
\href{http://arxiv.org/abs/0904.1584}{{\ttfamily arXiv:0904.1584 [hep-ph]}}.

\bibitem{HVCP}
J.~J. Heckman and C.~Vafa, ``{CP Violation and F-theory GUTs},''
  \href{http://dx.doi.org/10.1016/j.physletb.2010.10.034}{{\em Phys. Lett.}
  {\bfseries B694} (2011) 482--484},
\href{http://arxiv.org/abs/0904.3101}{{\ttfamily arXiv:0904.3101 [hep-th]}}.

\bibitem{Dudas:2009hu}
E.~Dudas and E.~Palti, ``{Froggatt-Nielsen models from $E_8$ in F-theory
  GUTs},'' {\em JHEP} {\bfseries 01} (2010) 127,
\href{http://arxiv.org/abs/arXiv:0912.0853 [hep-th]}{{\ttfamily arXiv:0912.0853
  [hep-th]}}.

\bibitem{FGUTSNC}
S.~Cecotti, M.~C.~N. Cheng, J.~J. Heckman, and C.~Vafa, ``{Yukawa Couplings in
  F-theory and Non-Commutative Geometry},'' {\em Surv. in Diff. Geom.}
  {\bfseries 15} (2010) ,
\href{http://arxiv.org/abs/0910.0477}{{\ttfamily arXiv:0910.0477 [hep-th]}}.

\bibitem{Marchesano:2009rz}
F.~Marchesano and L.~Martucci, ``{Non-perturbative effects on seven-brane
  Yukawa couplings},''
  \href{http://dx.doi.org/10.1103/PhysRevLett.104.231601}{{\em Phys. Rev.
  Lett.} {\bfseries 104} (2010) 231601},
\href{http://arxiv.org/abs/0910.5496}{{\ttfamily arXiv:0910.5496 [hep-th]}}.

\bibitem{Font:2012wq}
A.~Font, L.~E. Ibanez, F.~Marchesano, and D.~Regalado, ``{Non-perturbative
  effects and Yukawa hierarchies in F-theory $SU(5)$ Unification},''
  \href{http://dx.doi.org/10.1007/JHEP03(2013)140,
  10.1007/JHEP07(2013)036}{{\em JHEP} {\bfseries 1303} (2013) 140},
\href{http://arxiv.org/abs/1211.6529}{{\ttfamily arXiv:1211.6529 [hep-th]}}.

\bibitem{Pawelczyk:2013tza}
J.~Pawelczyk, ``{A F-GUT inspired model of Yukawa couplings with
  matter-messenger unification},''
\href{http://arxiv.org/abs/1305.5162}{{\ttfamily arXiv:1305.5162 [hep-ph]}}.

\bibitem{Font:2013ida}
A.~Font, F.~Marchesano, D.~Regalado, and G.~Zoccarato, ``{Up-type quark masses
  in $SU(5)$ F-theory models},''
\href{http://arxiv.org/abs/1307.8089}{{\ttfamily arXiv:1307.8089 [hep-th]}}.

\bibitem{HVLHC}
J.~J. Heckman and C.~Vafa, ``{From F-theory GUTs to the LHC},''
\href{http://arxiv.org/abs/0809.3452}{{\ttfamily arXiv:0809.3452 [hep-ph]}}.

\bibitem{Heckman:2010bq}
J.~J. Heckman, ``{Particle Physics Implications of F-theory},'' {\em Ann. Rev.
  Nuc. Part. Sci.} {\bfseries 60} (2010) 237,
\href{http://arxiv.org/abs/1001.0577}{{\ttfamily arXiv:1001.0577 [hep-th]}}.

\bibitem{Weigand:2010wm}
T.~Weigand, ``{Lectures on F-theory compactifications and model building},''
  {\em Class. Quant. Grav.} {\bfseries 27} (2010) 214004,
\href{http://arxiv.org/abs/arXiv:1009.3497 [hep-th]}{{\ttfamily arXiv:1009.3497
  [hep-th]}}.

\bibitem{Maharana:2012tu}
A.~Maharana and E.~Palti, ``{Models of Particle Physics from Type IIB String
  Theory and F-theory: A Review},''
  \href{http://dx.doi.org/10.1142/S0217751X13300056}{{\em Int. J. Mod. Phys.}
  {\bfseries A28} (2013) 1330005},
\href{http://arxiv.org/abs/1212.0555}{{\ttfamily arXiv:1212.0555 [hep-th]}}.

\bibitem{Wijnholt:2012fx}
M.~Wijnholt, ``{Higgs Bundles and String Phenomenology},''
\href{http://arxiv.org/abs/1201.2520}{{\ttfamily arXiv:1201.2520 [math.AG]}}.

\bibitem{Berenstein:2001nk}
D.~Berenstein, V.~Jejjala, and R.~G. Leigh, ``{The Standard model on a
  D-brane},'' \href{http://dx.doi.org/10.1103/PhysRevLett.88.071602}{{\em Phys.
  Rev. Lett.} {\bfseries 88} (2002) 071602},
\href{http://arxiv.org/abs/hep-ph/0105042}{{\ttfamily arXiv:hep-ph/0105042
  [hep-ph]}}.

\bibitem{Antoniadis:2002qm}
I.~Antoniadis, E.~Kiritsis, J.~Rizos, and T.~N. Tomaras, ``{D-branes and the
  standard model},''
  \href{http://dx.doi.org/10.1016/S0550-3213(03)00256-6}{{\em Nucl. Phys.}
  {\bfseries B660} (2003) 81--115},
\href{http://arxiv.org/abs/hep-th/0210263}{{\ttfamily arXiv:hep-th/0210263}}.

\bibitem{UrangaBottomUp}
G.~Aldazabal, L.~E. Ibanez, F.~Quevedo, and A.~M. Uranga, ``{D-branes at
  singularities: A bottom-up approach to the string embedding of the standard
  model},'' {\em JHEP} {\bfseries 08} (2000) 002,
\href{http://arxiv.org/abs/hep-th/0005067}{{\ttfamily arXiv:hep-th/0005067}}.

\bibitem{VerlindeWijnholtBottomUp}
H.~Verlinde and M.~Wijnholt, ``{Building the Standard Model on a D3-brane},''
  {\em JHEP} {\bfseries 01} (2007) 106,
\href{http://arxiv.org/abs/hep-th/0508089}{{\ttfamily arXiv:hep-th/0508089}}.

\bibitem{FUZZ}
J.~J. Heckman and H.~Verlinde, ``{Evidence for F(uzz) Theory},''
  \href{http://dx.doi.org/10.1007/JHEP01(2011)044}{{\em JHEP} {\bfseries 01}
  (2011) 044},
\href{http://arxiv.org/abs/1005.3033}{{\ttfamily arXiv:1005.3033 [hep-th]}}.

\bibitem{DWIII}
R.~Donagi and M.~Wijnholt, ``{Higgs Bundles and UV Completion in F-Theory},''
\href{http://arxiv.org/abs/0904.1218}{{\ttfamily arXiv:0904.1218 [hep-th]}}.

\bibitem{Collinucci:2008zs}
A.~Collinucci, ``{New F-theory lifts},''
  \href{http://dx.doi.org/10.1088/1126-6708/2009/08/076}{{\em JHEP} {\bfseries
  0908} (2009) 076},
\href{http://arxiv.org/abs/0812.0175}{{\ttfamily arXiv:0812.0175 [hep-th]}}.

\bibitem{Collinucci:2009uh}
A.~Collinucci, ``{New F-theory lifts II: Permutation orientifolds and enhanced
  singularities},'' {\em JHEP} {\bfseries 1004} (2010) 076,
\href{http://arxiv.org/abs/0906.0003}{{\ttfamily arXiv:0906.0003 [hep-th]}}.

\bibitem{Marsano:2009gv}
J.~Marsano, N.~Saulina, and S.~{Sch\"afer}-Nameki, ``{Monodromies, Fluxes, and
  Compact Three-Generation F-theory GUTs},'' {\em JHEP} {\bfseries 08} (2009)
  046,
\href{http://arxiv.org/abs/0906.4672}{{\ttfamily arXiv:0906.4672 [hep-th]}}.

\bibitem{Blumenhagen:2009up}
R.~Blumenhagen, T.~W. Grimm, B.~Jurke, and T.~Weigand, ``{F-theory uplifts and
  GUTs},'' \href{http://dx.doi.org/10.1088/1126-6708/2009/09/053}{{\em JHEP}
  {\bfseries 09} (2009) 053},
\href{http://arxiv.org/abs/0906.0013}{{\ttfamily arXiv:0906.0013 [hep-th]}}.

\bibitem{Blumenhagen:2009yv}
R.~Blumenhagen, T.~W. Grimm, B.~Jurke, and T.~Weigand, ``{Global F-theory
  GUTs},'' \href{http://dx.doi.org/10.1016/j.nuclphysb.2009.12.013}{{\em Nucl.
  Phys.} {\bfseries B829} (2010) 325--369},
\href{http://arxiv.org/abs/0908.1784}{{\ttfamily arXiv:0908.1784 [hep-th]}}.

\bibitem{Grimm:2009yu}
T.~W. Grimm, S.~Krause, and T.~Weigand, ``{F-Theory GUT Vacua on Compact
  Calabi-Yau Fourfolds},''
  \href{http://dx.doi.org/10.1007/JHEP07(2010)037}{{\em JHEP} {\bfseries 07}
  (2010) 037},
\href{http://arxiv.org/abs/0912.3524}{{\ttfamily arXiv:0912.3524 [hep-th]}}.

\bibitem{Marsano:2009wr}
J.~Marsano, N.~Saulina, and S.~{Sch\"afer}-Nameki, ``{Compact F-theory GUTs
  with $U(1)_{PQ}$},'' \href{http://dx.doi.org/10.1007/JHEP04(2010)095}{{\em
  JHEP} {\bfseries 04} (2010) 095},
\href{http://arxiv.org/abs/0912.0272}{{\ttfamily arXiv:0912.0272 [hep-th]}}.

\bibitem{Marsano:2010ix}
J.~Marsano, N.~Saulina, and S.~{Sch\"afer}-Nameki, ``{A Note on G-Fluxes for
  F-theory Model Building},''
  \href{http://dx.doi.org/10.1007/JHEP11(2010)088}{{\em JHEP} {\bfseries 11}
  (2010) 088},
\href{http://arxiv.org/abs/1006.0483}{{\ttfamily arXiv:1006.0483 [hep-th]}}.

\bibitem{Marsano:2011hv}
J.~Marsano and S.~Schafer-Nameki, ``{Yukawas, G-flux, and Spectral Covers from
  Resolved Calabi-Yau's},''
  \href{http://dx.doi.org/10.1007/JHEP11(2011)098}{{\em JHEP} {\bfseries 11}
  (2011) 098},
\href{http://arxiv.org/abs/1108.1794}{{\ttfamily arXiv:1108.1794 [hep-th]}}.

\bibitem{Esole:2011sm}
M.~Esole and S.-T. Yau, ``{Small resolutions of $SU(5)$-models in F-theory},''
\href{http://arxiv.org/abs/1107.0733}{{\ttfamily arXiv:1107.0733 [hep-th]}}.

\bibitem{Esole:2011cn}
M.~Esole, J.~Fullwood, and S.-T. Yau, ``{$D_5$ elliptic fibrations: non-Kodaira
  fibers and new orientifold limits of F-theory},''
\href{http://arxiv.org/abs/1110.6177}{{\ttfamily arXiv:1110.6177 [hep-th]}}.

\bibitem{Marsano:2012yc}
J.~Marsano, H.~Clemens, T.~Pantev, S.~Raby, and H.-H. Tseng, ``{A Global
  $SU(5)$ F-theory model with Wilson line breaking},''
  \href{http://dx.doi.org/10.1007/JHEP01(2013)150}{{\em JHEP} {\bfseries 01}
  (2013) 150},
\href{http://arxiv.org/abs/1206.6132}{{\ttfamily arXiv:1206.6132 [hep-th]}}.

\bibitem{Cvetic:2012ts}
M.~Cvetic, R.~Donagi, J.~Halverson, and J.~Marsano, ``{On Seven-Brane Dependent
  Instanton Prefactors in F-theory},''
  \href{http://dx.doi.org/10.1007/JHEP11(2012)004}{{\em JHEP} {\bfseries 1211}
  (2012) 004},
\href{http://arxiv.org/abs/1209.4906}{{\ttfamily arXiv:1209.4906 [hep-th]}}.

\bibitem{Marsano:2012bf}
J.~Marsano, N.~Saulina, and S.~Schafer-Nameki, ``{Global Gluing and G-flux},''
\href{http://arxiv.org/abs/1211.1097}{{\ttfamily arXiv:1211.1097 [hep-th]}}.

\bibitem{Donagi:2012ts}
R.~Donagi, S.~Katz, and M.~Wijnholt, ``{Weak Coupling, Degeneration and Log
  Calabi-Yau Spaces},''
\href{http://arxiv.org/abs/1212.0553}{{\ttfamily arXiv:1212.0553 [hep-th]}}.

\bibitem{Braun:2013cb}
A.~P. Braun and T.~Watari, ``{On Singular Fibres in F-Theory},''
  \href{http://dx.doi.org/10.1007/JHEP07(2013)031}{{\em JHEP} {\bfseries 07}
  (2013) 031},
\href{http://arxiv.org/abs/1301.5814}{{\ttfamily arXiv:1301.5814 [hep-th]}}.

\bibitem{Braun:2013yti}
V.~Braun, T.~W. Grimm, and J.~Keitel, ``{New Global F-theory GUTs with U(1)
  symmetries},''
\href{http://arxiv.org/abs/1302.1854}{{\ttfamily arXiv:1302.1854 [hep-th]}}.

\bibitem{Heckman:2013kza}
J.~J. Heckman, ``{Statistical Inference and String Theory},''
\href{http://arxiv.org/abs/1305.3621}{{\ttfamily arXiv:1305.3621 [hep-th]}}.

\bibitem{Hebecker:2013eba}
A.~Hebecker, ``{AdS/CFT for Accelerator Physics or Building the Tower of
  Babel},''
\href{http://arxiv.org/abs/1305.6311}{{\ttfamily arXiv:1305.6311 [hep-th]}}.

\bibitem{BershadskyFOURD}
M.~Bershadsky, A.~Johansen, T.~Pantev, and V.~Sadov, ``{On four-dimensional
  compactifications of F-theory},''
  \href{http://dx.doi.org/10.1016/S0550-3213(97)00393-3}{{\em Nucl. Phys.}
  {\bfseries B505} (1997) 165--201},
\href{http://arxiv.org/abs/hep-th/9701165}{{\ttfamily arXiv:hep-th/9701165}}.

\bibitem{Hayashi:2009ge}
H.~Hayashi, T.~Kawano, R.~Tatar, and T.~Watari, ``{Codimension-3 Singularities
  and Yukawa Couplings in F-theory},''
  \href{http://dx.doi.org/10.1016/j.nuclphysb.2009.07.021}{{\em Nucl. Phys.}
  {\bfseries B823} (2009) 47--115},
\href{http://arxiv.org/abs/0901.4941}{{\ttfamily arXiv:0901.4941 [hep-th]}}.

\bibitem{Hayashi:2010zp}
H.~Hayashi, T.~Kawano, Y.~Tsuchiya, and T.~Watari, ``{More on Dimension-4
  Proton Decay Problem in F-theory -- Spectral Surface, Discriminant Locus and
  Monodromy},''
\href{http://arxiv.org/abs/1004.3870}{{\ttfamily arXiv:1004.3870 [hep-th]}}.

\bibitem{EPOINT}
J.~J. Heckman, A.~Tavanfar, and C.~Vafa, ``{The Point of $E_8$ in F-theory
  GUTs},'' \href{http://dx.doi.org/10.1007/JHEP08(2010)040}{{\em JHEP}
  {\bfseries 08} (2010) 040},
\href{http://arxiv.org/abs/0906.0581}{{\ttfamily arXiv:0906.0581 [hep-th]}}.

\bibitem{TBRANES}
S.~Cecotti, C.~Cordova, J.~J. Heckman, and C.~Vafa, ``{T-Branes and
  Monodromy},'' \href{http://dx.doi.org/10.1007/JHEP07(2011)030}{{\em JHEP}
  {\bfseries 07} (2011) 030},
\href{http://arxiv.org/abs/1010.5780}{{\ttfamily arXiv:1010.5780 [hep-th]}}.

\bibitem{glueI}
R.~Donagi and M.~Wijnholt, ``{Gluing Branes, I},''
  \href{http://dx.doi.org/10.1007/JHEP05(2013)068}{{\em JHEP} {\bfseries 05}
  (2013) 068},
\href{http://arxiv.org/abs/1104.2610}{{\ttfamily arXiv:1104.2610 [hep-th]}}.

\bibitem{glueII}
R.~Donagi and M.~Wijnholt, ``{Gluing Branes II: Flavour Physics and String
  Duality},'' \href{http://dx.doi.org/10.1007/JHEP05(2013)092}{{\em JHEP}
  {\bfseries 05} (2013) 092},
\href{http://arxiv.org/abs/1112.4854}{{\ttfamily arXiv:1112.4854 [hep-th]}}.

\bibitem{Donagi:2003hh}
R.~Donagi, S.~Katz, and E.~Sharpe, ``{Spectra of D-branes with Higgs vevs},''
  {\em Adv. Theor. Math. Phys.} {\bfseries 8} (2005) 813--859,
\href{http://arxiv.org/abs/hep-th/0309270}{{\ttfamily arXiv:hep-th/0309270}}.

\bibitem{Chiou:2011js}
C.-C. Chiou, A.~E. Faraggi, R.~Tatar, and W.~Walters, ``{T-branes and Yukawa
  Couplings},'' \href{http://dx.doi.org/10.1007/JHEP05(2011)023}{{\em JHEP}
  {\bfseries 05} (2011) 023},
\href{http://arxiv.org/abs/1101.2455}{{\ttfamily arXiv:1101.2455 [hep-th]}}.

\bibitem{Donagi:1998vw}
R.~Donagi, ``{Heterotic / F theory duality: ICMP lecture},'' {\em Math. Phys.
  Proc., 12th Inter. Congress, ICMP '97 Brisbane Australia} (1997) 206--213,
  \href{http://arxiv.org/abs/hep-th/9802093}{{\ttfamily arXiv:hep-th/9802093}}.

\bibitem{MorrisonVafaI}
D.~R. Morrison and C.~Vafa, ``{Compactifications of F-Theory on Calabi--Yau
  Threefolds -- I},''
  \href{http://dx.doi.org/10.1016/0550-3213(96)00242-8}{{\em Nucl. Phys.}
  {\bfseries B473} (1996) 74--92},
\href{http://arxiv.org/abs/hep-th/9602114}{{\ttfamily arXiv:hep-th/9602114}}.

\bibitem{MorrisonVafaII}
D.~R. Morrison and C.~Vafa, ``{Compactifications of F-Theory on Calabi--Yau
  Threefolds -- II},''
  \href{http://dx.doi.org/10.1016/0550-3213(96)00369-0}{{\em Nucl. Phys.}
  {\bfseries B476} (1996) 437--469},
\href{http://arxiv.org/abs/hep-th/9603161}{{\ttfamily arXiv:hep-th/9603161}}.

\bibitem{BershadskyPLUS}
M.~Bershadsky, K.~Intriligator, S.~Kachru, D.~Morrison, V.~Sadov, and C.~Vafa,
  ``{Geometric singularities and enhanced gauge symmetries},''
  \href{http://dx.doi.org/10.1016/S0550-3213(96)90131-5}{{\em Nucl. Phys.}
  {\bfseries B481} (1996) 215--252},
\href{http://arxiv.org/abs/hep-th/9605200}{{\ttfamily arXiv:hep-th/9605200}}.

\bibitem{Aspinwall:1998he}
P.~S. Aspinwall and R.~Y. Donagi, ``{The Heterotic String, The Tangent Bundle,
  and Derived Categories},'' {\em Adv. Theor. Math. Phys.} {\bfseries 2} (1998)
  1041--1074,
\href{http://arxiv.org/abs/hep-th/9806094}{{\ttfamily arXiv:hep-th/9806094}}.

\bibitem{Bershadsky:1997zv}
M.~Bershadsky, T.~Chiang, B.~R. Greene, A.~Johansen, and C.~Lazaroiu,
  ``{$F$-theory and linear sigma models},''
  \href{http://dx.doi.org/10.1016/S0550-3213(98)00429-5}{{\em Nucl.Phys.}
  {\bfseries B527} (1998) 531--570},
\href{http://arxiv.org/abs/hep-th/9712023}{{\ttfamily arXiv:hep-th/9712023}}.

\bibitem{Anderson:2007nc}
L.~B. Anderson, Y.-H. He, and A.~Lukas, ``{Heterotic Compactification, An
  Algorithmic Approach},''
  \href{http://dx.doi.org/10.1088/1126-6708/2007/07/049}{{\em JHEP} {\bfseries
  07} (2007) 049},
\href{http://arxiv.org/abs/hep-th/0702210}{{\ttfamily arXiv:hep-th/0702210}}.

\bibitem{Anderson:2008uw}
L.~B. Anderson, Y.-H. He, and A.~Lukas, ``{Monad Bundles in Heterotic String
  Compactifications},''
  \href{http://dx.doi.org/10.1088/1126-6708/2008/07/104}{{\em JHEP} {\bfseries
  07} (2008) 104},
\href{http://arxiv.org/abs/0805.2875}{{\ttfamily arXiv:0805.2875 [hep-th]}}.

\bibitem{Anderson:2009mh}
L.~B. Anderson, J.~Gray, Y.-H. He, and A.~Lukas, ``{Exploring Positive Monad
  Bundles And A New Heterotic Standard Model},''
  \href{http://dx.doi.org/10.1007/JHEP02(2010)054}{{\em JHEP} {\bfseries 1002}
  (2010) 054},
\href{http://arxiv.org/abs/0911.1569}{{\ttfamily arXiv:0911.1569 [hep-th]}}.

\bibitem{Alexandrov:2013yva}
S.~Alexandrov, J.~Manschot, D.~Persson, and B.~Pioline, ``{Quantum
  Hypermultiplet Moduli Spaces in $\mathcal{N} = 2$ String Vacua: A Review},''
\href{http://arxiv.org/abs/1304.0766}{{\ttfamily arXiv:1304.0766 [hep-th]}}.

\bibitem{HitchinSelf}
N.~J. Hitchin, ``The self-duality equations on a riemann surface,'' {\em Proc.
  Math. Soc.} {\bfseries (3) 55} (1987) 59.

\bibitem{Funparticles}
J.~J. Heckman and C.~Vafa, ``{An Exceptional Sector for F-theory GUTs},''
  \href{http://dx.doi.org/10.1103/PhysRevD.83.026006}{{\em Phys. Rev.}
  {\bfseries D83} (2011) 026006},
\href{http://arxiv.org/abs/1006.5459}{{\ttfamily arXiv:1006.5459 [hep-th]}}.

\bibitem{FCFT}
J.~J. Heckman, Y.~Tachikawa, C.~Vafa, and B.~Wecht, ``{$\mathcal{N} = 1$ SCFTs
  from Brane Monodromy},''
  \href{http://dx.doi.org/10.1007/JHEP11(2010)132}{{\em JHEP} {\bfseries 11}
  (2010) 132},
\href{http://arxiv.org/abs/1009.0017}{{\ttfamily arXiv:1009.0017 [hep-th]}}.

\bibitem{D3gen}
J.~J. Heckman and S.-J. Rey, ``{Baryon and Dark Matter Genesis from Strongly
  Coupled Strings},'' \href{http://dx.doi.org/10.1007/JHEP06(2011)120}{{\em
  JHEP} {\bfseries 06} (2011) 120},
\href{http://arxiv.org/abs/1102.5346}{{\ttfamily arXiv:1102.5346 [hep-th]}}.

\bibitem{HVW}
J.~J. Heckman, C.~Vafa, and B.~Wecht, ``{The Conformal Sector of F-theory
  GUTs},'' \href{http://dx.doi.org/10.1007/JHEP07(2011)075}{{\em JHEP}
  {\bfseries 07} (2011) 075}, \href{http://arxiv.org/abs/1103.3287}{{\ttfamily
  arXiv:1103.3287 [hep-th]}}.

\bibitem{DSSM}
J.~J. Heckman, P.~Kumar, C.~Vafa, and B.~Wecht, ``{Electroweak Symmetry
  Breaking in the DSSM},''
  \href{http://dx.doi.org/10.1007/JHEP01(2012)156}{{\em JHEP} {\bfseries 01}
  (2012) 156},
\href{http://arxiv.org/abs/1108.3849}{{\ttfamily arXiv:1108.3849 [hep-ph]}}.

\bibitem{HoloHiggs}
J.~J. Heckman, P.~Kumar, and B.~Wecht, ``{The Higgs as a Probe of
  Supersymmetric Extra Sectors},''
  \href{http://dx.doi.org/10.1007/JHEP07(2012)118}{{\em JHEP} {\bfseries 07}
  (2012) 118},
\href{http://arxiv.org/abs/1204.3640}{{\ttfamily arXiv:1204.3640 [hep-ph]}}.

\bibitem{Heckman:2012jm}
J.~J. Heckman, P.~Kumar, and B.~Wecht, ``{$S$ and $T$ for SCFTs},''
  \href{http://dx.doi.org/10.1103/PhysRevD.88.065016}{{\em Phys. Rev.}
  {\bfseries D88} (2013) 065016},
\href{http://arxiv.org/abs/1212.2979}{{\ttfamily arXiv:1212.2979 [hep-th]}}.

\bibitem{Katz:1996ht}
S.~H. Katz, D.~R. Morrison, and M.~R. Plesser, ``{Enhanced Gauge Symmetry in
  Type II String Theory},''
  \href{http://dx.doi.org/10.1016/0550-3213(96)00331-8}{{\em Nucl. Phys.}
  {\bfseries B477} (1996) 105--140},
\href{http://arxiv.org/abs/hep-th/9601108}{{\ttfamily arXiv:hep-th/9601108}}.

\bibitem{Argyres:1996eh}
P.~C. Argyres, M.~R. Plesser, and N.~Seiberg, ``{The Moduli Space of
  $\mathcal{N} = 2$ SUSY QCD and Duality in $\mathcal{N} = 1$ SUSY QCD},''
  \href{http://dx.doi.org/10.1016/0550-3213(96)00210-6}{{\em Nucl. Phys.}
  {\bfseries B471} (1996) 159--194},
\href{http://arxiv.org/abs/hep-th/9603042}{{\ttfamily arXiv:hep-th/9603042}}.

\bibitem{Diaconescu:2005jw}
D.-E. Diaconescu, R.~Donagi, R.~Dijkgraaf, C.~Hofman, and T.~Pantev,
  ``{Geometric transitions and integrable systems},''
  \href{http://dx.doi.org/10.1016/j.nuclphysb.2006.04.016}{{\em Nucl. Phys.}
  {\bfseries B752} (2006) 329--390},
\href{http://arxiv.org/abs/hep-th/0506196}{{\ttfamily arXiv:hep-th/0506196}}.

\bibitem{Diaconescu:2006ry}
D.-E. Diaconescu, R.~Donagi, and T.~Pantev, ``{Intermediate Jacobians and ADE
  Hitchin Systems},''
\href{http://arxiv.org/abs/hep-th/0607159}{{\ttfamily arXiv:hep-th/0607159}}.

\bibitem{Clemens}
H.~C. Clemens, ``{Double Solids},'' {\em Adv. in Math} {\bfseries 47} (1983)
  107--230.

\bibitem{Friedman}
R.~Friedman, ``{Simultaneous Resolution of Threefold Double Points},'' {\em
  Math. Ann.} {\bfseries 274} (1986) 671--689.

\bibitem{Tian}
G.~Tian, {\em Smoothing $3$-folds with trivial canonical bundle and ordinary
  double points in ``Essays on Mirror Manifolds''}.
\newblock Internat. Press, Hong Kong, 1992.

\bibitem{del}
R.~Donagi, L.~Ein, and R.~Lazarsfeld, ``{Nilpotent cones and sheaves on $K3$
  surfaces},'' {\em Contemp. Math.} {\bfseries 207} (1997) 51--61.

\bibitem{KatzMorrison}
S.~Katz and D.~R. Morrison, ``{Gorenstein Threefold Singularities with Small
  Resolutions via Invariant Theory for Weyl Groups},'' {\em J.Alg.Geom.}
  {\bfseries 1} (1992) 449,
  \href{http://arxiv.org/abs/alg-geom/9202002}{{\ttfamily
  arXiv:alg-geom/9202002}}.

\bibitem{KatzVafa}
S.~H. Katz and C.~Vafa, ``{Matter from geometry},''
  \href{http://dx.doi.org/10.1016/S0550-3213(97)00280-0}{{\em Nucl. Phys.}
  {\bfseries B497} (1997) 146--154},
\href{http://arxiv.org/abs/hep-th/9606086}{{\ttfamily arXiv:hep-th/9606086}}.

\bibitem{ParabolicHiggs}
O.~Garcia-Prada, P.~B. Gothen, and V.~Mu\~{n}oz, {\em Betti Numbers of the
  Moduli Space of Rank $3$ Parabolic Higgs Bundles}, vol.~Memoirs of the Amer.
  Math. Soc. Vol. 187 No. 879.
\newblock Amer. Math. Soc., Providence, RI, 2007.
\newblock \href{http://arxiv.org/abs/math/0411242}{{\ttfamily math/0411242}}.

\bibitem{Becker:1996gj}
K.~Becker and M.~Becker, ``{$\mathcal{M}$-theory on Eight-Manifolds},''
  \href{http://dx.doi.org/10.1016/0550-3213(96)00367-7}{{\em Nucl. Phys.}
  {\bfseries B477} (1996) 155--167},
\href{http://arxiv.org/abs/hep-th/9605053}{{\ttfamily arXiv:hep-th/9605053}}.

\bibitem{Witten:1996md}
E.~Witten, ``{On Flux Quantization in $M$-Theory and the Effective Action},''
  \href{http://dx.doi.org/10.1016/S0393-0440(96)00042-3}{{\em J. Geom. Phys.}
  {\bfseries 22} (1997) 1--13},
\href{http://arxiv.org/abs/hep-th/9609122}{{\ttfamily arXiv:hep-th/9609122}}.

\bibitem{Grimm:2010ez}
T.~W. Grimm and T.~Weigand, ``{On Abelian Gauge Symmetries and Proton Decay in
  Global F-theory GUTs},''
  \href{http://dx.doi.org/10.1103/PhysRevD.82.086009}{{\em Phys. Rev.}
  {\bfseries D82} (2010) 086009},
\href{http://arxiv.org/abs/1006.0226}{{\ttfamily arXiv:1006.0226 [hep-th]}}.

\bibitem{Grimm:2011tb}
T.~W. Grimm, M.~Kerstan, E.~Palti, and T.~Weigand, ``{Massive Abelian Gauge
  Symmetries and Fluxes in F-theory},''
  \href{http://dx.doi.org/10.1007/JHEP12(2011)004}{{\em JHEP} {\bfseries 12}
  (2011) 004},
\href{http://arxiv.org/abs/1107.3842}{{\ttfamily arXiv:1107.3842 [hep-th]}}.

\bibitem{Grassi:2013kha}
A.~Grassi, J.~Halverson, and J.~L. Shaneson, ``{Matter From Geometry Without
  Resolution},'' \href{http://dx.doi.org/10.1007/JHEP10(2013)205}{{\em JHEP}
  {\bfseries 1310} (2013) 205},
\href{http://arxiv.org/abs/1306.1832}{{\ttfamily arXiv:1306.1832 [hep-th]}}.

\bibitem{Heckman:2013sfa}
J.~J. Heckman, H.~Lin, and S.-T. Yau, ``{Building Blocks for Generalized
  Heterotic/F-theory Duality},''
\href{http://arxiv.org/abs/1311.6477}{{\ttfamily arXiv:1311.6477 [hep-th]}}.

\bibitem{Friedman:1997yq}
R.~Friedman, J.~Morgan, and E.~Witten, ``{Vector Bundles and $F$ Theory},''
  \href{http://dx.doi.org/10.1007/s002200050154}{{\em Comm. Math. Phys.}
  {\bfseries 187} (1997) 679--743},
\href{http://arxiv.org/abs/hep-th/9701162}{{\ttfamily arXiv:hep-th/9701162}}.

\bibitem{Friedman:1997ih}
R.~Friedman, J.~W. Morgan, and E.~Witten, ``{Vector Bundles Over Elliptic
  Fibrations},''
\href{http://arxiv.org/abs/alg-geom/9709029}{{\ttfamily
  arXiv:alg-geom/9709029}}.

\bibitem{Curio:1998bva}
G.~Curio and R.~Y. Donagi, ``{Moduli in $N = 1$ heterotic / F-theory
  duality},'' \href{http://dx.doi.org/10.1016/S0550-3213(98)00185-0}{{\em Nucl.
  Phys.} {\bfseries B518} (1998) 603--631},
\href{http://arxiv.org/abs/hep-th/9801057}{{\ttfamily arXiv:hep-th/9801057}}.

\bibitem{Aspinwall:1997ye}
P.~S. Aspinwall and D.~R. Morrison, ``{Point-like instantons on K3
  orbifolds},'' \href{http://dx.doi.org/10.1016/S0550-3213(97)00516-6}{{\em
  Nucl.Phys.} {\bfseries B503} (1997) 533--564},
\href{http://arxiv.org/abs/hep-th/9705104}{{\ttfamily arXiv:hep-th/9705104}}.

\bibitem{Markman}
E.~Markman, ``{Spectral Curves and Integrable Systems},''
  \href{http://dx.doi.org/10.1007/JHEP11(2012)004}{{\em Comp. Math.} {\bfseries
  93} (1994) 255--290}.

\bibitem{Donagi:1995am}
R.~Donagi and E.~Markman, ``{Spectral curves, algebraically completely
  integrable Hamiltonian systems, and moduli of bundles},''
\href{http://arxiv.org/abs/alg-geom/9507017}{{\ttfamily
  arXiv:alg-geom/9507017}}.

\bibitem{SimpsonMartin}
A.~C. L\'{o}pez~Mart\'{i}n, ``{Simpson Jacobians of Reducible Curves},'' {\em
  J. Fur Die Reine und Ang. Math.} {\bfseries 582} (2005) 1--39,
\href{http://arxiv.org/abs/math/041039}{{\ttfamily arXiv:math/041039}}.

\bibitem{WatariTATARHETF}
H.~Hayashi, R.~Tatar, Y.~Toda, T.~Watari, and M.~Yamazaki, ``{New Aspects of
  Heterotic--F Theory Duality},''
  \href{http://dx.doi.org/10.1016/j.nuclphysb.2008.07.031}{{\em Nucl. Phys.}
  {\bfseries B806} (2009) 224--299},
\href{http://arxiv.org/abs/0805.1057}{{\ttfamily arXiv:0805.1057 [hep-th]}}.

\bibitem{Aspinwall:1998bw}
P.~S. Aspinwall, ``{Aspects of the hypermultiplet moduli space in string
  duality},'' {\em JHEP} {\bfseries 9804} (1998) 019,
\href{http://arxiv.org/abs/hep-th/9802194}{{\ttfamily arXiv:hep-th/9802194}}.

\bibitem{0853.14016}
D.~Bayer and D.~Eisenbud, ``{Ribbons and their canonical embeddings},'' {\em
  Trans. Am. Math. Soc.} {\bfseries 347} (1995) 719--756.

\bibitem{huybrechts2010geometry}
D.~Huybrechts and M.~Lehn, {\em The Geometry of Moduli Spaces of Sheaves}.
\newblock Cambridge Mathematical Library. Cambridge University Press,
  Cambridge, 2010.

\bibitem{Lukas:1999nh}
A.~Lukas and K.~Stelle, ``{Heterotic anomaly cancellation in
  five-dimensions},'' {\em JHEP} {\bfseries 01} (2000) 010,
\href{http://arxiv.org/abs/hep-th/9911156}{{\ttfamily arXiv:hep-th/9911156}}.

\bibitem{Anderson:2009nt}
L.~B. Anderson, J.~Gray, A.~Lukas, and B.~Ovrut, ``{Stability Walls in
  Heterotic Theories},''
  \href{http://dx.doi.org/10.1088/1126-6708/2009/09/026}{{\em JHEP} {\bfseries
  09} (2009) 026},
\href{http://arxiv.org/abs/0905.1748}{{\ttfamily arXiv:0905.1748 [hep-th]}}.

\bibitem{Louis:2011hp}
J.~Louis, M.~Schasny, and R.~Valandro, ``{6D Effective Action of Heterotic
  Compactification on K3 with Nontrivial Gauge Bundles},''
  \href{http://dx.doi.org/10.1007/JHEP04(2012)028}{{\em JHEP} {\bfseries 1204}
  (2012) 028},
\href{http://arxiv.org/abs/1112.5106}{{\ttfamily arXiv:1112.5106 [hep-th]}}.

\bibitem{Morrison:2012np}
D.~R. Morrison and W.~Taylor, ``{Classifying bases for 6D F-theory models},''
  \href{http://dx.doi.org/10.2478/s11534-012-0065-4}{{\em Centr. Eur. J. Phys.}
  {\bfseries 10} (2012) 1072--1088},
\href{http://arxiv.org/abs/1201.1943}{{\ttfamily arXiv:1201.1943 [hep-th]}}.

\bibitem{VafaWitten}
C.~Vafa and E.~Witten, ``{A Strong coupling test of S duality},''
  \href{http://dx.doi.org/10.1016/0550-3213(94)90097-3}{{\em Nucl. Phys.}
  {\bfseries B431} (1994) 3--77},
\href{http://arxiv.org/abs/hep-th/9408074}{{\ttfamily arXiv:hep-th/9408074}}.

\bibitem{landman}
A.~Landman, ``{On the Picard-Lefschetz transformation for algebraic manifolds
  acquiring general singularities},'' {\em Trans. AMS} {\bfseries 181} (1973)
  89--126.

\bibitem{morrisoncs}
D.~R. Morrison, {\em The Clemens-Schmid exact sequence and applications, in
  ``Topics in Transcendental algebraic geometry'' (Ann. Math. Stud.)},
  vol.~106.
\newblock Princeton Univ. Press, Princeton, 1984.

\bibitem{carlsonext}
J.~Carlson, {\em Extensions of mixed Hodge structures, in ``Algebraic Geometry,
  (Angers 1979)''}.
\newblock Sijthoff \& Noordhoff, Princeton, 1980.

\bibitem{kmptoappear}
S.~Katz, D.~R. Morrison, and M.~R. Plesser, ``to appear,''.

\bibitem{DeligneReview}
H.~Ensault and E.~Viehweg, ``{Deligne-Beilinson Cohomology},'' {\em preprint} .

\bibitem{GriffithsI}
P.~A. Griffiths, ``{Periods of integrals on algebraic manifolds I. Construction
  and properties of the modular varieties},'' {\em AJM} {\bfseries 90 (2)}
  (1968) 281--356.

\bibitem{GriffithsII}
P.~A. Griffiths, ``{Periods of integrals on algebraic manifolds II.
  Construction and properties of the modular varieties},'' {\em AJM} {\bfseries
  90 (3)} (1968) 805--865.

\bibitem{GriffithsClemens}
C.~H. Clemens and P.~A. Griffiths, ``{The Intermediate Jacobian of the Cubic
  Threefold},'' {\em Ann. of Math.} {\bfseries 95} (1972) 281--356.

\bibitem{GAJER}
P.~Gajer, ``{Geometry of Deligne Cohomology},'' {\em Invent. Math.} {\bfseries
  127 (1)} (1997) 155--207,
  \href{http://arxiv.org/abs/alg-geom/9601025}{{\ttfamily alg-geom/9601025}}.

\end{thebibliography}\endgroup

\end{document}